\documentclass{article}
\PassOptionsToPackage{numbers,compress}{natbib}
\usepackage[preprint]{neurips_2026}

\usepackage[T1]{fontenc}
\usepackage[utf8]{inputenc}
\usepackage[english]{babel}

\usepackage{microtype}
\usepackage{setspace}
\usepackage{lscape}
\usepackage{float}

\usepackage{amsmath}
\usepackage{amssymb}
\usepackage{mathtools}
\usepackage{amsthm}
\usepackage{bm}

\usepackage{makecell}
\usepackage{booktabs}
\usepackage{longtable}
\usepackage{multirow}
\usepackage{array}
\usepackage{tabularx}
\usepackage{siunitx}
\usepackage{wrapfig}

\usepackage{graphicx}
\usepackage{subcaption}

\usepackage{enumitem}
\usepackage{tcolorbox}
\usepackage{dashrule}
\usepackage{etoolbox}
\usepackage{lipsum}
\usepackage{pifont}

\tcbuselibrary{breakable}
\tcbuselibrary{skins,breakable}

\usepackage{enumitem}
\usepackage[dvipsnames]{xcolor}
\usepackage{hyperref}
\usepackage[capitalize,noabbrev]{cleveref}

\hypersetup{
    colorlinks=true,
    linkcolor=red!50!black,
    urlcolor=linkpurple,
    citecolor=green!50!black 
}

\definecolor{Flagging}{HTML}{D97C7C}
\definecolor{Counts}{HTML}{8BB3E6}
\definecolor{Extraction}{HTML}{8FD3B0}
\definecolor{linkgrey}{RGB}{90,90,90}
\definecolor{linkpurple}{RGB}{102,45,145}
\definecolor{stdgray}{gray}{0.55}

\newcommand{\titlelink}[3]{%
  \href{#1}{%
    {\color{linkpurple}%
    \raisebox{-0.2\height}{\includegraphics[height=10pt]{#2}}%
    \hspace{8pt}\texttt{#3}}%
  }%
}

\definecolor{OliveGreenDark}{HTML}{556B2F}
\definecolor{RedDark}{HTML}{8B1A1A}
\definecolor{flagred}{RGB}{214,39,40}
\definecolor{flagblue}{RGB}{31,119,180}
\definecolor{flaggreen}{RGB}{44,160,44}

\usepackage{xcolor}
\definecolor{myred}{RGB}{200,0,0}

\usepackage{etoc}
\usepackage[toc,page,header]{appendix}

\renewcommand{\thepart}{}
\renewcommand{\partname}{}

\usepackage{enumitem}

\usepackage[textsize=tiny]{todonotes}

\usepackage{listings}
\usepackage{minted}

\let\oldaddcontentsline\addcontentsline

\let\addcontentsline\oldaddcontentsline

\newcommand{\published}{\textcolor{OliveGreen!75}{\Large$\bullet$}}

\newcommand{\extractionphase}{\textcolor{YellowOrange!75}{\Large$\bullet$}}
\newcommand{\notstarted}{\textcolor{Bittersweet!75}{\Large$\bullet$}}

\newcommand{\deflagging}{\textcolor{Flagging}{\Large$\bullet$}}
\newcommand{\decounts}{\textcolor{Counts}{\Large$\bullet$}}
\newcommand{\deextraction}{\textcolor{Extraction}{\Large$\bullet$}}

\newcommand{\bestM}[1]{\textbf{#1}}       
\newcommand{\secondM}[1]{\underline{#1}}  
\newcommand{\highP}[1]{{\color{OliveGreen}#1}} 
\newcommand{\lowP}[1]{{\color{red!70!black}#1}} 

\newcommand{\IN}{\textcolor{OliveGreenDark}{\textbf{\checkmark}}}
\newcommand{\EX}{\textcolor{RedDark}{\textbf{\texttimes}}}

\newcommand{\name}{AgentSLR}

\newcommand{\dashtext}[2][\linewidth]{%
  \par\medskip
  \noindent
  \makebox[#1]{%
    \hdashrule{0.45\linewidth}{0.4pt}{3pt 3pt}%
    \hspace{0.6em}%
    \textit{#2}%
    \hspace{0.6em}%
    \hdashrule{0.45\linewidth}{0.4pt}{3pt 3pt}%
  }%
  \medskip\par
}

\theoremstyle{plain}

\theoremstyle{definition}

\theoremstyle{remark}

\title{Evaluating AI-based Scientific Knowledge Synthesis with Epidemiological Systematic Reviews}

\author{%
\textbf{Shreyansh Padarha}$^{1,}$\thanks{%
Equal contribution. \\ Correspondence to:
\texttt{shreyansh.padarha@oii.ox.ac.uk},
\texttt{adam.mahdi@oii.ox.ac.uk}.} \quad
\textbf{Ryan Othniel Kearns}$^{1,*}$ \quad
\textbf{Tristan Naidoo}$^{2}$ \\
\textbf{Lingyi Yang}$^{3}$ \quad
\textbf{{\L}ukasz Borchmann}$^{4}$ \quad
\textbf{Piotr B{\l}aszczyk}$^{5}$ \\
\textbf{Christian Morgenstern}$^{2}$ \quad
\textbf{Ruth McCabe}$^{2}$ \quad
\textbf{Sangeeta Bhatia}$^{2}$\\
\textbf{Philip H.~Torr}$^{1}$ \quad
\textbf{Jakob Foerster}$^{1}$ \quad
\textbf{Scott A.~Hale}$^{1}$\\
\textbf{Thomas Rawson}$^{1}$\quad
\textbf{Anne Cori}$^{2}$ \quad
\textbf{Elizaveta Semenova}$^{2}$ \quad
\textbf{Adam Mahdi}$^{1}$\\[0.6em]
\normalfont
$^{1}$University of Oxford \quad
$^{2}$Imperial College London \\
$^{3}$University of Nottingham\quad
$^{4}$Snowflake AI Research \quad
$^{5}$Independent
}

\begin{document}

\maketitle

\vspace{-1.5em}

\begin{abstract}
Systematic literature reviews (SLRs) are a demanding and high-stakes form of scientific knowledge synthesis that remains underspecified as an evaluation setting for large language models (LLMs). We introduce \textit{\name{}}, a large-scale evaluation harness comprising an SLR automation workflow and an expert annotated dataset covering $16{,}248$ articles, designed to test LLM capabilities across the stages of SLRs in epidemiology. Reference annotations were derived from peer-reviewed studies on WHO priority pathogens and produced by domain experts. The harness evaluates each review stage as a separate unit with dedicated metrics enabling targeted failure analysis. We evaluated five frontier reasoning models and found that no single model dominated across all tasks, showing sub-task specialisation often hidden by aggregate benchmarks. Structured data extraction is a major bottleneck, with no model exceeding an average field-level $F1$ of $0.67$. Estimated costs vary substantially, by up to $96\times$ across evaluated models.  Documented failure modes suggest that the evaluated models are not yet reliable enough for unsupervised deployment in epidemiology, where findings can inform public policy.
\end{abstract}

\begin{center}
\vspace{-0.5em}
\small
\titlelink{https://oxrml.com/agent-slr}
          {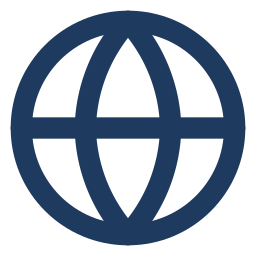}
          {oxrml.com/agent-slr}
\quad\quad
\titlelink{https://huggingface.co/datasets/OxRML/AgentSLR}
          {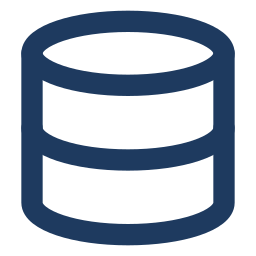}
          {Dataset}
\quad\quad
\titlelink{https://github.com/OxRML/AgentSLR}
          {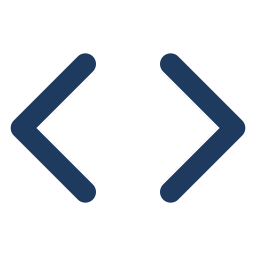}
          {Harness Code}
\vspace{-0.5em}
\end{center}

\section{Introduction}
Large language models (LLMs) now answer expert-level scientific questions \cite{reinGPQAGraduatelevelGoogleproof2024}, reason over extended research tasks \cite{kwaMeasuringAIAbility2025} and interpret scientific figures~\cite{robertsSciFIBenchBenchmarkingLarge2024}. 
These capabilities motivate applications beyond isolated question answering, towards scientific workflows where evidence must be retrieved, judged, organised and interpreted under domain constraints \cite{wangScientificDiscoveryAge2023, zhang2025exploring, lu2024aiscientist}. Systematic literature reviews (SLRs) are among the most important examples of such workflows. They require screening, extraction and synthesis of thousands of scientific articles.

LLM-based evidence synthesis workflows show benefits, but their reliability remains limited. Models can reduce the time required for title and abstract screening~\citep{oami2024performance}, while still producing false inclusions and false exclusions that require expert review. Beyond screening, LLM summaries can omit details that limit the scope of study findings, making conclusions appear more general than the source article support~\citep{peters2025generalization}. These problems get worse when models are chained together in longer workflows, where errors compound and result in failures \citep{pan2025multiagent}.  

\begin{figure*}[t!]
\centering
\includegraphics[width=\linewidth]{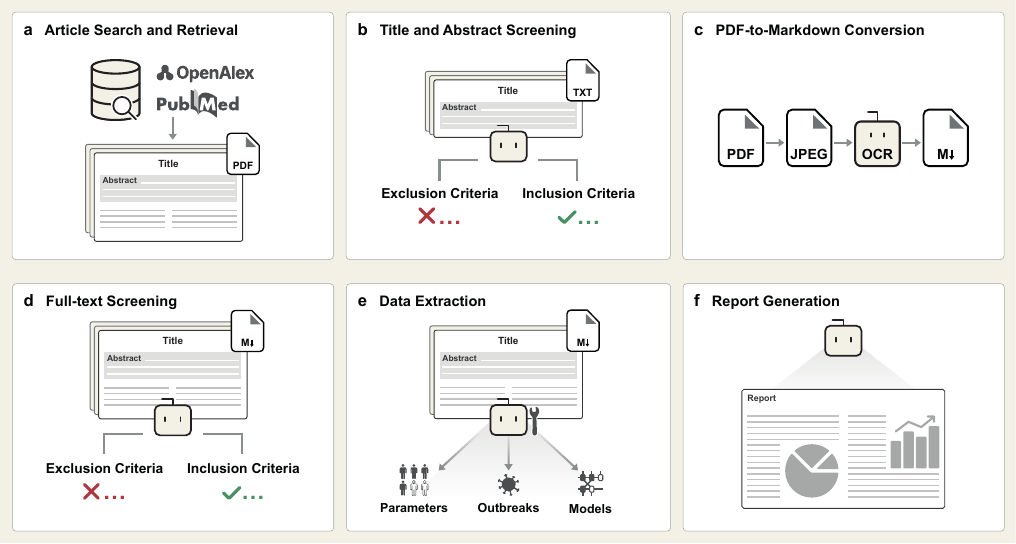}
\caption{\textbf{\name{} evaluation harness (LLM-assisted workflow).} (a) \textit{Article Search and Retrieval} queries bibliographic databases with domain-specific Boolean searches and obtains PDFs from open-access sources. (b) \textit{Title and Abstract Screening} uses LLMs to filter articles using expert-designed inclusion and exclusion criteria. (c) \textit{PDF-to-Markdown Conversion} uses OCR to convert PDFs to markdown files. (d) \textit{Full-text Screening} applies the same approach as in Title and Abstract Screening but with stricter criteria. (e) \textit{Data Extraction} employs multi-stage tool-calling with schema validation to extract structured epidemiological data (parameters, models, outbreaks). (f) \textit{Report Generation} synthesises extracted data through artefact generation followed by LLM refinement. 
} 
\vspace{-1.5em}
\label{fig:flashfig}
\end{figure*}

Infectious disease epidemiology provides a particularly rigorous evaluation setting for LLM knowledge synthesis. Epidemiological SLRs require careful extraction of key parameters such as the basic reproduction number, serial interval and case-fatality ratio. The same numbers may be reported differently across studies, split by age group, geography or clinical severity \citep{he2020estimation,WARD2026100882}. This variation in reporting is not just a difficulty. It is the core test. A model that can only retrieve numbers but cannot match them to the right context will fail at the task. 

In this work, we ask whether LLMs can carry out the core evidence handling tasks of epidemiological SLRs. We conduct this study in collaboration with the Pathogen Epidemiology Review Group (PERG), whose verified expert annotations from WHO-designated priority pathogen reviews serve as the human reference standard for evaluation.

Our contributions are as follows:
\begin{itemize} [leftmargin=0.4cm,itemsep=4pt, topsep=2pt, parsep=0pt, partopsep=0pt]
\item We release an \textit{expert-annotated evaluation dataset} for scientific evidence synthesis in infectious disease epidemiology, constructed from SLRs on WHO-designated priority pathogens. The dataset contains $16{,}248$ article records matched to expert screening labels, $3{,}808$ parameter extractions, $687$ transmission model extractions and $189$ outbreak extractions.

\item We introduce \name{}, an \textit{evaluation harness} for LLM-assisted epidemiological SLR workflows covering retrieval, screening and structured extraction. \name{} scores extraction at the record level, including cases where the same evidence can be recorded in more than one correct way.

\item We evaluate five frontier reasoning LLMs and find that \textit{no single model consistently dominates} across stages of the workflow. Failure modes such as refusal behaviour in closed systems show that open, locally reproducible infrastructure is critical for scientific evidence synthesis.  

\end{itemize}
\section{Related Work}
Evaluating AI systems on systematic literature reviews requires operationalising the full review workflow, spanning retrieval, screening and evidence structuring~\citep{page2021prisma,marshall2019toward}. LLMs can transfer screening logic across title, abstract and full-text stages without task-specific fine-tuning, achieving high sensitivity and specificity across multiple systematic reviews~\citep{ottoSRPromptPaper,homiar2025development}. For data extraction, LLMs perform well on constrained schemas but degrade on complex fields, with human-incorporated workflows generally outperforming LLM-only approaches~\citep{gartlehner2024data,mahmoudi2025critical,lai2025language}. Building on stage-specific advances, recent work has shifted towards aggregate workflow-level pipelines~\citep{scherbakov2025emergence,ottosrCao2025automation,parkinson2025metabeeai,lee2025heor}. Yet these pipelines tend to be tailored to individual domains. For example, \citet{parkinson2025metabeeai} target bee ecotoxicology and \citet{ottosrCao2025automation} focus on Cochrane-style clinical reviews. Epidemiology poses a distinct challenge requiring its own dedicated pipeline or evaluation harness.


Recent evaluation work examines which review and research-synthesis capabilities LLMs reliably support. \citet{wang2025accelerating} evaluate clinical evidence synthesis using pre-defined stable review questions and test whether extracted values match reference annotation values from published systematic reviews. \citet{madeyski2025llm4screenlit} recommend recall-focused screening metrics, full confusion matrices and cost-sensitive analysis under class imbalance. \citet{wang2025evidencebench} evaluate within paper sentence retrieval for biomedical hypotheses using expert written evidence summaries. \citet{polzak2025largelanguagemodelsmatch} test whether models match Cochrane review conclusions from source studies. 



These studies motivate direct evaluation of review workflows. \name{} differs in its evaluation construct and introduces additional degrees of complexity.  A model must identify the evidence family present in a paper (parameters, transmission models or outbreaks reported), use the appropriate schema and recover values with uncertainty bounds and population context. The same paper may report multiple estimates across age groups, locations or time periods, with the values implicit in disaggregated tables. Records of the same evidence family often appear in one paper with no canonical alignment to human annotations, requiring design of metrics beyond exact match. To our knowledge, \name{} is the first evaluation of LLMs on epidemiological SLRs against expert annotations from peer-reviewed WHO priority pathogen reviews. 
See Appendix~\ref{app:related_work} for an extended related work comparison.



\section{Evaluation Harness}\label{sec:pipeline}
Here, we describe \name{}, the proposed open-source evaluation harness. The harness has two parts. The first is a workflow built to replicate the main evidence-handling stages of a human-conducted epidemiological SLR, as shown in Figure~\ref{fig:flashfig}. The second is an evaluation protocol that maps workflow outputs to expert labels through stage-isolated metrics.

\subsection{Building SLR Workflows}
\paragraph{Article Search and Retrieval}\label{sec:search}
\name{} queries three bibliographic databases (OpenAlex, PubMed, and Europe PMC) using domain-specific Boolean search strategies covering seven core epidemiological domains (Figure~\ref{fig:flashfig}a, Appendix~\ref{app:search_queries}). 
Retrieved records are first deduplicated using identifier and bibliographic metadata-based matching, then full texts are automatically retrieved from open-access sources. 
The download pipeline incorporates caching, streaming and file validation, parallel execution and checkpointing. 

\paragraph{Title and Abstract Screening}
Initial screening is conducted using titles and abstracts based on predefined inclusion and exclusion criteria (Figure~\ref{fig:flashfig}b, Appendix~\ref{app:article_screening_prompts_criteria}). Following the ScreenPrompt methodology \citep{ottoSRPromptPaper}, we structure screening with five components: study objectives, inclusion/exclusion criteria, chain-of-thought reasoning instructions, article abstract and structured output format. 

\paragraph{PDF-to-Markdown Conversion}\label{sec:pdf_to_markdown_conversion}
Each downloaded PDF is rendered page-wise into high-resolution images, then processed with an OCR model to recover text while preserving document hierarchy, equations (LaTeX) and tables (Figure~\ref{fig:flashfig}c). The process produces one Markdown file per article.
\paragraph{Full-text Screening}
Converted articles undergo full-text screening using the tested model with a prompt structure analogous to abstract screening but with stricter criteria, requiring extractable quantitative epidemiological parameters (e.g. transmission rates, incubation periods and severity outcomes) while excluding literature reviews, meta-analyses and case studies describing fewer than 10 infected individuals (Figure~\ref{fig:flashfig}d, Appendix~\ref{app:article_screening_prompts_criteri_fulltext}).

\paragraph{Data Extraction}\label{sec:data_extraction}
We extract data for three categories (epidemiological parameters, transmission models and concluded outbreaks) using a multi-stage, schema-constrained framework (Figure~\ref{fig:flashfig}e, Appendix~\ref{app:data_extraction_details}). Extraction assumes the use of agentic models and provides tools that enforce schematised and structured outputs, mimicking human annotators extracting relevant data and filling out survey forms. 
For each data category, the pipeline first conducts \textit{presence flagging} to identify articles containing relevant data, followed by targeted extraction using parameter-specific, model-specific, or outbreak-specific tool calls for validated outputs. For epidemiological parameters, extraction also involves population tagging (e.g. age groups, geographic locations and clinical severity), which enables subsequent aggregation of parameter estimates into summary statistics across population contexts. 



\paragraph{Report Generation}
Extracted data are converted into a report using LLM self-refinement on top of standardised code producing SLR-like artefacts, including figures, tables and aggregate statistics (Figure~\ref{fig:flashfig}f, Appendix~\ref{app:report_generation_methods}). Report generation sits outside the evaluation construct studied. Our metrics target evidence handling, whereas evaluating generated reports would require exact data-matched reference reports and a separate construct for narrative synthesis. LLM interpretation of aggregate public-health statistics also introduces safety risks beyond the scope of this study. 


\subsection{Evaluation Protocol}
\label{sec:evaluation_protocol}
\paragraph{Screening Metrics} We evaluate screening as a binary article decision $y\in\{\text{include},\text{exclude}\}$ against PERG (human) labels, reporting precision, recall and macro $F_1$. Consistent with PRISMA style study selection~\citep{page2021prisma}, \name{} supports the usual abstract to full-text funnel and two ablations: (a) AI abstract$\rightarrow$AI full-text, (b) human abstract$\rightarrow$AI full-text and (c) AI direct full-text.


\paragraph{Extraction metrics}
Structured extraction cannot be evaluated by a single exact match. An article may contain zero, one or many records of the same data type, and the records have no unique identifiers that tell us which model output corresponds to which expert annotation. A generated record can also be partly useful, it may identify the correct outbreak but miss a date, or recover the correct parameter value but attach incomplete population context. We therefore separate three constructs: whether relevant data are detected, whether the correct number of records is produced and whether the matched records contain the correct field values.


\textit{Flagging} measures detection. For each $\langle\mathrm{article},\mathrm{data\_type}\rangle$ pair, we compare a PERG-derived presence label (at least one extraction) to the LLM’s flag and compute precision and recall over these binary labels. \textit{Count} measures volume. If the reference contains $n$ records of a data type and the LLM generates $\hat{n}$ records, we assign partial credit as
\[
\mathrm{TP}=\min(n,\hat{n}),\qquad
\mathrm{FP}=\max(0,\hat{n}-n),\qquad
\mathrm{FN}=\max(0,n-\hat{n}).
\]
Thus, if the reference has two records and the LLM generates five, the count score gives two true positives and three false positives. It does not decide whether the two TP are field correct.
\textit{Extraction} measures field-level content. We first establish correspondences between reference and generated records within the same article by computing a pairwise similarity score. For a reference extraction $E$, generated extraction $\hat{E}$ and key fields we use
\[
s(E,\hat{E})=\sum_{k\in\mathcal{F}}w_k d_k(E[k],\hat{E}[k]),
\qquad
d_k(v,\hat{v})=\frac{|v\cap\hat{v}|}{|v\cup\hat{v}|}.
\]
where $v=E[k]$ and $\hat{v}=\hat{E}[k]$ are the sets of values for field $k$. Here $w_k$ is a normalised field weight and $d_k$ is Jaccard similarity, with single-value fields treated as singleton sets. We use the modified Jonker-Volgenant assignment algorithm \citep{jonker1987shortest} on the cost matrix $1-s$ to find the one-to-one matching that maximises total similarity. Field-level precision and recall are computed after this matching step. This design separates errors from over-extraction, missed records and inaccurate fields, which better reflects how expert reviewers inspect structured records. See Appendix~\ref{app:additional_evaluation_details} for details.


\section{Experiment Settings}

\subsection{Dataset}
\label{sec:methods_data}
The SLR workflow part of the harness (\Cref{sec:search}) allows for automatic and repeated retrieval of articles in real time. However, to evaluate LLM extractions we use human-reference labels gathered from SLRs conducted by epidemiologists at the Pathogen Epidemiology Review Group (PERG). PERG has undertaken an initiative to conduct SLRs for nine priority pathogens identified by the WHO as having high epidemic or pandemic potential \cite{worldhealthorganizationPathogensPrioritizationScientific2024}. The group has published five peer-reviewed SLRs \cite{cuomo-dannenburgMarburgVirusDisease2024, doohanLassaFeverOutbreaks2024, nashEbolaVirusDisease2024, morgensternSevereAcuteRespiratory2025,mccain2026systematic} with two more in the data extraction phase (\cref{tab:article_count_summary_ours_pergs}). We evaluated each workflow stage with all pathogen data available, so seven pathogens are evaluated for screening and four (Ebola, Lassa, SARS, and Zika) are evaluated for data extraction. Within data extraction, parameter and transmission model extraction are evaluated on all four pathogens, while outbreak extraction is evaluated on Lassa and Zika only, as PERG's published SLRs for Ebola and SARS did not include outbreak extraction. After correspondence with PERG, we exclude Marburg due to inconsistencies in data format, and MERS and Nipah because PERG's extraction phase is still in progress.

\begin{wraptable}{r}{0.42\textwidth}
\centering
\footnotesize
\caption{\textbf{Released dataset (benchmark) record counts.} AgentSLR Matched denotes the downloaded article subset that matched PERG labelled records and was released on HuggingFace. The coloured symbols represent the progress of PERG's SLRs per priority pathogen: published \published{}, conducting data extraction \extractionphase{}, and yet to begin screening \notstarted{}, as of March 2026.}
\setlength{\tabcolsep}{3pt}
\renewcommand{\arraystretch}{1}
\label{tab:article_count_summary_ours_pergs}
\begin{tabular}{lrr}
\toprule
\textbf{Pathogen} & \textbf{PERG\textsuperscript{*}} & \textbf{\name{} Matched} \\
\midrule
\published{} Marburg virus & 2,593 & 801 (30.9\%) \\
\published{} Ebola virus & 11,605 & 4,119 (35.5\%) \\
\published{} Lassa fever & 2,131 & 667 (31.3\%) \\
\published{} SARS-CoV-1 & 12,280 & 2,047 (16.7\%) \\
\published{} Zika virus & 10,510 & 2,164 (20.6\%) \\
\extractionphase{} MERS-CoV & 19,656 & 5,714 (29.1\%) \\
\extractionphase{} Nipah virus & 1,458 & 736 (50.5\%) \\
\notstarted{} RVF virus & -- & -- \\
\notstarted{} CCHF virus & -- & -- \\
\midrule
\textbf{Total} & 60,233 & 16,248 (27.0\%) \\
\bottomrule
\end{tabular}
\par{\scriptsize
\textsuperscript{*}Articles post deduplication and empty abstract removal.}
\end{wraptable}

The data labelled and extracted by humans for the peer-reviewed SLRs have been partially available as open-source resources through the \texttt{epireview} and \texttt{priority-pathogen} R packages \cite{naidooEpireviewToolsUpdate2025, nashPrioritypathogens2026}. Through \name{} we release an extended version of this dataset on \href{https://huggingface.co/datasets/OxRML/AgentSLR}{HuggingFace}, with associated code on \href{https://github.com/OxRML/AgentSLR}{GitHub}. Our released dataset contains $16{,}248$ downloaded article records matched to PERG linked labels, based on article Covidence ID. It also contains $3{,}808$ parameter extractions, $687$ transmission model extractions and $189$ outbreak extractions.\footnote{The data released includes metadata, URLs and structured annotations. The supplementary harness code can be used to download articles and conduct OCR, with instructions and single command scripts provided in the README.} For reproducibility, we evaluate the downloaded open-access subset retrieved through the bibliographic databases we queried (Appendix~\ref{app:search_queries}). The harness can retrieve a larger article corpus, but only articles with accessible full text can be downloaded, processed and matched to PERG labels. This yields $16{,}248$ matched records, corresponding to approximately $27.0\%$ of the PERG corpus. To assess whether the open-access subset is representative of the broader PERG corpus, we manually retrieve 1,004 closed-access articles from the broader PERG corpus and run the \name{} harness on a matched comparison sample. We find broadly comparable performance across stages (see Appendix~\ref{app:data_representativeness}).


\subsection{Models}
\label{sec:models}
\name{} is compatible with both open and closed weight models, with tool calls and requests schematised through OpenAI's Responses and Chat Completions APIs. Due to complex calculations required, especially during data extraction, we evaluate LLMs that can reason and scale at inference-time. We conduct comparative evaluations using OpenAI's \texttt{GPT-5.2} (closed-source) and \texttt{gpt-oss-120b} (open-source), Moonshot AI's \texttt{Kimi K2.5}, Z.AI's \texttt{GLM-4.7} and DeepSeek's \texttt{DeepSeek-V3.2}. Attempts to evaluate Claude Opus 4.5 and Sonnet 4.5 resulted in streaming refusals.\footnote{We experienced this refusal problem with all Claude models above version 4.0 (See \href{https://platform.claude.com/docs/en/test-and-evaluate/strengthen-guardrails/handle-streaming-refusals}{Documentation} from Anthropic). Potential causes and implications are discussed in \Cref{sec:discussion}.} All models have reasoning set to high where possible, with a maximum generation limit of 64K tokens per pass. Open-source models are hosted with \texttt{vllm} \cite{kwon2023efficient} on an NVIDIA H200 cluster node. For PDF-to-Markdown conversion, we use the \texttt{mistral-ocr-2512} API endpoint \cite{mistralaiMistralOCR32025}, a state-of-the-art OCR model suited to scanned documents with complex mathematical and tabular content. For reproducibility, \name{} is also configured to run with open-weight OCR models. We find no statistically significant change in performance when running OCR (10\% sample) with DeepSeek-OCR \cite{wei2025deepseekocrcontextsopticalcompression} or PaddleOCR \cite{cui2025paddleocr}.

\section{Results}\label{sec:pipeline_experimental_results}

\subsection{Scientific Synthesis Is Not a Single Capability}
\begin{figure*}[t]
    \centering
    \includegraphics[width=\textwidth]{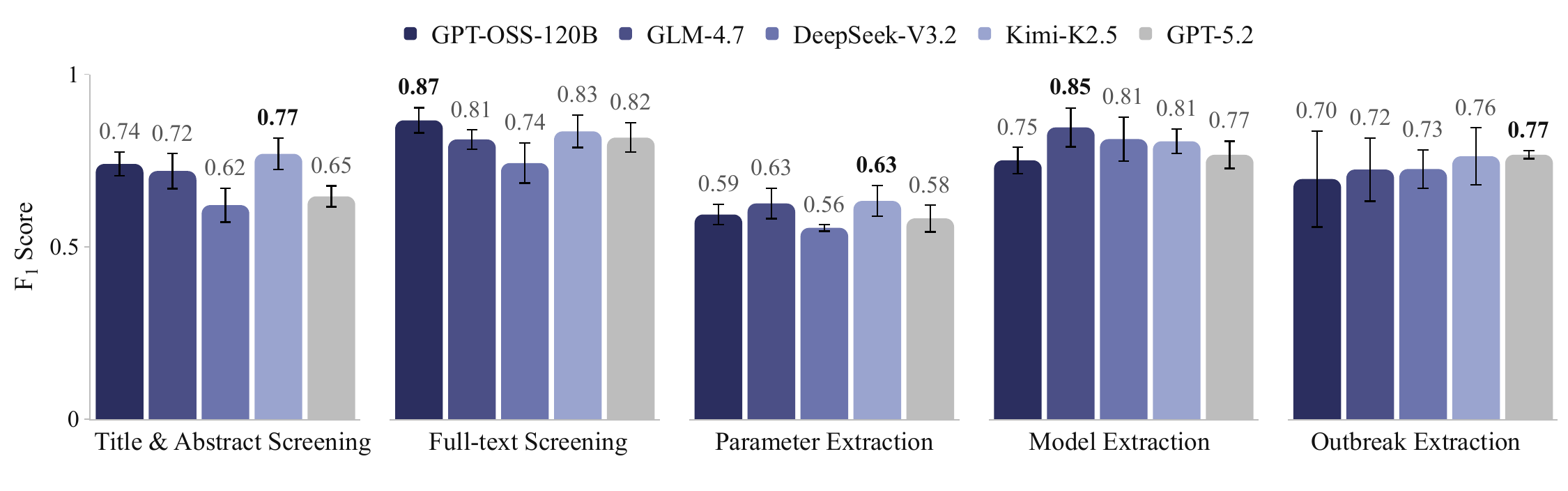}
    \caption{\textbf{Stage-wise model performance on \name{} harness.} Each model is run on individual pathogens with \name{} harness.  Averages are computed over the pathogens evaluated at each stage, following reference data availability described in Section~\ref{sec:methods_data}. Error bars indicate one standard deviation across pathogens. No single model dominates across all stages: \texttt{Kimi-K2.5} and \texttt{gpt-oss-120b} lead screening, while extraction leaders vary by data type. Full metrics are provided in Appendix~\ref{app:model_ablations}.}
    \label{fig:model_ablation}
    \vspace{-1em}
\end{figure*}

Evaluating five LLMs using the \name{} harness across pathogens to produce epidemiological SLRs, we find that no single model dominates the full synthesis workflow (\Cref{fig:model_ablation}). At the article screening stages, \texttt{Kimi-K2.5} and \texttt{gpt-oss-120b} perform the best, with the former leading title and abstract screening ($F_1 = 0.77$) and the latter full-text screening ($F_1=0.87$). All models struggle with parameter extraction, where \texttt{Kimi-K2.5 again achieves the highest performance} ($F_1=0.63$). \texttt{GLM-4.7} performs best on model extraction, while \texttt{GPT-5.2} leads on outbreak extraction. \texttt{DeepSeek-V3.2} exhibits the most variable performance, ranking last in article screening but becoming competitive in extraction, where function calling is enabled. Across all extraction stages, the gap between the best and worst model is considerably narrower than at screening. These disparities suggest scientific synthesis comprises separable sub-tasks, each posing distinct challenges.

\begin{table}[b!]
\vspace{-1em}
\centering
\caption{\textbf{Structured evidence extraction performance.}
F\textsubscript{1} scores are averaged across pathogens for five frontier models and three extraction tasks: \deflagging{} Flagging, \decounts{} Counts, and \deextraction{} field-level Extraction. Standard deviations ($\pm$) are reported across pathogens.}
\label{tab:data_extraction_f1}
\footnotesize
\setlength{\tabcolsep}{3pt}
\renewcommand{\arraystretch}{1.1}
\newcommand{\cell}[2]{\makecell{#1 \\ {\color{stdgray}\scriptsize$\pm$#2}}}
\begin{tabular}{l ccc ccc ccc ccc}
\toprule
& \multicolumn{3}{c}{\textbf{Parameters}}
& \multicolumn{3}{c}{\textbf{Models}}
& \multicolumn{3}{c}{\textbf{Outbreaks}}
& \multicolumn{3}{c}{\textbf{Average}} \\
\cmidrule(lr){2-4}\cmidrule(lr){5-7}\cmidrule(lr){8-10}\cmidrule(lr){11-13}
\textbf{Model}
  & \deflagging{} & \decounts{} & \deextraction{}
  & \deflagging{} & \decounts{} & \deextraction{}
  & \deflagging{} & \decounts{} & \deextraction{}
  & \deflagging{} & \decounts{} & \deextraction{} \\
\midrule
\texttt{gpt-oss-120b}
  & \cell{0.66}{0.07} & \cell{0.59}{0.08} & \cell{0.54}{0.03}
  & \cell{0.91}{0.05} & \cell{0.68}{0.05} & \cell{0.67}{0.03}
  & \cell{0.61}{0.13} & \cell{0.69}{0.31} & \cell{0.79}{0.02}
  & \cell{0.75}{0.15} & \cell{0.65}{0.13} & \cell{0.64}{0.10} \\
\addlinespace[2pt]
\texttt{GPT-5.2}
  & \cell{0.66}{0.07} & \cell{0.50}{0.07} & \cell{0.59}{0.03}
  & \cell{0.90}{0.05} & \cell{0.72}{0.04} & \cell{0.67}{0.03}
  & \cell{0.66}{0.12} & \cell{0.80}{0.05} & \cell{0.84}{0.04}
  & \cell{0.76}{0.14} & \cell{0.65}{0.14} & \cell{0.67}{0.10} \\
\addlinespace[2pt]
\texttt{DeepSeek-V3.2}
  & \cell{0.60}{0.07} & \cell{0.56}{0.04} & \cell{0.50}{0.03}
  & \cell{0.87}{0.04} & \cell{0.92}{0.08} & \cell{0.65}{0.09}
  & \cell{0.65}{0.04} & \cell{0.78}{0.19} & \cell{0.75}{0.02}
  & \cell{0.72}{0.14} & \cell{0.75}{0.19} & \cell{0.61}{0.11} \\
\addlinespace[2pt]
\texttt{Kimi-K2.5}
  & \cell{0.72}{0.09} & \cell{0.62}{0.05} & \cell{0.56}{0.02}
  & \cell{0.92}{0.04} & \cell{0.81}{0.04} & \cell{0.68}{0.04}
  & \cell{0.64}{0.07} & \cell{0.87}{0.11} & \cell{0.78}{0.07}
  & \cell{0.78}{0.14} & \cell{0.75}{0.12} & \cell{0.65}{0.10} \\
\addlinespace[2pt]
\texttt{GLM-4.7}
  & \cell{0.72}{0.09} & \cell{0.61}{0.05} & \cell{0.54}{0.02}
  & \cell{0.93}{0.07} & \cell{0.93}{0.06} & \cell{0.68}{0.05}
  & \cell{0.68}{0.03} & \cell{0.72}{0.29} & \cell{0.77}{0.02}
  & \cell{0.80}{0.13} & \cell{0.76}{0.18} & \cell{0.64}{0.10} \\
\bottomrule
\end{tabular}
\end{table}

\subsection{Structured Evidence Extraction Is the Bottleneck Across Models}
\Cref{tab:data_extraction_f1} isolates the type and sub-task of the structured data extracted. Flagging the presence of a data type is the most reliable construct on average, with mean $F_1$ ranging from $0.72$ to $0.80$ across models, while field-level extraction of evidence remains lower and shows less variance, ranging from $0.61$ to $0.67$. This bottleneck is particularly evident for epidemiological parameter values, where  no model reaches $F_1=0.60$ for parameter extraction, despite stronger parameter Flagging for \texttt{Kimi-K2.5} and \texttt{GLM-4.7} ($0.72$).  Transmission model flagging is consistently high ($0.87$ to $0.93$), yet field-level model extraction remains bounded at $0.65$ to $0.68$, indicating that recognising a relevant article is substantially easier than producing the correct structured record. Outbreak extraction is the main exception, with field-level scores reaching $0.84$ for \texttt{GPT-5.2}, but this result is evaluated only on Lassa and Zika and shows high variance in Count performance. With no LLM exceeding an average extraction (\deextraction{}) $F_1$ of $0.67$, the limiting factor extends beyond model choice to a fundamental capability gap in grounding heterogeneous scientific reporting conventions onto a single validated schema.

\subsection{Screening Strategy Controls What Evidence Survives}
\begin{figure*}[t!]
    \centering
    \includegraphics[width=\linewidth]{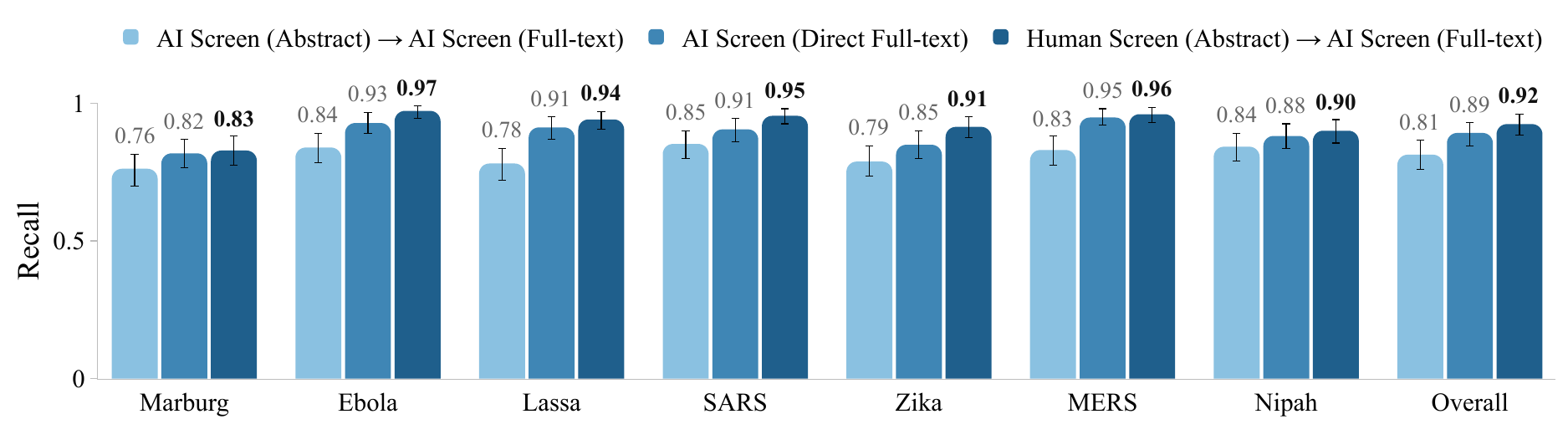}
    \caption{\textbf{Recall of article screening strategies across pathogens.} Two ablation strategies (direct full-text and human-conditioned screening) with \name{} (\texttt{GPT-OSS-120B}) offer better recall (or `fetch rate') than standard AI-based two-stage screening, with confidence intervals overlapping across most pathogens. Full article screening metrics results are reported in Appendix~\ref{app:extended_results_article_screening}.}
    \label{fig:fulltext_recall_barchart}
\end{figure*}

Using \texttt{gpt-oss-120b}, we evaluate three full-text screening strategies to quantify how review (and workflow) design affects evidence survival (\cref{fig:fulltext_recall_barchart}). The fully automated two-stage strategy (mimicking humans), where abstract screening filters articles before full-text screening, achieves recall $0.81$ and $F_1=0.77$ against PERG decisions. Conditioning full-text screening on human abstract decisions raises recall to $0.92$ and gives the strongest overall classification performance ($F_1=0.87$). This shows how early human triage can protect relevant studies from upstream model errors.  Direct AI full-text screening without abstract gating also improves recall to $0.89$, but lowers $F_1$ to $0.73$. This configuration avoids abstract information loss without human input, but it increases screening runtime by $2.3\times$ ($9.55$ versus $4.16$ hours) and raises OCR costs from USD 36.6 to USD 303.2. Screening strategy therefore controls the false negative risk and operating cost of the evaluation beyond the aggregate score.

\subsection{Expert Review Separates Utility from Autonomy}

\begin{table}[b!]
\centering
\footnotesize
\caption{\textbf{Expert-rated holistic evaluation of data extraction quality.} We report expert-rated flagging precision, field-level extraction accuracy, and perceived  competence (AgentSLR with \texttt{gpt-oss-120b}) for parameter, model, and outbreak extractions, aggregated across six epidemiologists. Values show mean $\pm$ standard error. Competence is rated on a 1--7 Likert scale (1: Incompetent, 7: Autonomous).}
\label{tab:human_validation}
\setlength{\tabcolsep}{5pt}
\renewcommand{\arraystretch}{1.1}
\begin{tabular}{lccc}
\toprule
\textbf{Data Type} & \textbf{Flagging Precision} & \textbf{Extraction Accuracy} & \textbf{Competence (1--7)} \\
\midrule
\textbf{Parameters} & $0.66$ {\color{stdgray}\scriptsize$\pm$ 0.06} & $0.77$ {\color{stdgray}\scriptsize$\pm$ 0.05} & $4.23$ {\color{stdgray}\scriptsize$\pm$ 0.20} \\
\addlinespace[2pt]
\textbf{Models} & $0.40$ {\color{stdgray}\scriptsize$\pm$ 0.07} & $0.83$ {\color{stdgray}\scriptsize$\pm$ 0.05} & $2.80$ {\color{stdgray}\scriptsize$\pm$ 0.32} \\
\addlinespace[2pt]
\textbf{Outbreaks} & $0.61$ {\color{stdgray}\scriptsize$\pm$ 0.09} & $0.80$ {\color{stdgray}\scriptsize$\pm$ 0.07} & $3.90$ {\color{stdgray}\scriptsize$\pm$ 0.45} \\
\bottomrule
\end{tabular}
\end{table}

We validated \texttt{gpt-oss-120b} extraction quality across all three data types using a survey completed by six expert epidemiologists (\Cref{tab:human_validation}). Each expert reviewed a random subset of system outputs alongside the corresponding articles (details in Appendix~\ref{app:expert_validaation_data}). This validation audits  output quality rather than re-annotating the PERG reference set. Expert-rated flagging precision is highest for parameters ($0.66$) and outbreaks ($0.61$), but substantially lower for models ($0.40$), indicating that a majority of model extractions are flagged by the system, but deemed irrelevant by experts. Field-level extraction accuracy is more consistent across types ($0.77$--$0.83$), suggesting that when an extraction is correctly flagged, populated fields are largely accurate. The competence ratings make the deployment boundary clear: parameters score $4.23$ and outbreaks $3.90$ on the $1$--$7$ scale, close to the survey threshold for a tool usable under moderate supervision, while models score only $2.80$. Qualitative feedback from epidemiologists is consistent with this split. Recurring failure modes include limited use of document structure, failure to distinguish newly reported findings from cited prior work, and insufficient contextual reasoning for fields that require holistic inference. The system can reduce human workload by providing a correctable starting point, and the false positives are typically easy to identify, but this warrants human review rather than full autonomous use.


\subsection{Performance Does Not Scale Reliably with Cost}

\begin{wrapfigure}{r}{0.45\textwidth}
    \centering
    \includegraphics[width=\linewidth]{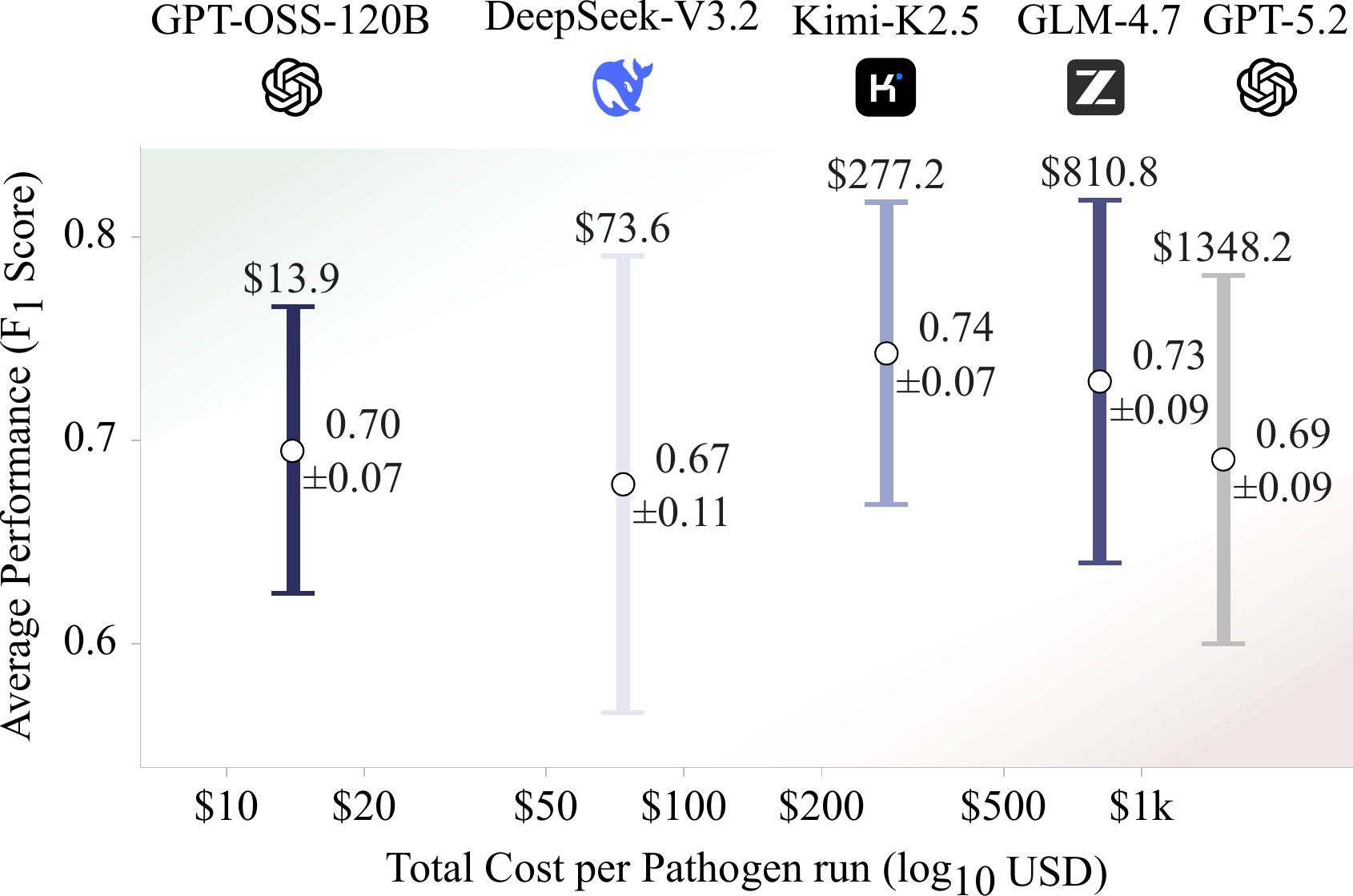}
    \caption{\textbf{Cost vs.\ average performance per \name{} pathogen run.} Each point shows a model's mean macro $F_1$ across all pipeline stages against estimated total cost per run (USD, $\log_{10}$ scale), with error bars denoting one standard deviation across stages. Full cost and token breakdowns are provided in Appendix~\ref{app:pipeline_statistics_costs}.} 
    \label{fig:cost_vs_performance}
\vspace{-0.5em}
\end{wrapfigure}

Running the SLR workflow at scale requires models to process a large and narrowing evidence funnel. Across pathogens on average $9{,}132$ articles reach title and abstract screening, $1{,}102$ reach full-text screening, and $395$ reach data extraction. At this scale, model choice affects both evidence quality and operating cost, but higher cost does not reliably predict higher performance (\cref{fig:cost_vs_performance}). \texttt{gpt-oss-120b} reaches competitive average performance ($F_1=0.70$) at the lowest estimated cost per pathogen run (USD 13.9), while \texttt{GPT-5.2} costs over $96\times$ more (USD 1{,}348.2) and has slightly lower performance ($F_1=0.69$). The best-performing model overall, \texttt{Kimi-K2.5} ($F_1 = 0.74$), falls in the mid-cost range (\$277), while \texttt{GLM-4.7} incurs the second-highest cost (\$811) for a comparable $F_1$ of $0.73$. \texttt{DeepSeek-V3.2} is cheaper than \texttt{Kimi-K2.5} at USD 73.6, but has lower average performance ($F_1=0.67$). The cost differences are driven by per article token use during parameter extraction, where \texttt{GPT-5.2} produces $91.1K$ output tokens per article\footnote{While maximum generation length is capped at 64K, a model can have multiple extractions per article} compared with $3.0K$ for \texttt{DeepSeek-V3.2}. Cost and quality therefore need to be evaluated at the subtask level before selecting a model for SLR deployment. 

\section{Discussion}
\label{sec:discussion}

\subsection{Key Findings}
Evaluating five frontier LLMs using \name{} reveals that epidemiological SLRs expose scientific synthesis as a set of separable and unsaturated capabilities. No evaluated model dominates across the workflow. Across all five LLMs, no model exceeds an average field-level extraction $F_1$ of $0.67$. Critically, this performance ceiling is not attributable to model choice alone. The performance spread between models narrows markedly at extraction relative to screening. This convergence under tool-calling highlights the difficulty of schema-grounded extraction from heterogeneous scientific reporting, where the same quantities are presented through different conventions and must be normalised across study designs rather than simply identified.


\subsection{Deployment Risk}
\label{sec:discussion_deployment_risks}
Deployment of LLMs in public health research workflows faces two distinct risk categories, namely infrastructure risk and error structure risk. Infrastructure risk encompasses model availability, content restrictions, version instability and operating cost. This risk applies primarily to closed-source models. Attempts to evaluate Claude Opus~4.5 and Sonnet~4.5 consistently produced streaming refusals, which we attribute to content filters on epidemiological terminology as these may indicate dangerous intent.\footnote{See the \href{https://support.claude.com/en/articles/12449294-understanding-sonnet-4-5-s-api-safety-filters}{Anthropic documentation} on Sonnet~4.5 API safety filters.} When applied too broadly, such restrictions render entire model families unavailable for legitimate public-health research and reinforce the case for open-weight alternatives. We find that \texttt{gpt-oss-120b} reaches competitive average performance at over $96\times$ lower cost than \texttt{GPT-5.2}. Such open-source models allow for version pinning and local hosting, both scientific requirements for long-running living reviews where reproducibility matters. At the error structure level, our stage-isolated evaluation shows why average scores, or replications of papers (like PaperBench~\cite{starace2025paperbench}), are insufficient in high-stakes SLRs. False negatives in screening remove evidence permanently, provenance and context failures can attach correct values to the wrong claim. Similarly, pathogen-specific performance gaps, e.g. when LLMs struggle on Zika extraction, signal systematic bias that can propagate through any synthesis drawing on AI-assisted extraction.

\subsection{Assisted Evidence Synthesis}
\label{sec:discussion_assisted_sys}

\begin{wrapfigure}{r}{0.4\textwidth}
\vspace{-2em}
\centering\includegraphics[width=0.4\textwidth]{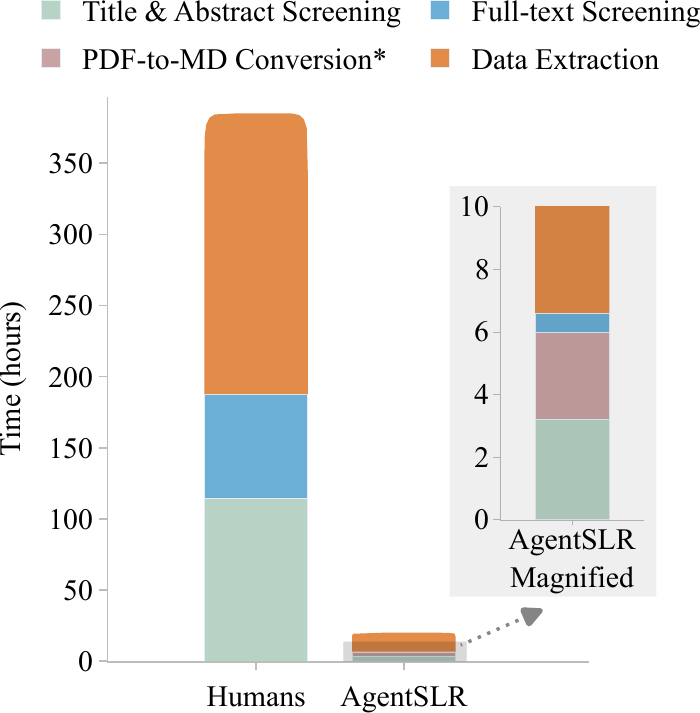}

\caption{\textbf{SLR completion time (Human vs.\ \name{}).} Using \texttt{gpt-oss-120b} across the full SLR creation pipeline takes $20$ hours versus $385$ hours for manual review, a $19.3\times$ wall-clock speed-up and $58\times$ reduction in calendar days. Full runtime statistics are reported in Appendix~\ref{app:pipeline_statistics}.}
    \label{fig:time_comparison}
\vspace{-1em}
\end{wrapfigure}

\name{} substantially reduces the cost of running review stages compared with humans (\Cref{fig:time_comparison}), but these numbers should be interpreted as conditional efficiency rather than evidence of expert replacement. With \texttt{gpt-oss-120b}, the per-pathogen workflow processes an average funnel of $9{,}132$ articles at title and abstract screening, $1{,}102$ at full-text screening and $395$ at data extraction in $20$ hours. The corresponding human estimate is $385$ labour hours, giving a $19.3\times$ efficiency gain. As the system runs continuously, this is also a reduction from $48.1$ eight-hour workdays to $0.83$ calendar days, or $58\times$ fewer calendar days. Full-text screening is the clearest example, taking under two seconds per article compared with a four minute human estimate, or $118\times$ faster. These runtime figures are reported separately from quality metrics because speed is only meaningful once a stage meets an acceptable quality threshold. Therefore, LLMs are not at the capability threshold to replace expert judgment, but they can substantially reduce the time required for evidence triage. 


The clearest short-term role for LLMs is to accelerate human-led review at the points where missed evidence is most costly. This is especially true for full-text screening after human abstract triage, where manual review limits SLR scalability, and for direct full-text screening when abstracts do not fully report relevance. In these settings, false inclusions can be removed during validation, while false exclusions are much harder to recover. Expert validation extends this picture to data extraction, where parameter and outbreak competence sit near the threshold for use under moderate supervision and experts consistently report that outputs reduce workload by providing a correctable starting point. Applied to living systematic reviews, where the cost of each update cycle is the primary barrier to continuous monitoring, this combination can make routine updates feasible in priority pathogen surveillance~\citep{bergstrom2026screening}.

\subsection{Limitations and Future Work}
\label{sec:discussion_)limitations_future_work}
We identify key limitations of our evaluation study. Our evaluation is restricted to articles available through open access routes in the queried sources, covering approximately $27.0\%$ of the PERG corpus. However, a manually retrieved closed-access subset yields consistent results. The workflow follows English-only screening criteria, and misses multilingual evidence. Our metrics prioritise evidence survival, which is appropriate for high-stakes screening but this can shift some burden to downstream filtering when precision is low. Finally, we do not evaluate meta-analysis or final review writing, where statistical modelling and narrative claims would require separate expert-grounded tests. Future work should measure human uplift directly, develop the annotation interface described in Appendix~\ref{appendix:annotation_tool} into a production tool, extend the evaluation to other scientific domains and study heterogeneous model routing across tasks where different LLMs show complementary strengths.

\section{Conclusion}
We introduce \name{} as an evaluation harness to study LLM-assisted systematic literature review in infectious disease epidemiology. Epidemiological SLRs provide an expert-grounded, high-stakes setting for evaluating AI scientific knowledge synthesis. The released dataset and harness decompose review production into retrieval, screening and schema-grounded extraction, enabling stage-level diagnosis against human reference annotations rather than aggregate claims of automation. Across current LLMs, we find clear assistive value, especially for screening and correctable preliminary extraction, but also persistent limitations in structured evidence extraction, error structure, cost and model availability. These results argue for human-supervised deployment in epidemiology, and for treating SLRs as a demanding evaluation setting for measuring whether future AI systems can synthesise scientific evidence reliably.
\section*{Acknowledgement}
S.P. and A.M. are supported in part by the Engineering and Physical Sciences Research Council (EPSRC) under Grant EP/X028909/1 and Oxford Internet Institute’s Research Programme funded by the Dieter Schwarz Foundation. R.O.K. is supported by the Clarendon Scholarship and the Jesus College Old Members’ Scholarship. E.S. acknowledges support in part from the AI2050 program at Schmidt Sciences (Grant No. G-22-64476). E.S. and A.C. acknowledge that this study is funded by the National Institute for Health Research (NIHR) Health Protection Research Unit in Health Analytics \& Modelling (NIHR207404), a partnership between UK Health Security Agency (UKHSA), London School of Hygiene \& Tropical Medicine, and Imperial College of Science, Technology, \& Medicine. The views expressed are those of the author(s) and not necessarily those of the NIHR, UKHSA, or the Department of Health and Social Care. The authors thank Saverio Trioni for helpful conversations. The authors would like to thank the Pathogen Epidemiology Review Group (PERG), School of Public Health, Imperial College London, for their support and eagerness to contribute to this project.




\clearpage
\appendix

\renewcommand{\partname}{}
\renewcommand{\thepart}{}
\part{Appendix} 

\etocsettocstyle{\setlength{\parskip}{0pt}}{}
\etocsetnexttocdepth{2}
\localtableofcontents

\clearpage
\section{Extended Related Work}\label{app:related_work}
We structure the related work for our evaluation harness into two parts. First is the workflow, and the attempt at automating SLRs for a particular domain, the need and challenge in doing that. Second, is benchmark and evaluation studies within scientific evidence synthesis and SLRs in particular.

\subsection{SLR Automation and Workflows}
Prior LLM-based SLR work has addressed individual pipeline stages in isolation~\citep{ottoSRPromptPaper,gartlehner2024data}, and more recent systems have demonstrated feasibility in clinical and biological domains~\citep{ottosrCao2025automation,parkinson2025metabeeai,lee2025heor}. Each existing system is bespoke to its target domain, uses incompatible reference annotation schemas, and cannot be transferred to epidemiological parameter extraction. Direct quantitative comparison across systems is therefore methodologically ill-posed. Table~\ref{tab:related_work_comparison} provides a structured qualitative comparison across domain, methodology, and evaluation rigour.

\begin{table}[h!]
\vspace{-1.em}
\centering
\footnotesize
\setlength{\tabcolsep}{3pt}
\renewcommand{\arraystretch}{1.3}
\caption{\textbf{Comparison against related work.} LLM-based SLR pipelines differ substantially in domain, extraction methodology, and evaluation rigour, making direct quantitative comparison methodologically ill-posed. $\checkmark$ and $\times$ denote presence and absence of each property. \textit{Open code}: source code is publicly released. \textit{Open weight}: compatible with open-weight models. \textit{Human reference}: evaluation uses expert-curated annotations as reference labels (proxy). \textit{Independent evaluation}: performance is not assessed using LLM-as-a-judge. \textit{Stage evaluation}: evaluation is decomposed into stage-specific metrics enabling failure attribution. \name{} workflow is the only framework satisfying all five properties.}
\label{tab:related_work_comparison}
\begin{tabular}{l p{2.5cm} p{2.5cm} p{2.5cm} p{2.5cm}}
\toprule
& \textbf{otto-SR~\citep{ottosrCao2025automation}}
& \textbf{A4SLR~\citep{lee2025heor}}
& \textbf{MetaBeeAI~\citep{parkinson2025metabeeai}}
& \textbf{AgentSLR (ours)} \\
\midrule
\textbf{Domain}
  & Clinical evidence
  & Health economics
  & Biology
  & Epidemiology \\
\textbf{Methodological focus}
  & \makecell[l]{Cochrane-style \\ intervention reviews}
  & \makecell[l]{HEOR and \\ HTA use cases}
  & Bee ecotoxicology
  & \makecell[l]{WHO-priority pathogen \\ epidemiology} \\
\textbf{Extraction method}
  & \makecell[l]{LLM extraction with \\ post-hoc human \\ correction}
  & \makecell[l]{Prompt-based with \\ domain templates}
  & \makecell[l]{Multi-pass \\ chunk retrieval}
  & \makecell[l]{Tool-calling with \\ JSON schema constraints} \\
\textbf{Open code}
  & $\times$ & $\times$ & $\checkmark$ & $\checkmark$ \\
\textbf{Open weight}
  & $\times$ & $\times$ & $\times$     & $\checkmark$ \\
\textbf{Human reference}
  & $\checkmark$ & $\checkmark$ & $\checkmark$ & $\checkmark$ \\
\textbf{Independent evaluation}
  & $\times$     & $\checkmark$ & $\checkmark$ & $\checkmark$ \\
\textbf{Stage evaluation}
  & $\times$     & $\checkmark$ & $\times$     & $\checkmark$ \\
\bottomrule
\end{tabular}
\end{table}

\name{} workflow is distinguished on three axes within LLM-assisted automation of SLRs that no existing pipeline jointly satisfies. It is the first system to target epidemiological SLRs on WHO-priority pathogens, the first pipeline compatible with open-weight models with disclosed source code, and the first to report stage-isolated evaluation against expert-curated annotations without LLM-as-a-judge. Each property has direct operational consequences. Open-weight compatibility enables version pinning and local deployment, which are requirements for living reviews where reproducibility is a scientific constraint. As we observe in the main text, refusal to process prompts can render entire closed-source model families unavailable for legitimate public-health research. Finally, stage-isolated evaluation enables failure attribution at the component level, supporting targeted improvement without confounding from cross-stage error propagation.

\subsection{Comparison with Benchmarks}
Recent benchmarks evaluate parts of scientific evidence synthesis, but they target different objects. Some test clinical review workflows, others test evidence sentence retrieval, conclusion matching, or long form research reports. Table~\ref{tab:benchmark_comparison} summarises these differences before we discuss the closest comparators.

\begin{table}[t!]
\centering
\footnotesize
\setlength{\tabcolsep}{3pt}
\renewcommand{\arraystretch}{1.3}
\caption{\textbf{Evaluation studies adjacent to \name{}.} We compare each study by the capability evaluated, the reference signal used, and its relation to epidemiological SLR evidence handling.}
\label{tab:benchmark_comparison}
\begin{tabularx}{\linewidth}{>{\raggedright\arraybackslash}p{2.4cm} >{\raggedright\arraybackslash}p{3.2cm} >{\raggedright\arraybackslash}p{3.1cm} >{\raggedright\arraybackslash}X}
\toprule
\textbf{Study} & \textbf{Evaluated capability} & \textbf{Reference signal} & \textbf{Relation to \name{}} \\
\midrule
\makecell[l]{TrialReview\\Bench~\citep{wang2025accelerating}}
& Clinical SR search, screening, extraction and synthesis
& Included studies and study characteristics from 100 cancer treatment reviews
& Broad clinical workflow evaluation, focused on intervention evidence rather than epidemiological parameters, transmission models and outbreaks. \\
\makecell[l]{LLM4\\SCREENLIT~\citep{madeyski2025llm4screenlit}}
& Screening evaluation methodology
& Reanalysis of confusion matrix based screening studies
& Directly motivates recall focused and cost sensitive screening metrics, but does not provide a biomedical extraction dataset. \\
\makecell[l]{Evidence\\Bench~\citep{wang2025evidencebench}}
& Evidence sentence retrieval within biomedical papers
& Expert written evidence summaries and sentence annotations
& Strong evidence grounding benchmark, but it tests locating supporting sentences rather than study selection or schema grounded records. \\
\makecell[l]{Med\\Evidence~\citep{polzak2025largelanguagemodelsmatch}}
& Matching Cochrane review conclusions from source studies
& 284 human curated conclusion questions from 100 Cochrane reviews
& Tests cross study conclusion reasoning, while abstracting away article screening and structured extraction. \\
\makecell[l]{DeepScholar\\Bench~\citep{patel2025deepscholar}}
& Generative related work synthesis from live scholarly search
& Human written related work sections and automated synthesis, retrieval and citation metrics
& Evaluates long form academic synthesis, but not clinical or epidemiological review labels. \\
Mushtaq et al.~\citep{mushtaq2025can}
& Judging completed SLR manuscripts
& PRISMA aligned scores on five SLRs
& Evaluates manuscript quality after a review exists, not evidence retrieval, screening or extraction. \\
\bottomrule
\end{tabularx}
\vspace{-1.em}
\end{table}

\paragraph{Clinical workflow evaluation}
The closest workflow level comparator is TrialReviewBench~\citep{wang2025accelerating}. It evaluates search, screening, extraction and synthesis for clinical oncology reviews. Its strength is breadth across PRISMA aligned stages and human AI collaboration. Its construct differs from ours because the extraction targets are intervention centred study characteristics and outcomes. \name{} instead evaluates epidemiological parameters, transmission models and outbreaks across WHO priority pathogen reviews.

\paragraph{Evidence reasoning}
MedEvidence~\citep{polzak2025largelanguagemodelsmatch} and EvidenceBench~\citep{wang2025evidencebench} isolate complementary evidence reasoning skills. MedEvidence asks whether models can match Cochrane conclusions from the same source studies, and shows failures around uncertainty and scientific scepticism. EvidenceBench evaluates whether models can retrieve the sentences that support biomedical hypotheses. Both are important controlled evaluations, but neither measures evidence survival through screening nor the conversion of heterogeneous paper level evidence into structured epidemiological records.

\paragraph{Screening metrics}
LLM4SCREENLIT~\citep{madeyski2025llm4screenlit} is primarily methodological. Its recommendations on recall, lost evidence, full confusion matrices and cost sensitive screening are directly aligned with high stakes SLR evaluation. We follow the same risk aware motivation for screening, then extend the evaluation construct to extraction through flagging, count and field level matching metrics.

\paragraph{SciLitBench}
SciLitBench~\citep{zabaletascilitbench} is a review automation dataset on AI assisted literature reviews, but we omit detailed comparison because the public draft contains unresolved placeholders, including ``TODOXXX'', and undisclosed evaluation labels and design.

\paragraph{\name{} positioning}
Taken together, these studies occupy three neighbouring regimes: clinical SR workflow automation, biomedical evidence retrieval and conclusion matching, and open ended deep research report evaluation. \name{} is situated between the first two. It is not a general benchmark for deep research agents, and it does not evaluate final meta analysis or narrative report writing. Its contribution is a high stakes evaluation dataset and harness for epidemiological SLR evidence handling, using PERG expert annotations from peer reviewed WHO priority pathogen reviews, stage isolated screening metrics, and record matching metrics for parameters, transmission models and outbreaks. This fills a gap left by settings that stop at binary screening, sentence retrieval or conclusion classification, because public health reviews require models to preserve evidence through screening and then normalise heterogeneous estimates into schema validated records with uncertainty and population context.
\clearpage
\section{Data Representativeness \& Ecological Validity}\label{app:data}
\label{app:data_representativeness}
We evaluate LLMs against systematic literature reviews (SLRs) produced by the Pathogen Epidemiology Review Group (PERG), whose curated article sets and extracted data are made available through the \texttt{epireview} and \texttt{priority-pathogen} R packages \cite{naidooEpireviewToolsUpdate2025, nashPrioritypathogens2026}. The evaluation spans up to seven WHO priority pathogens \cite{worldhealthorganizationPathogensPrioritizationScientific2024}, with screening evaluated across all seven and structured extraction evaluated across four (Ebola, Lassa, SARS-CoV-1, and Zika) for which PERG's published SLRs and extraction artefacts were both complete and consistently formatted. This section details the article corpus underlying these evaluations and provides evidence that the open-access subset used for benchmarking is representative of the broader PERG population corpus.

A potential concern with the data used is whether the 27.0\% open-access overlap (mentioned in \Cref{tab:article_count_summary_ours_pergs}) constitutes a biased sample of the broader PERG corpus, which would affect the validity of reported performance estimates. To assess this directly, we manually retrieved (through institutional access) and processed 1,004 closed-access articles across four pathogens (Ebola, Lassa, SARS-CoV-1, and Zika) that appear in the broader PERG population corpus but were not accessible through OpenAlex at the time of retrieval. We then constructed a matched stratified random sample from the \name{} open-access subset and ran the full \texttt{gpt-oss-120b} pipeline on both groups under identical conditions. 

\begin{table}[h!]
\centering
\footnotesize
\caption{\textbf{Representativeness of the open-access evaluation subset across pipeline stages.} Macro F1 with 95\% bootstrap confidence intervals for the \name{} open-access evaluation sample and a matched sample of 1,004 closed-access articles from the broader PERG population corpus, retrieved via institutional access across Ebola, Lassa, SARS-CoV-1, and Zika. Both groups are processed using \texttt{gpt-oss-120b} under identical conditions. $\Delta$ F1 denotes the signed difference (open-access minus population). Differences are small across stages (range: $-5.8$ to $+3.1$ pp). The full-text screening interval excludes zero, indicating lower performance for the open-access subset at this stage. The abstract screening and parameter extraction intervals include zero. Overall, these results suggest broad comparability between the open-access subset and the broader PERG population corpus, with a modest full-text screening gap.}
\setlength{\tabcolsep}{6pt}
\renewcommand{\arraystretch}{1.2}
\label{tab:representativeness}
\begin{tabular}{lccc}
\toprule
\textbf{Stage} & \textbf{\name{} (Open-Access)} & \textbf{PERG Population} & \textbf{$\Delta$ F1} \\
\midrule
Abstract Screening   & 0.877\ [0.853, 0.899] & 0.906\ [0.878, 0.930] & $-$0.028\ [$-$0.062,\ \ 0.010] \\
Full-text Screening  & 0.860\ [0.830, 0.886] & 0.918\ [0.890, 0.942] & $-$0.058\ [$-$0.093, $-$0.017] \\
Parameter Extraction & 0.728\ [0.654, 0.799] & 0.697\ [0.607, 0.781] & $+$0.031\ [$-$0.079,\ \ 0.142] \\
\bottomrule
\end{tabular}
\end{table}

\Cref{tab:representativeness} reports stage-level macro F1 with 95\% bootstrap confidence intervals for the \name{} open-access sample and the PERG population sample (closed-access). Performance is broadly comparable across evaluated stages. The largest difference occurs in full-text screening ($-5.8$ percentage points), where the confidence interval excludes zero, indicating slightly reduced sensitivity on the open-access subset at the stage. However, abstract screening and parameter extraction show overlapping confidence intervals and similar effect sizes, indicating that the main workflow trends remain qualitatively consistent across accessible and inaccessible subsets. Absolute performance estimates should therefore be interpreted relative to the lawfully retrievable evaluation corpus.\footnote{OpenAlex notes that cached PDFs retain their original copyright and grant no additional reuse rights (\url{https://developers.openalex.org/download/full-text-pdfs}). We therefore release identifiers, metadata and annotations, but not OCR text or PDF-derived Markdown unless redistribution is explicitly permitted by the source licence.}
\clearpage
\section{Evaluation Constructs}\label{app:additional_evaluation_details}


Construct validity asks whether the metrics measure the phenomenon a benchmark is designed to test, with limited contamination from confounding sources of variance~\citep{bean2025measuring}. We name the construct of \name{} as \textit{evidence handling fidelity}: the ability of an LLM-driven pipeline to identify, retain and structurally represent claims from primary scientific literature in a manner consistent with expert review practice. This construct is narrower than scientific knowledge synthesis. Synthesis additionally requires aggregation across studies, weighting of heterogeneous evidence, and the production of meta-analytic conclusions, none of which we evaluate. Stating the construct explicitly clarifies what \name{} can and cannot test, and bounds the population of claims we are entitled to make from its results.

Each metric operationalises one facet of this construct. Screening precision, recall and macro $F_1$ measure inclusion fidelity at the article level. Macro averaging is chosen because the construct treats correct inclusions and correct exclusions as both meaningful, and the class imbalance in PERG screening would otherwise allow the majority class to dominate the score. We report recall alongside $F_1$ because the operationally costly screening error is the missed relevant article, which propagates irrecoverably to all downstream stages. The Flagging, Counts and Extraction decomposition maps onto three separable failure modes that an expert reviewer distinguishes when inspecting machine-generated records: whether the system saw this kind of evidence, whether it produced the right number of records, and whether the field values were correct. The bipartite matching step encodes the construct decision that reference and predicted records have no canonical alignment within an article, and that a record which is partially correct (correct outbreak, wrong date) should contribute partial credit rather than be discarded. Field weights and the choice of Jaccard similarity for multi-value fields are explicit modelling choices and not neutral measurement, and we therefore treat cross-model rankings as more robust than absolute scores when comparing systems.

Three threats to construct validity remain in the present harness. \textit{Construct under-representation} arises because we do not evaluate meta-analytic synthesis or report writing, and our results consequently bound only the evidence handling component of an SLR. \textit{Construct-irrelevant variance} arises from OCR error, prompt formulation and the choice of field weights in the matching step, each of which introduces variation that is not part of the construct. We mitigate this by reporting standard deviations across pathogens and by relying on cross-model rankings rather than absolute scores when drawing comparative conclusions. \textit{Reference-label noise} arises because PERG annotations, although produced by trained reviewers under a documented protocol, are not free of disagreement. The expert validation in \Cref{app:human_expert_validation} provides a partial check on this by recovering field-level accuracy estimates from a fresh pool of reviewers, and we read the resulting numbers as a ceiling for the automated metrics rather than as a pure measure of system error.

\subsection{Article Screening}

We evaluate screening as a binary article-level decision $y\in\{\IN,\EX\}$ (\IN\ = include; \EX\ = exclude) against the PERG reference label. Let $\mathrm{TP}$ be articles correctly labelled as \IN, $\mathrm{FP}$ be those incorrectly labelled as \IN and $\mathrm{FN}$ be those incorrectly labelled as \EX; then
\[
\mathrm{Precision}=\frac{\mathrm{TP}}{\mathrm{TP}+\mathrm{FP}},\qquad
\mathrm{Recall}=\frac{\mathrm{TP}}{\mathrm{TP}+\mathrm{FN}},\qquad
F_1=\frac{2PR}{P+R},
\]
where $\mathrm{Precision}$ measures the reliability of \IN\ decisions and $\mathrm{Recall}$ measures how well we avoid assigning \EX\ to PERG-\IN\ papers. We report macro-$F_1$ to weight \IN\ and \EX\ performance equally, rather than letting the majority class dominate.

By default, article screening happens in two subsequent stages: first on the abstract and then on the full text. Full-text screening is therefore evaluated with different ablations so we can quantify both the stage-specific and holistic performance. We use three code-defined evaluation configurations for the ablations: (i) \textbf{AI abstract$\rightarrow$AI full-text}, where any abstract \EX\ forces final \EX; (ii) \textbf{Human abstract$\rightarrow$AI full-text} (PERG-conditioned), where any PERG abstract \EX\ forces final \EX; and (iii) \textbf{AI direct full-text}, which evaluates the AI full-text decision without filtering by abstract screening decisions.

\subsection{Data Extraction}\label{app:data_extraction_extended_methods}

\paragraph{Schema validation and data quality} Prior to evaluation, we validated and filtered our human reference extractions to ensure that only properly formatted annotations were compared. For fields typed as \texttt{Enum}s in the schemas outlined in \Cref{app:data_extraction_details}, we defined acceptable values based on the PERG REDCap survey schema, which standardises entries through dropdown lists. Other fields provide a multi-select option --- these we handled as \texttt{List[Enum]} types in the tool call schemas. We filter any reference extractions where \texttt{Enum}-typed values do not agree with the schema, in order to avoid penalising LLMs for extractions it is not allowed to produce. Because of schema verification applied in the tool-calling stage, LLMs produce no such invalid extractions.

After validation, we aligned articles using shared identifiers, retaining only articles labelled \IN\ by both PERG and LLMs (through \name{} workflow). This intersection matches the reference-labelled data to our article pool (\Cref{tab:article_count_summary_ours_pergs}), and thus avoids counting errors due to paper availability from the article screening stages.

\paragraph{Evaluation Framework} We evaluated extraction performance according to three measures: \textit{Flagging}, \textit{Count}, and \textit{Extraction}. All three measures are operationalised with standard classification metrics, specifically, we define and collect precision and recall for each.


\medskip
\noindent\textbf{Flagging}\enskip measures whether LLMs correctly identify the relevant data types to extract from each article. This measure considers all $\langle\mathrm{article}, \mathrm{data\_type}\rangle$ pairs, assigning labels with the functions
\begin{subequations}\label{eq:flagging}
\begin{align*}
    y(\langle\mathrm{article}, \mathrm{data\_type}\rangle) &= \begin{cases}
        1 & \text{There is a human extraction of }\mathrm{data\_type}\text{ from }\mathrm{article} \\
        0 & \text{otherwise}
    \end{cases} \\
    \hat{y}(\langle\mathrm{article}, \mathrm{data\_type}\rangle) &= \begin{cases}
        1 & \text{LLM identifies }\mathrm{data\_type}\text{ as relevant in }\mathrm{article} \\
        0 & \text{otherwise}
    \end{cases}
\end{align*}
\end{subequations}
and calculating precision and recall on these labels as in a standard binary classification task.

\medskip
\noindent\textbf{Count}\enskip measures whether the overall volume of LLM extractions agrees with those in the human reference-labelled data, irrespective of any agreement between the extraction contents. We operationalise this measure using a partial credit scheme: if an article had $n$ models in the reference and our extractor identified $\hat{n}$ models, we counted true positives as correctly matching counts
$$\mathrm{TP} = \min(n, \hat{n}),$$
false positives as excess extractions
$$\mathrm{FP} = \max(0, \hat{n} - n),$$
and false negatives as missed extractions
$$\mathrm{FN} = \max(0, n - \hat{n}).$$
For example, if the reference contained 2 data points but we extracted 5, we would receive credit for the 2 correct extractions ($\mathrm{TP}=2$), be penalised for 3 spurious models ($\mathrm{FP}=3$), and would receive no penalty for missed models ($\mathrm{FN}=0$). We sum all counts across all common articles and calculate precision and recall as standard.

\medskip
\noindent\textbf{Extractions}\enskip faced a more complex matching challenge: while extractions can be trivially compared by raw count, they consist of many metadata fields, and lack unique identifiers to establish canonical correspondence. Matching every field value exactly is an unreasonably challenging task, and it provides no measure beyond absolute correspondence. To assess the field-level quality of our extractions, we first established optimal one-to-one correspondences between human reference annotations and LLM extractions within each article by computing pairwise similarity.

For each extraction pair, we defined a subset of key fields $\mathcal{F}$ from the fields defined in \Cref{app:data_extraction_details} and compared these using normalised weights. The similarity between a true extraction $E$ and an LLM extraction $\hat{E}$ was computed as
\begin{equation*}
s(E, \hat{E}) = \sum_{k \in \mathcal{F}} w_k \cdot d_k(E[k], \hat{E}[k]),
\end{equation*}
where $w_k$ is the normalised weight for field $k$ (with $\sum_{k \in \mathcal{F}} w_k = 1$) and $d_k$ is the Jaccard similarity between fields in the extractions
\begin{equation*}
d_k(v, \hat{v}) = J(v, \hat{v}) = \dfrac{|v \cap \hat{v}|}{|v \cup \hat{v}|}.
\end{equation*}

We then applied the modified Jonker--Volgenant algorithm~\citep{jonker1987shortest} using SciPy's \texttt{scipy.optimize.linear\_sum\_assignment()}\footnote{\url{https://docs.scipy.org/doc/scipy/reference/generated/scipy.optimize.linear_sum_assignment.html}} function to the cost matrix (cost $= 1 - s$), finding the matching that maximised total similarity. 

\Cref{tab:matching_example} illustrates this optimal bipartite matching on an example. Suppose a single article has two reference models extracted by expert epidemiologists (PERG), while the LLM produces three extractions. Because the sets differ in size, no perfect bijection exists, and the algorithm must leave at least one LLM extraction unmatched. For explanation purposes we restrict to two fields: \texttt{model\_type}, a single-value field scored by exact match ($\delta_{\mathrm{type}}\in\{0,1\}$), and \texttt{interventions}, a multi-value field scored by Jaccard similarity ($J_{\mathrm{int}} = |v\cap\hat{v}|/|v\cup\hat{v}|$). With equal weights, each pairwise cell reduces to $s_{ij} = 0.5\,\delta_{\mathrm{type}} + 0.5\,J_{\mathrm{int}}$. 

The algorithm correctly recovers both reference correspondences -- achieving total similarity $2.00$ out of a maximum possible $2.00$ -- while the spurious LLM-extraction M$_2$ is left unmatched and counted as a false positive under the Count metric. Crucially, this unmatched extraction incurs a Count penalty only; it does not contaminate field-level Extraction scores, ensuring over-extraction and extraction inaccuracy are penalised independently.

Once optimal correspondences are established, we evaluated each field
within each matched pair to compute field-level precision and recall. For single-value fields, we counted true positives as sets of equal values, false positives as all LLM-extracted values with no or an unequal match, and false negatives as all human reference values with no or an unequal match.

For multi-value fields, we defined
\begin{align*}
    \mathrm{TP} &= |v \cap \hat{v}| \\
    \mathrm{FP} &= |\hat{v} \setminus v| \\
    \mathrm{FN} &= |v \setminus \hat{v}|
\end{align*}
where $v\in E$ and $\hat{v}\in\hat{E}$ are sets of values. Aggregating across all matched pairs and articles, we computed precision and recall as standard.


\begin{table}[t!]
\centering
\caption{%
  \textbf{Optimal bipartite matching example: 2 PERG reference models,
  3 LLM-extracted models.}
  \textbf{(a)}~Input field values (two fields shown for illustration).
  \textbf{(b)}~Pairwise similarity matrix $\mathbf{S}$.
  \textbf{(c)}~Optimal matching; LLM M$_2$ is unmatched (FP).
  \textbf{(d)}~Per-cell similarity calculations for entries of $\mathbf{S}$.
}
\label{tab:matching_example}

\vspace{0.4cm}

\begin{center}
  \small\textbf{(a)\;Model Field Values}\\[7pt]
  \footnotesize
  \begin{tabular}{@{}lll@{}}
    \toprule
    \textbf{Model} & \textbf{Type} & \textbf{Interventions}\\
    \midrule
    \multicolumn{3}{@{}l@{}}{\textit{PERG Reference}}\\[2pt]
    \hspace{5pt}PERG M$_1$ & SIR  & Vaccination\\
    \hspace{5pt}PERG M$_2$ & SEIR & Quarantine; Vaccination\\
    \addlinespace[5pt]
    \multicolumn{3}{@{}l@{}}{\textit{AI-Extracted}}\\[2pt]
    \hspace{5pt}LLM M$_1$ & SIR  & Vaccination\\
    \hspace{5pt}LLM M$_2$ & SIR  & Treatment\\
    \hspace{5pt}LLM M$_3$ & SEIR & Quarantine; Vaccination\\
    \bottomrule
  \end{tabular}
\end{center}
\vspace{0.55cm}
\begin{minipage}[t]{0.49\linewidth}
  \centering
  \small\textbf{(b)\;Pairwise Similarity Matrix $\mathbf{S}$}\\[1pt]
  \footnotesize
  \setlength{\abovedisplayskip}{0pt}
  \[
    \mathbf{S}
    \;=\;
    \overset{%
      \begin{array}{@{}ccc@{}}
        \scriptstyle\text{LLM M}_{1} &
        \scriptstyle\text{LLM M}_{2} &
        \scriptstyle\text{LLM M}_{3}
      \end{array}}{%
      \left[\begin{array}{@{\hspace{6pt}}c@{\hspace{10pt}}c@{\hspace{10pt}}c@{\hspace{6pt}}}
        1.00 & 0.50 & 0.25 \\[5pt]
        0.25 & 0.00 & 1.00
      \end{array}\right]}
    \;\begin{array}{@{}l@{}}
        \scriptstyle\leftarrow\;\text{PERG M}_{1}\\[5pt]
        \scriptstyle\leftarrow\;\text{PERG M}_{2}
      \end{array}
  \]
\end{minipage}%
\hfill
\begin{minipage}[t]{0.49\linewidth}
  \centering
  \small\textbf{(c)\;Optimal Matching}\\[6pt]
  \footnotesize
  \begin{tabular}{@{}l@{\hspace{12pt}}r@{}}
    PERG M$_1\;\leftrightarrow\;$LLM M$_1$ & $s=1.00$\\[2pt]
    PERG M$_2\;\leftrightarrow\;$LLM M$_3$ & $s=1.00$\\[2pt]
    LLM M$_2$\;:\;unmatched\quad(FP)        & \multicolumn{1}{c}{---}\\
    \cmidrule(l){2-2}
    Total similarity                             & $2.00$\\
  \end{tabular}
\end{minipage}

\vspace{0.55cm}

\begin{center}
\small\textbf{(d)\;Similarity Calculations}
\end{center}
\begin{tcolorbox}[
  enhanced,
  colback=white,
  colframe=white,
  boxrule=0pt,
  arc=2pt,
  left=8pt, right=8pt, top=5pt, bottom=6pt,
  title={
    \normalfont\small
    $s_{ij}
      = \underbrace{0.5\,\delta_{\mathrm{type}}}_{\text{exact match on type}}
      + \underbrace{0.5\,J_{\mathrm{int}}}_{\text{Jaccard on interventions}}$},
  fonttitle=\sffamily\small,
  coltitle=black,
  colbacktitle=gray!14,
  attach boxed title to top center={yshift=-2mm, xshift=4mm},
  boxed title style={boxrule=0.4pt, arc=1pt}
]
\footnotesize
\renewcommand{\arraystretch}{1.35}
\centering
\begin{tabular}{@{\hspace{2pt}}c @{\hspace{10pt}} l @{\hspace{10pt}} l @{\hspace{10pt}} r @{\hspace{2pt}}}
  & \textbf{Type match\;$\delta_{\mathrm{type}}$}
  & \textbf{Jaccard\;$J_{\mathrm{int}}$}
  & $\boldsymbol{s_{ij}}$\\
  \hline
  $\mathbf{S}_{1,1}$ & $\mathbf{SIR = SIR}$:\;1.0
       & $J(\{V\},\{V\})     =1/1=1.0$
       & $\mathbf{1.00}$\\
  $\mathbf{S}_{1,2}$ & $\mathbf{SIR = SIR}$:\;1.0
       & $J(\{V\},\{T\})     =0/2=0.0$
       & $\mathbf{0.50}$\\
  $\mathbf{S}_{1,3}$ & $\mathbf{SIR \neq SEIR}$:\;0.0
       & $J(\{V\},\{Q,V\})   =1/2=0.5$
       & $\mathbf{0.25}$\\
  $\mathbf{S}_{2,1}$ & $\mathbf{SEIR \neq SIR}$:\;0.0
       & $J(\{Q,V\},\{V\})   =1/2=0.5$
       & $\mathbf{0.25}$\\
  $\mathbf{S}_{2,2}$ & $\mathbf{SEIR \neq SIR}$:\;0.0
       & $J(\{Q,V\},\{T\})   =0/3=0.0$
       & $\mathbf{0.00}$\\
  $\mathbf{S}_{2,3}$ & $\mathbf{SEIR = SEIR}$:\;1.0
       & $J(\{Q,V\},\{Q,V\}) =2/2=1.0$
       & $\mathbf{1.00}$\\
\end{tabular}

\vspace{4pt}
{\footnotesize
V\,=\,Vaccination,\enspace Q\,=\,Quarantine,\enspace T\,=\,Treatment;\enspace
$J(A,B) = |A\cap B|\,/\,|A\cup B|$.}
\end{tcolorbox}

\end{table}

\subsubsection*{Data Extraction: Parameters}

Parameter extraction is more varied than model and outbreak extraction. While there is only one $\mathrm{data\_type}$ for each of model and outbreak extraction, parameters are broken down into nine distinct parameter \textit{classes} (listed in \Cref{app:parameters_extended_extraction_process}) each with different fields to extract. Therefore, we resolve nine parameter $\mathrm{data\_type}$s at the level of parameter classes and calculate Flagging and Count metrics for each of these separately.

We defined our key parameter fields as
\begin{align*}
    \mathcal{F} = \{&\texttt{parameter\_class}, \texttt{parameter\_type}, \texttt{value}, \texttt{unit}, \texttt{method}, \texttt{value\_type}, \\& \texttt{statistical\_approach}, \texttt{paired\_uncertainty}, \texttt{single\_type\_uncertainty}, \\
    &\texttt{population\_sex}, \texttt{population\_group}, \texttt{population\_sample\_type}\},
\end{align*}
ensuring that each sub-stage of value extraction, uncertainty extraction, and population context extraction are represented by multiple fields common across parameter classes. We normalise weights $w_k$ so as to make each sub-stage equally important in determining similarity. This reflects the multi-stage structure of extraction (Appendix~\ref{app:data_extraction_details}), where each stage is gated on the previous and therefore warrants equal evaluative weight.  Moreover perturbing within-group field weights by $\pm10\%$ leaves cross-model rankings unchanged. Fields are grouped as follows:
\vspace{-8pt}
\begin{itemize}[leftmargin=*, itemsep=-3pt]
\item \textit{Categorical fields} (2 fields): parameter class; parameter type;
\item \textit{Value fields} (3 fields): value; unit; method;
\item \textit{Uncertainty fields} (4 fields): value type; statistical approach; single type uncertainty; paired uncertainty;
\item \textit{Population fields} (3 fields): population sex; population group; population sample type.
\end{itemize}

\subsubsection*{Data Extraction: Transmission Models}

Table~\ref{tab:filtering_stats} shows the filtering statistics across pathogens: human reference datasets had between 3.85\% (Lassa) and 23.14\% (Zika) invalid entries removed.

\begin{table}[h]
\footnotesize
\centering
\caption{\textbf{Validation statistics for PERG reference data and AI-extracted transmission model annotations across four pathogens.} PERG entries contained invalid field values due to manual data entry inconsistencies, while AI-extracted values showed no invalid entries due to structured schema enforcement during extraction. The AI-extracted numbers below are with \texttt{gpt-oss-120b}, but we find across all LLMs the number of invalid entries are 0.}
\label{tab:filtering_stats}
\begin{tabular}{lcccc}
\toprule
\textbf{Pathogen} & \textbf{PERG Total} & \textbf{PERG Invalid} & \textbf{Invalid (\%)} & \textbf{AI-extracted Total} \\
\midrule
Lassa & 52 & 2 & 3.85 & 19 \\
Ebola & 294 & 46 & 15.7 & 239 \\
SARS & 112 & 8 & 7.14 & 85 \\
Zika & 229 & 53 & 23.1 & 132 \\
\bottomrule
\end{tabular}
\end{table}

For data extraction for models, we defined our key fields as
\begin{align*}
    \mathcal{F} = \{&\texttt{model\_type}, \texttt{compartmental\_type}, \texttt{stoch\_deter}, \texttt{theoretical\_model},\\ &\texttt{assumptions}, \texttt{interventions\_type}, \texttt{transmission\_route}\}.
\end{align*}

\subsubsection*{Data Extraction: Outbreaks}
Table~\ref{tab:outbreak_filtering_stats} shows the filtering statistics across pathogens: PERG datasets had between 0\% (Lassa) and 9.43\% (Zika) invalid entries removed.

\begin{table}[h]
\footnotesize
\centering
\caption{\textbf{Validation statistics for PERG reference data and AI-extracted outbreak annotations across two pathogens.} PERG entries contained invalid field values due to manual data entry inconsistencies, while AI-extracted values showed 0\% invalid entries due to structured schema enforcement during extraction.}
\label{tab:outbreak_filtering_stats}
\begin{tabular}{lcccc}
\toprule
\textbf{Pathogen} & \textbf{PERG Total} & \textbf{PERG Invalid} & \textbf{Invalid (\%)} & \textbf{AI-Extracted Total} \\
\midrule
Lassa & 30 & 0 & 0.00 & 62 \\
Zika & 159 & 15 & 9.43 & 240 \\
\bottomrule
\end{tabular}
\end{table}

For data extraction for outbreaks, we defined our key fields as
\begin{align*}
        \mathcal{F} = \{&\texttt{outbreak\_start\_day}, \texttt{outbreak\_start\_month}, \texttt{outbreak\_start\_year},\\&
        \texttt{outbreak\_end\_day}, \texttt{outbreak\_end\_month}, \texttt{outbreak\_end\_year},\\&
        \texttt{cases\_confirmed}, \texttt{deaths}, \texttt{outbreak\_country}, \texttt{outbreak\_location},\\& \texttt{detection\_mode}, \texttt{pre\_outbreak\_status}\}.
\end{align*}

Weights $w_k$ were determined by the discriminative power of each field $k$ for identifying unique outbreak events. \texttt{outbreak\_country}, \texttt{outbreak\_start\_year}, \texttt{cases\_confirmed}, and \texttt{deaths} received weights of $1.0$, while supporting temporal fields (\texttt{outbreak\_start\_month}, \texttt{outbreak\_end\_year}) received weights of $0.6$--$0.8$, and contextual fields (\texttt{outbreak\_location}, \texttt{mode\_of\_detection}) received weights of $0.5$--$0.7$. Given the limited number of outbreak records ($n=169$), we verified that setting all weights to $1.0$ does not produce a statistically significant change in extraction $F_1$ at the $95\%$ level. The reported weights reflect a cascading importance ordering.

To provide interpretable summaries of extraction performance, we grouped the 17 outbreak fields into four categories based on their epidemiological function:
\vspace{-8pt}
\begin{itemize}[leftmargin=*, itemsep=-3pt]
\item \textit{Temporal Features} (7 fields): outbreak start/end dates (year, month, day) and duration;
\item \textit{Geographic and Spatial Features} (2 fields): outbreak country and specific location;
\item \textit{Case Burden} (5 fields): confirmed, suspected, asymptomatic, and severe case counts, plus deaths;
\item \textit{Epidemiological Context and Metadata} (3 fields): mode of detection, pre-outbreak status, and asymptomatic transmission description.
\end{itemize}

\subsection{Human Expert Validation}\label{app:human_expert_validation}

For human expert validation, we recruited six epidemiologists to complete a series of form submissions to grade \name{} workflow generated data extractions (with \texttt{gpt-oss-120b}). Each epidemiologist was on-boarded with the expectation to spend up to $10$ hours on the validation process over $1$ to $2$ weeks as their availability permitted. We did not assign experts randomly across parameters, models, and outbreaks -- instead, we considered expertise with specific pathogens and familiarity with specific SLR workflows when making assignments. Our assignments resulted in three epidemiologists completing validation solely for parameters, two solely for models, and the final sixth epidemiologist completing validation across all three data modalities.

For each data type in parameters, models, and outbreaks, and for each pathogen in Lassa, Ebola, SARS, and Zika, we sample screened articles randomly without replacement until we generate subsamples guaranteed to exceed the time commitment from each expert. The experts are then instructed to proceed through their assigned extractions in order. Despite normalising for counts across different pathogens, different articles may have varying numbers of extractions, and these extractions may take varying amounts of time to grade. Thus, our experimental setup does not guarantee parity across pathogens or across data types.

Experts are onboarded with a private GitHub repository that contains the Markdown extractions from our OCR model (\Cref{sec:pdf_to_markdown_conversion}), along with Markdown documents rendering the structured data extractions in a readable format. Submissions are collected through Google Forms. Each form proceeds through groups of questions in the same order. Each question contains an optional free-text field for providing context, which we use to collect and synthesise qualitative impressions of the pipeline as well as specific error patterns.

The groups of questions in each form cover the following:
\begin{enumerate}
    \item The expert records the article identifier, pathogen identifier, and the pathogen.
    \item The expert assesses whether the Markdown document has any significant issues that would affect data extraction.
    \item Before looking at the \name{}-extracted data, the expert determines whether there is any relevant data in the article to extract.
    \item The expert rates their particular extraction for overall relevance.
    \item The expert answers a series of yes-or-no questions to validate the accuracy of each extracted field.
    \item The expert grades the overall pipeline competence using a Likert scale rating between $1$ and $7$. We provide these particular descriptions to calibrate the Likert scale:\begin{itemize}
        \item ``1" means ``the system gets nothing right; I couldn't use it to speed up my process at all."
        \item ``4" means ``the system identifies some things but struggles with edge cases; I could use it with moderate supervision / secondary screening."
        \item ``7" means ``the system is perfectly capable of doing all parameter extraction for me."
    \end{itemize}
    \item The expert provides a self-reported estimate of the time they took to complete the survey.
\end{enumerate}

\clearpage

\clearpage
\section{Pipeline Statistics: Data Processed \& Time}\label{app:pipeline_statistics}

\subsection{Runtime Statistics}
\label{runtime_stats}
\subsubsection*{Article counts across SLR stages}
To contextualise runtime estimates for both human and automated pipelines, we first summarise the approximate number of articles processed at each stage of the systematic literature review (SLR). These counts are intended to reflect annotator workload rather than final inclusion totals, and correspond to successive filtering stages commonly used in SLR workflows. In particular, counts decrease substantially between title and abstract screening, full-text screening, and data extraction as relevance criteria are progressively applied.

\begin{table}[H]
\footnotesize
    \centering
    \caption{
    \textbf{Estimated number of articles reviewed by human annotators at PERG across successive stages of each systematic literature review (SLR).}
    Counts for \emph{Title and Abstract Screening} correspond to records remaining after deduplication and exclusion of entries with missing or empty abstract metadata.
    \emph{Full-text Screening} includes articles flagged as potentially relevant during abstract screening and advanced for full-text review.
    \emph{Data Extraction} represents articles deemed suitable for extracting structured, task-relevant evidence.
    All values are estimates intended to reflect annotator workload at each phase rather than finalised inclusion totals.
    }
    \label{tab:perg_slr_article_counts}
    \begin{tabular}{lrrr}
        \toprule
        \bf Pathogen & \bf Title \& Abstract Screening & \bf Full-text Screening & \bf Data Extraction \\
        \midrule
        Ebola   & 11{,}605 & 1{,}674 & 522 \\
        Lassa   & 2{,}131  & 512     & 193 \\
        SARS    & 12{,}280 & 878     & 289 \\
        Zika    & 10{,}510 & 1{,}343 & 574 \\
        \midrule
        Average & 9{,}132 & 1{,}102 & 395 \\
        \bottomrule
    \end{tabular}
\end{table}

\subsubsection*{Runtime estimation methodology}
Using the average article counts from \Cref{tab:perg_slr_article_counts}, we estimate total processing time for both the PERG human SLR workflow and the \name{} automated pipeline. Per-article time estimates for PERG were obtained through consultation with a Research Associate at PERG who routinely contributes to SLR projects. \name{} workflow runtimes (with \texttt{gpt-oss-120b}) were measured directly from pipeline execution logs. All per-article times are converted to hours and multiplied by the average number of articles processed at each stage. We observe differing run-times based on latency and throughput of models, \texttt{gpt-oss-120b} represents the upper-bound across the spectrum.

\begin{table}[H]
\footnotesize
\centering
\caption{\textbf{Comparison of average human time investment (PERG) versus automated processing time (\name{} workflow) across systematic literature review stages.}
The table reports average articles processed across the four pathogens (Ebola, Lassa, SARS, Zika), average per-article time (in seconds), and total processing time (in hours), highlighting efficiency gains from automation. \name{} workflow timings are computed using \texttt{gpt-oss-120b} as the underlying model. PDF-to-Markdown conversion is applied to all 9{,}132 retrieved articles to preserve the option of direct full-text screening across the complete corpus.}
\label{tab:perg_vs_\name{}_runtime}

\renewcommand{\arraystretch}{1.15}
\begin{tabular}{l|r|rr|rr}
\toprule
\textbf{Stage} 
& \makecell{\textbf{Articles}\\\textbf{(Avg.)}} 
& \makecell{\textbf{\name{}}\\\textbf{(s/article)}} 
& \makecell{\textbf{PERG}\\\textbf{(s/article)}} 
& \makecell{$\overline{\textbf{\name{}}}$\\\textbf{(Hours)}} 
& \makecell{$\overline{\textbf{PERG}}$\\\textbf{(Hours)}} \\
\midrule
Article Retrieval 
& 9{,}132 
& 0.63 
& 0 
& 1.6 
& 0.00 \\
Title \& Abstract Screening 
& 9{,}132 
& 0.63 
& 45 
& 1.6 
& 114.2 \\
PDF-to-Markdown Conversion 
& 9{,}132 
& 1.1 
& 0 
& 2.8 
& 0.00 \\
Full-text Screening 
& 1{,}102 
& 2.0 
& 240 
& 0.62 
& 73.5 \\
Data Extraction 
& 395 
& 122.1 
& 1{,}800 
& 13.4 
& 197.5 \\
\midrule
\textbf{Total} 
& -- 
& -- 
& -- 
& \textbf{20.0} 
& \textbf{385.1} \\
\bottomrule
\end{tabular}
\end{table}

\subsubsection*{PERG runtime calculations} 
\textbf{Title and abstract screening} at PERG is estimated at $30$ to $60$ seconds per article. Assuming an average of $45$ seconds ($0.0125$ hours) per article and $9{,}132$ articles screened on average, the estimated time is $9{,}132 \times 0.0125 = 114.15$ hours. 

\textbf{Full-text screening} is estimated at $2$ to $6$ minutes per article. Assuming an average of $4$ minutes ($0.066\overline{6}$ hours) per article and $1{,}102$ articles screened on average, the estimated time is $1{,}102 \times 0.066\overline{6} = 73.47$ hours. 

\textbf{Data extraction} is estimated at a median of $30$ minutes ($0.5$ hours) per article. With $395$ articles processed on average, the estimated time is $395 \times 0.5 = 197.50$ hours.

\subsubsection*{\name{} workflow (\texttt{gpt-oss-120b}) runtime calculations}
\textbf{Article retrieval} in \name{} requires $0.63$ seconds per article. With $9{,}132$ articles retrieved on average, the estimated time is $1.6$ hours.

\textbf{Title and abstract screening} requires $0.63$ seconds per article. With $9{,}132$ articles screened on average, the estimated time is $1.6$ hours.

\textbf{PDF-to-Markdown conversion} requires $1.1$ seconds per article. PDF-to-Markdown conversion is applied to all $9{,}132$ retrieved articles to preserve the option of direct full-text screening across the complete corpus, giving an estimated time of $2.8$ hours. This estimate reflects parallel execution with $14$ concurrent requests, yielding an average processing time of $0.05$ seconds per page and $1.1$ seconds per document. Under sequential execution, the measured processing time increases substantially to an average of $0.95$ seconds per page and $16.47$ seconds per document.

\textbf{Full-text screening} requires $2.0$ seconds per article. With $1{,}102$ articles screened on average, the estimated time is $0.62$ hours.

\textbf{Data extraction} requires $122.1$ seconds per article, comprising outbreak identification, model extraction, and parameter extraction. With $395$ articles processed on average, the estimated time is $13.4$ hours.

\subsection{Token Usage and Operational Cost of \name{} harness}
\label{app:pipeline_statistics_costs}

\begin{figure*}[h]
    \centering
    \includegraphics[width=\linewidth]{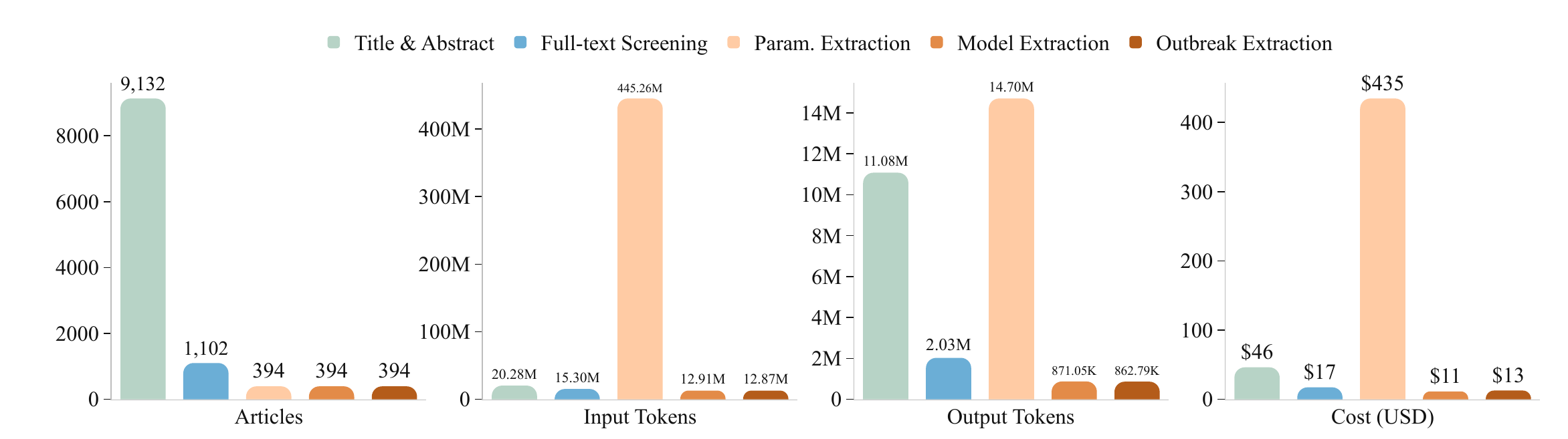}
    \caption{
\textbf{Articles processed, scaled token and costs by pipeline stage across models.}
We report the average number of articles reaching each stage, and the corresponding total input tokens, output tokens, and USD cost per stage averaged across models. Token totals are computed by multiplying per article token usage (Table~\ref{tab:token_usage_by_stage}) by the average article counts per stage (Table~\ref{tab:perg_slr_article_counts}).
Parameter extraction dominates overall compute, with substantially higher input and output token totals than other stages.
Title and abstract screening processes the largest volume of articles but contributes comparatively less to total cost.
}
    \label{fig:tokens_costs_articles_split}
\end{figure*}

Using the average article counts reported in Table~\ref{tab:perg_slr_article_counts}, we estimate total token usage and USD cost
across pipeline stages by combining per-article token statistics with model-specific
pricing. Figure~\ref{fig:tokens_costs_articles_split} summarises the resulting
distribution of articles processed, aggregate input tokens, output tokens, and total
cost by stage, averaged across models. Per-article input and output token usage by
stage and model is reported in Table~\ref{tab:token_usage_by_stage}. Total stage costs
are computed by multiplying mean per-article token usage by the average number of
articles reaching each stage, and then applying published per-million-token pricing
for both input and output tokens. All prices used in these calculations are retrieved
directly from the primary API pricing documentation of each model provider at the time
of evaluation.
Under this pricing regime, parameter extraction dominates overall compute and cost due
to substantially higher input and output token volumes, while title and abstract
screening processes the largest number of articles but contributes comparatively
little to total cost. All reported costs reflect managed API usage; alternative cost
estimates could be derived for deployments hosted on dedicated GPU nodes, where pricing
would depend on hardware configuration, utilisation, and amortisation assumptions
rather than per-token billing.

\footnotetext{
\textbf{GPT-OSS-120B:} \url{https://openrouter.ai/openai/gpt-oss-120b}. \\
\textbf{GPT-5.2:} \url{https://developers.openai.com/api/docs/pricing/}. \\
\textbf{DeepSeek-V3.2:} \url{https://api-docs.deepseek.com/quick_start/pricing}. \\
\textbf{Kimi-K2.5:} \url{https://platform.moonshot.ai/docs/pricing/chat}. \\
\textbf{GLM-4.7:} \url{https://docs.z.ai/guides/overview/pricing}.
}

\begin{table*}[h]
\centering
\caption{
\textbf{Per article token usage and estimated cost by stage and model.}
We report mean input tokens, output tokens, and USD cost for processing a \textbf{single article} at each stage.
\highP{Green} marks the minimum and \lowP{Red} marks the maximum within each row and subcolumn.
}
\label{tab:token_usage_by_stage}
\setlength{\tabcolsep}{3.5pt}
\renewcommand{\arraystretch}{1.2}
{\footnotesize
\begin{tabular}{llccccccc}
\toprule
\textbf{Model} & &
\shortstack[c]{\textbf{Title \&}\\\textbf{Abstract}\\\textbf{Screening}} &
\shortstack[c]{\textbf{Article}\\\textbf{Screening}\\\textbf{(AI Cond.)}} &
\shortstack[c]{\textbf{Parameter}\\\textbf{Extraction}} &
\shortstack[c]{\textbf{Model}\\\textbf{Extraction}} &
\shortstack[c]{\textbf{Outbreak}\\\textbf{Extraction}} &
\textbf{Overall} \\
\midrule
\multirow{3}{*}{\texttt{GPT-OSS-120B (High)}}
  & Input Tok.  & \lowP{2.3K}         & \lowP{16.9K}        & \highP{510.2K}       & 35.9K               & \lowP{49.9K}         & 615.3K  \\
  & Output Tok. & 1.2K                & 1.1K                & 19.8K                & 1.9K                & 2.6K                 & 26.7K   \\
  & Cost (USD)  & \highP{${<}0.01$}   & \highP{${<}0.01$}   & \highP{0.02}         & \highP{${<}0.01$}   & \highP{${<}0.01$}    & \highP{0.02} \\
\midrule
\multirow{3}{*}{\texttt{GPT-5.2 (High)}}
  & Input Tok.  & 2.2K                & 13.1K               & 961.1K               & 31.5K               & 32.4K                & 1040.4K \\
  & Output Tok. & \highP{0.6K}        & 1.4K                & \lowP{91.1K}         & 2.1K                & \lowP{3.5K}          & \lowP{98.7K} \\
  & Cost (USD)  & \lowP{0.01}         & \lowP{0.04}         & \lowP{2.95}          & \lowP{0.08}         & \lowP{0.10}          & \lowP{3.20} \\
\midrule
\multirow{3}{*}{\texttt{DeepSeek-V3.2}}
  & Input Tok.  & \highP{2.2K}        & \highP{12.9K}       & 523.2K               & \lowP{37.2K}        & 26.1K                & \highP{601.6K} \\
  & Output Tok. & 0.6K                & \highP{0.8K}        & \highP{3.0K}         & \lowP{3.1K}         & \highP{0.2K}         & \highP{7.8K} \\
  & Cost (USD)  & ${<}0.01$           & ${<}0.01$           & 0.14                 & 0.01                & ${<}0.01$            & 0.17 \\
\midrule
\multirow{3}{*}{\texttt{Kimi-K2.5}}
  & Input Tok.  & 2.3K                & 13.1K               & 605.5K               & \highP{29.3K}       & 29.5K                & 679.8K \\
  & Output Tok. & \lowP{2.0K}         & 2.7K                & 40.0K                & 2.2K                & 2.7K                 & 49.6K  \\
  & Cost (USD)  & ${<}0.01$           & 0.01                & 0.48                 & 0.02                & 0.02                 & 0.55   \\
\midrule
\multirow{3}{*}{\texttt{GLM-4.7}}
  & Input Tok.  & 2.2K                & 13.4K               & \lowP{3050.4K}       & 29.7K               & \highP{25.4K}        & \lowP{3121.1K} \\
  & Output Tok. & 1.6K                & \lowP{3.1K}         & 32.7K                & \highP{1.7K}        & 2.0K                 & 41.1K  \\
  & Cost (USD)  & ${<}0.01$           & 0.01                & 1.90                 & 0.02                & 0.01                 & 1.96   \\
\bottomrule
\end{tabular}
}
\end{table*}

\clearpage

\clearpage
\section{Extended Evaluation Results}\label{app:model_ablations}\label{app:pipeline_stage_eval}
In this section we report the disaggregated and extended version of results presented in the main text. The section is structured into three parts. First, we report model-level results from evaluating five LLMs with the \name{} harness workflow. Second, we report field-level results for \texttt{gpt-oss-120b}, showing how the harness supports detailed failure analysis beyond aggregate model comparisons. Third, we report expanded article screening strategy (ablation) results for the \texttt{gpt-oss-120b} run.

\subsection{Results Across Models}
Tables~\ref{tab:ta_screening_model_ablation}--\ref{tab:outbreak_extraction_model_ablation} report pathogen-level metrics for the five models summarised in \Cref{fig:model_ablation}. Results are grouped by article screening, parameter extraction, transmission model extraction and outbreak extraction.

\subsubsection*{Title \& Abstract Screening}
At title and abstract screening (Table~\ref{tab:ta_screening_model_ablation}), the spread in overall $F_1$ from $0.62$ (\texttt{DeepSeek-V3.2}) to $0.77$ (\texttt{Kimi-K2.5}) is driven almost entirely by recall rather than precision. \texttt{DeepSeek-V3.2} and GPT-5.2 are the two most precise models ($0.83$ and $0.82$ respectively) yet rank last and fourth on $F_1$, with recalls of $0.59$ and $0.61$ against \texttt{Kimi-K2.5}'s $0.75$. \texttt{Kimi-K2.5} is the best-performing model for all seven pathogens. Nipah is the worst-performing pathogen for three models and sits at or below $0.72$ for all five. This is consistent with it being one of the smallest and most heterogeneous corpora in the PERG dataset. The same pathogen has the narrowest precision-recall gap across models, suggesting that the difficulty is intrinsic to the articles rather than a model-specific calibration issue.

\begin{table*}[h!]
\centering
\caption{\textbf{Title and abstract screening metrics across LLMs.} \highP{Green} and \lowP{Red} denote the best- and worst-performing pathogens for each model (in terms of $F_1$ score). \bestM{Bold} indicates the best-performing model for each pathogen, and \secondM{Underline} indicates the second-best. $P=\text{precision}$; $R=\text{recall}$; $F_1=\text{F1-Score}$.}
\label{tab:ta_screening_model_ablation}
\setlength{\tabcolsep}{3pt}
\renewcommand{\arraystretch}{1.1}
{\footnotesize
\begin{tabular}{l|ccc|ccc|ccc|ccc|ccc}
\toprule
\textbf{Pathogen} & \multicolumn{3}{c|}{\texttt{gpt-oss-120b}} & \multicolumn{3}{c|}{\texttt{GPT-5.2}} & \multicolumn{3}{c|}{\textbf{\texttt{DeepSeek-V3.2}}} & \multicolumn{3}{c|}{\textbf{\texttt{Kimi-K2.5}}} & \multicolumn{3}{c}{\textbf{\texttt{GLM-4.7}}} \\
& $P$ & $R$ & $F1$ & $P$ & $R$ & $F1$ & $P$ & $R$ & $F1$ & $P$ & $R$ & $F1$ & $P$ & $R$ & $F1$ \\
\midrule
\textbf{Marburg} & 0.80 & 0.64 & \lowP{\secondM{0.69}} & 0.97 & 0.58 & 0.62 & 0.97 & 0.55 & 0.58 & 0.79 & 0.65 & \lowP{\bestM{0.69}} & 0.88 & 0.61 & 0.66 \\
\textbf{Ebola} & 0.74 & 0.75 & 0.75 & 0.76 & 0.64 & \highP{0.68} & 0.80 & 0.61 & 0.64 & 0.79 & 0.79 & \bestM{0.79} & 0.88 & 0.72 & \secondM{0.77} \\
\textbf{Lassa} & 0.82 & 0.72 & \secondM{0.75} & 0.78 & 0.63 & 0.66 & 0.84 & 0.60 & 0.63 & 0.84 & 0.77 & \bestM{0.80} & 0.88 & 0.68 & 0.73 \\
\textbf{SARS} & 0.78 & 0.76 & 0.77 & 0.77 & 0.62 & 0.65 & 0.82 & 0.62 & \highP{0.66} & 0.80 & 0.78 & \bestM{0.79} & 0.89 & 0.73 & \highP{\secondM{0.78}} \\
\textbf{Zika} & 0.73 & 0.77 & \secondM{0.75} & 0.70 & 0.62 & 0.64 & 0.73 & 0.63 & 0.66 & 0.78 & 0.79 & \bestM{0.79} & 0.76 & 0.69 & 0.72 \\
\textbf{MERS} & 0.83 & 0.74 & \highP{\secondM{0.78}} & 0.87 & 0.62 & 0.67 & 0.86 & 0.60 & 0.65 & 0.86 & 0.78 & \highP{\bestM{0.81}} & 0.89 & 0.67 & 0.73 \\
\textbf{Nipah} & 0.84 & 0.66 & \secondM{0.70} & 0.92 & 0.58 & \lowP{0.59} & 0.81 & 0.54 & \lowP{0.53} & 0.85 & 0.68 & \bestM{0.72} & 0.90 & 0.61 & \lowP{0.65} \\
\midrule
\textbf{Overall} & 0.79 & 0.72 & \secondM{0.74} & 0.82 & 0.61 & 0.65 & 0.83 & 0.59 & 0.62 & 0.82 & 0.75 & \bestM{0.77} & 0.87 & 0.67 & 0.72 \\
\bottomrule
\end{tabular}
}
\end{table*}

\subsubsection*{Full-text Screening}
The ranking reorders substantially at full-text screening (Table~\ref{tab:fulltext_screening_model_ablation}), where \texttt{gpt-oss-120b} leads ($F_1$ $0.77$) and the two highest-precision abstract-stage models fall furthest. \texttt{DeepSeek-V3.2}'s precision drops from $0.83$ to $0.64$ and its recall from $0.59$ to $0.56$. It is the only model that loses ground on both measures at once, with its Marburg result ($F_1$ $0.42$, precision $0.37$) the single weakest pathogen-level score in either screening table. \texttt{gpt-oss-120b}'s advantage at this stage comes from recall. It achieves $0.81$ overall against the next-best \texttt{Kimi-K2.5} at $0.73$, and is one of only two models, alongside \texttt{GLM-4.7}, for which recall increases from abstract to full-text screening. The Nipah-to-Zika contrast also reverses. Nipah is the best-performing pathogen for \texttt{gpt-oss-120b} and \texttt{Kimi-K2.5} at full text ($F_1$ $0.85$ and $0.82$), whereas it was the worst for three models at the abstract stage. This suggests that the richer context of full texts resolves ambiguity that titles and abstracts leave open for this pathogen.

\begin{table*}[t]
\centering
\caption{\textbf{Full-text screening metrics across LLMs.}  \highP{Green} and \lowP{Red} denote the best- and worst-performing pathogens for each model (in terms of $F_1$ score). \bestM{Bold} indicates the best-performing model for each pathogen, and \secondM{Underline} indicates the second-best. $P=\text{precision}$; $R=\text{recall}$; $F_1=\text{F1-Score}$.}
\label{tab:fulltext_screening_model_ablation}
\footnotesize
\setlength{\tabcolsep}{3pt}
\renewcommand{\arraystretch}{1.1}
{\footnotesize
\begin{tabular}{l|ccc|ccc|ccc|ccc|ccc}
\toprule
\textbf{Pathogen} & \multicolumn{3}{c|}{\textbf{\texttt{gpt-oss-120b}}} & \multicolumn{3}{c|}{\texttt{GPT-5.2}} & \multicolumn{3}{c|}{\textbf{\texttt{DeepSeek-V3.2}}} & \multicolumn{3}{c|}{\textbf{\texttt{Kimi-K2.5}}} & \multicolumn{3}{c}{\textbf{\texttt{GLM-4.7}}} \\
& $P$ & $R$ & $F1$ & $P$ & $R$ & $F1$ & $P$ & $R$ & $F1$ & $P$ & $R$ & $F1$ & $P$ & $R$ & $F1$ \\
\midrule
\textbf{Marburg} & 0.75 & 0.76 & \secondM{0.75} & 0.76 & 0.59 & 0.59 & 0.37 & 0.49 & \lowP{0.42} & 0.66 & 0.66 & 0.66 & 0.86 & 0.72 & \highP{\bestM{0.76}} \\
\textbf{Ebola} & 0.73 & 0.84 & \bestM{0.77} & 0.61 & 0.60 & 0.60 & 0.68 & 0.59 & 0.55 & 0.72 & 0.74 & 0.71 & 0.75 & 0.75 & \secondM{0.75} \\
\textbf{Lassa} & 0.79 & 0.78 & \bestM{0.78} & 0.66 & 0.63 & 0.63 & 0.63 & 0.54 & 0.47 & 0.74 & 0.75 & \secondM{0.74} & 0.77 & 0.73 & 0.73 \\
\textbf{SARS} & 0.71 & 0.85 & \bestM{0.76} & 0.60 & 0.58 & 0.58 & 0.73 & 0.61 & \highP{0.59} & 0.67 & 0.69 & 0.66 & 0.73 & 0.72 & \secondM{0.72} \\
\textbf{Zika} & 0.66 & 0.79 & \lowP{\bestM{0.69}} & 0.50 & 0.50 & \lowP{0.50} & 0.61 & 0.55 & 0.52 & 0.63 & 0.64 & \lowP{\secondM{0.61}} & 0.60 & 0.59 & \lowP{0.59} \\
\textbf{MERS} & 0.76 & 0.83 & \bestM{0.79} & 0.66 & 0.61 & 0.61 & 0.74 & 0.60 & 0.58 & 0.76 & 0.79 & \secondM{0.77} & 0.76 & 0.68 & 0.69 \\
\textbf{Nipah} & 0.87 & 0.84 & \highP{\bestM{0.85}} & 0.80 & 0.63 & \highP{0.63} & 0.73 & 0.56 & 0.53 & 0.83 & 0.81 & \highP{\secondM{0.82}} & 0.72 & 0.61 & 0.61 \\
\midrule
\textbf{Overall} & 0.75 & 0.81 & \bestM{0.77} & 0.66 & 0.59 & 0.59 & 0.64 & 0.56 & 0.52 & 0.72 & 0.73 & \secondM{0.71} & 0.74 & 0.69 & 0.69 \\
\bottomrule
\end{tabular}
}
\end{table*}

\subsubsection*{Parameter Extraction}
Parameter extraction results are disaggregated by pathogen and extraction type in Table~\ref{tab:parameter_extraction_model_ablation}. \texttt{Kimi-K2.5} achieves the highest overall average $F_1$ ($0.63$), marginally ahead of \texttt{GLM-4.7} ($0.63$), with lower performance for \texttt{gpt-oss-120b} ($0.59$), GPT-5.2 ($0.58$) and \texttt{DeepSeek-V3.2} ($0.56$). Performance is most variable in the Counts sub-task. \texttt{gpt-oss-120b} attains strong precision ($0.83$) but low recall ($0.47$), reproducing the asymmetric pattern observed in the detailed field-level results below. GPT-5.2 shows the inverse pattern, with recall $0.83$ and precision $0.36$. At the field-level Extraction sub-task, GPT-5.2 achieves the highest overall $F_1$ ($0.59$), followed by \texttt{Kimi-K2.5} ($0.56$). The five models are broadly comparable, consistent with the interpretation that cross-model differences in average $F_1$ are driven by flagging and counting behaviour rather than the quality of individual field extractions. Zika is the weakest pathogen across most models and sub-tasks, while SARS is frequently the best-performing for \texttt{gpt-oss-120b}.

\begin{table*}[h]
\centering
\caption{\textbf{Parameter extraction metrics across LLMs.} \textbf{Average} denotes means across sub-tasks; \textbf{Overall} denotes means across pathogens. \highP{Green} and \lowP{Red} denote the best- and worst-performing pathogens for each model (in terms of $F_1$ score). \bestM{Bold} indicates the best-performing model for each pathogen, and \secondM{Underline} indicates the second-best. $P=\text{precision}$; $R=\text{recall}$; $F_1=\text{F1-Score}$.}
\label{tab:parameter_extraction_model_ablation}
\setlength{\tabcolsep}{3pt}
\renewcommand{\arraystretch}{1.1}
{\footnotesize
\begin{tabular}{ll|ccc|ccc|ccc|ccc|ccc}
\toprule
\textbf{Pathogen} & \textbf{Type} & \multicolumn{3}{c|}{\textbf{\texttt{gpt-oss-120b}}} & \multicolumn{3}{c|}{\texttt{GPT-5.2}} & \multicolumn{3}{c|}{\textbf{\texttt{DeepSeek-V3.2}}} & \multicolumn{3}{c|}{\textbf{\texttt{Kimi-K2.5}}} & \multicolumn{3}{c}{\textbf{\texttt{GLM-4.7}}} \\
& & $P$ & $R$ & $F1$ & $P$ & $R$ & $F1$ & $P$ & $R$ & $F1$ & $P$ & $R$ & $F1$ & $P$ & $R$ & $F1$ \\
\midrule
\multirow{4}{*}{\textbf{Ebola}} & Flagging & 0.60 & 0.92 & \highP{0.72} & 0.58 & 0.93 & 0.71 & 0.49 & 0.91 & 0.64 & 0.67 & 0.90 & \secondM{0.77} & 0.72 & 0.82 & \bestM{0.77} \\
 & Counts & 0.79 & 0.47 & 0.59 & 0.46 & 0.80 & \highP{0.58} & 0.57 & 0.59 & 0.58 & 0.52 & 0.73 & \secondM{0.61} & 0.59 & 0.65 & \bestM{0.62} \\
 & Extraction & 0.48 & 0.54 & \lowP{0.50} & 0.58 & 0.57 & \bestM{0.57} & 0.54 & 0.47 & 0.49 & 0.55 & 0.57 & \lowP{\secondM{0.55}} & 0.51 & 0.56 & \lowP{0.52} \\
 & Average & 0.62 & 0.64 & 0.60 & 0.54 & 0.77 & 0.62 & 0.54 & 0.65 & 0.57 & 0.58 & 0.73 & \bestM{0.64} & 0.61 & 0.68 & \secondM{0.64} \\
\midrule
\multirow{4}{*}{\textbf{Lassa}} & Flagging & 0.56 & 0.98 & 0.71 & 0.58 & 1.00 & \highP{0.73} & 0.54 & 0.91 & \highP{0.68} & 0.70 & 0.94 & \highP{\secondM{0.81}} & 0.77 & 0.87 & \highP{\bestM{0.82}} \\
 & Counts & 1.00 & 0.35 & \lowP{0.51} & 0.30 & 0.85 & 0.45 & 0.46 & 0.57 & \lowP{0.51} & 0.59 & 0.81 & \highP{\bestM{0.69}} & 0.74 & 0.59 & \highP{\secondM{0.66}} \\
 & Extraction & 0.58 & 0.54 & 0.55 & 0.66 & 0.63 & \highP{\bestM{0.63}} & 0.55 & 0.46 & \lowP{0.47} & 0.58 & 0.60 & \highP{\secondM{0.57}} & 0.58 & 0.56 & \highP{0.56} \\
 & Average & 0.71 & 0.62 & 0.59 & 0.51 & 0.83 & 0.61 & 0.52 & 0.65 & 0.55 & 0.62 & 0.79 & \bestM{0.69} & 0.70 & 0.67 & \secondM{0.68} \\
\midrule
\multirow{4}{*}{\textbf{SARS}} & Flagging & 0.50 & 0.81 & \bestM{0.62} & 0.47 & 0.83 & 0.60 & 0.39 & 0.78 & \lowP{0.52} & 0.56 & 0.69 & \lowP{0.62} & 0.58 & 0.67 & \lowP{\secondM{0.62}} \\
 & Counts & 0.80 & 0.61 & \highP{\bestM{0.69}} & 0.37 & 0.88 & 0.52 & 0.60 & 0.60 & \highP{\secondM{0.60}} & 0.45 & 0.72 & \lowP{0.56} & 0.51 & 0.59 & \lowP{0.55} \\
 & Extraction & 0.51 & 0.63 & \highP{0.56} & 0.56 & 0.63 & \bestM{0.58} & 0.58 & 0.54 & \highP{0.55} & 0.53 & 0.65 & \highP{\secondM{0.57}} & 0.51 & 0.62 & 0.55 \\
 & Average & 0.61 & 0.69 & \bestM{0.62} & 0.47 & 0.78 & 0.57 & 0.52 & 0.64 & 0.56 & 0.51 & 0.69 & \secondM{0.58} & 0.53 & 0.63 & 0.57 \\
\midrule
\multirow{4}{*}{\textbf{Zika}} & Flagging & 0.40 & 0.96 & \lowP{0.57} & 0.43 & 0.95 & \lowP{0.59} & 0.41 & 0.94 & 0.57 & 0.56 & 0.86 & \bestM{0.68} & 0.62 & 0.75 & \secondM{0.68} \\
 & Counts & 0.72 & 0.47 & 0.57 & 0.31 & 0.80 & \lowP{0.45} & 0.50 & 0.60 & 0.55 & 0.55 & 0.73 & \secondM{0.63} & 0.59 & 0.67 & \bestM{0.63} \\
 & Extraction & 0.52 & 0.57 & 0.53 & 0.56 & 0.59 & \lowP{\bestM{0.56}} & 0.55 & 0.50 & 0.51 & 0.54 & 0.59 & \secondM{0.55} & 0.52 & 0.58 & 0.54 \\
 & Average & 0.55 & 0.67 & 0.56 & 0.43 & 0.78 & 0.53 & 0.49 & 0.68 & 0.54 & 0.55 & 0.73 & \bestM{0.62} & 0.58 & 0.66 & \secondM{0.61} \\
\midrule
\multirow{4}{*}{\textbf{Overall}} & Flagging & 0.51 & 0.92 & 0.66 & 0.51 & 0.93 & 0.66 & 0.46 & 0.88 & 0.60 & 0.62 & 0.85 & \secondM{0.72} & 0.67 & 0.78 & \bestM{0.72} \\
 & Counts & 0.83 & 0.47 & 0.59 & 0.36 & 0.83 & 0.50 & 0.53 & 0.59 & 0.56 & 0.53 & 0.75 & \bestM{0.62} & 0.61 & 0.62 & \secondM{0.61} \\
 & Extraction & 0.52 & 0.57 & 0.54 & 0.59 & 0.61 & \bestM{0.59} & 0.56 & 0.49 & 0.50 & 0.55 & 0.60 & \secondM{0.56} & 0.53 & 0.58 & 0.54 \\
 & Average & 0.62 & 0.65 & 0.59 & 0.49 & 0.79 & 0.58 & 0.52 & 0.66 & 0.56 & 0.57 & 0.73 & \bestM{0.63} & 0.60 & 0.66 & \secondM{0.63} \\
\bottomrule
\end{tabular}
}
\end{table*}

\clearpage

\subsubsection*{Transmission Model Extraction}
Transmission model extraction results are presented in Table~\ref{tab:model_extraction_model_ablation}. \texttt{GLM-4.7} achieves the highest overall average $F_1$ ($0.85$), with strong performance across all three sub-tasks: Flagging ($0.93$), Counts ($0.93$), and Extraction ($0.68$). \texttt{DeepSeek-V3.2} ranks second overall ($0.81$), driven by notably high Counts performance ($0.92$), whilst \texttt{gpt-oss-120b} ranks last ($0.75$), held back by comparatively low Counts precision ($0.52$ overall). Lassa is the best-performing pathogen for all five models, with \texttt{GLM-4.7} achieving an overall average $F_1$ of $0.91$ for that pathogen alone, including perfect Flagging and Counts scores. SARS is consistently the most challenging pathogen. Flagging $F_1$ ranges from $0.82$ (\texttt{DeepSeek-V3.2}) to $0.87$ (\texttt{Kimi-K2.5}), and Extraction $F_1$ from $0.59$ to $0.66$. These patterns are consistent with the field-level difficulties in transmission route and assumption extraction reported below.

\begin{table*}[t!]
\centering
\caption{\textbf{Model extraction metrics across LLMs.}  \highP{Green} and \lowP{Red} denote the best- and worst-performing pathogens for each model (in terms of $F_1$ score). \textbf{Average} denotes means across sub-tasks; \textbf{Overall} denotes means across pathogens. \bestM{Bold} indicates the best-performing model for each pathogen, and \secondM{Underline} indicates the second-best. $P=\text{precision}$; $R=\text{recall}$; $F_1=\text{F1-Score}$.}
\label{tab:model_extraction_model_ablation}
\setlength{\tabcolsep}{3pt}
\renewcommand{\arraystretch}{1.1}
{\footnotesize
\begin{tabular}{ll|ccc|ccc|ccc|ccc|ccc}
\toprule
\textbf{Pathogen} & \textbf{Type} & \multicolumn{3}{c|}{\textbf{\texttt{gpt-oss-120b}}} & \multicolumn{3}{c|}{\texttt{GPT-5.2}} & \multicolumn{3}{c|}{\textbf{\texttt{DeepSeek-V3.2}}} & \multicolumn{3}{c|}{\textbf{\texttt{Kimi-K2.5}}} & \multicolumn{3}{c}{\textbf{\texttt{GLM-4.7}}} \\
& & $P$ & $R$ & $F1$ & $P$ & $R$ & $F1$ & $P$ & $R$ & $F1$ & $P$ & $R$ & $F1$ & $P$ & $R$ & $F1$ \\
\midrule
\multirow{4}{*}{\textbf{Ebola}} & Flagging & 0.92 & 0.92 & 0.92 & 0.89 & 0.90 & 0.89 & 0.87 & 0.86 & 0.86 & 0.93 & 0.93 & \secondM{0.93} & 0.95 & 0.94 & \bestM{0.95} \\
 & Counts & 0.50 & 1.00 & 0.67 & 0.56 & 1.00 & 0.71 & 0.81 & 0.99 & \secondM{0.89} & 0.63 & 0.99 & \lowP{0.77} & 0.88 & 0.99 & \bestM{0.93} \\
 & Extraction & 0.59 & 0.72 & 0.64 & 0.62 & 0.74 & \bestM{0.66} & 0.57 & 0.65 & 0.61 & 0.60 & 0.71 & \lowP{0.64} & 0.62 & 0.71 & \secondM{0.66} \\
 & Average & 0.67 & 0.88 & 0.74 & 0.69 & 0.88 & 0.76 & 0.75 & 0.83 & \secondM{0.79} & 0.72 & 0.88 & 0.78 & 0.81 & 0.88 & \bestM{0.84} \\
\midrule
\multirow{4}{*}{\textbf{Lassa}} & Flagging & 0.95 & 0.99 & \highP{\secondM{0.97}} & 0.95 & 0.99 & \highP{\secondM{0.97}} & 0.96 & 0.85 & 0.89 & 0.95 & 0.99 & \highP{\secondM{0.97}} & 1.00 & 1.00 & \highP{\bestM{1.00}} \\
 & Counts & 0.60 & 1.00 & \highP{0.75} & 0.60 & 1.00 & 0.75 & 1.00 & 1.00 & \highP{\bestM{1.00}} & 0.75 & 1.00 & \highP{\secondM{0.86}} & 1.00 & 1.00 & \highP{\bestM{1.00}} \\
 & Extraction & 0.68 & 0.73 & 0.70 & 0.68 & 0.79 & \highP{0.71} & 0.79 & 0.78 & \highP{\bestM{0.78}} & 0.70 & 0.78 & \highP{0.73} & 0.73 & 0.77 & \highP{\secondM{0.74}} \\
 & Average & 0.74 & 0.91 & 0.81 & 0.74 & 0.92 & 0.81 & 0.92 & 0.88 & \secondM{0.89} & 0.80 & 0.92 & 0.85 & 0.91 & 0.92 & \bestM{0.91} \\
\midrule
\multirow{4}{*}{\textbf{SARS}} & Flagging & 0.86 & 0.86 & \lowP{\secondM{0.86}} & 0.86 & 0.86 & \lowP{\secondM{0.86}} & 0.83 & 0.81 & \lowP{0.82} & 0.87 & 0.87 & \lowP{\bestM{0.87}} & 0.85 & 0.84 & \lowP{0.84} \\
 & Counts & 0.49 & 0.97 & 0.65 & 0.49 & 1.00 & \lowP{0.66} & 0.70 & 1.00 & \lowP{\secondM{0.82}} & 0.67 & 1.00 & 0.81 & 0.76 & 1.00 & \lowP{\bestM{0.86}} \\
 & Extraction & 0.60 & 0.71 & \lowP{\secondM{0.64}} & 0.61 & 0.73 & \lowP{0.64} & 0.55 & 0.64 & \lowP{0.59} & 0.63 & 0.74 & \bestM{0.66} & 0.59 & 0.68 & \lowP{0.62} \\
 & Average & 0.65 & 0.85 & 0.72 & 0.65 & 0.86 & 0.72 & 0.70 & 0.82 & 0.74 & 0.72 & 0.87 & \bestM{0.78} & 0.73 & 0.84 & \secondM{0.77} \\
\midrule
\multirow{4}{*}{\textbf{Zika}} & Flagging & 0.87 & 0.89 & 0.88 & 0.89 & 0.91 & 0.90 & 0.90 & 0.89 & \highP{0.90} & 0.90 & 0.92 & \secondM{0.91} & 0.93 & 0.93 & \bestM{0.93} \\
 & Counts & 0.48 & 0.98 & \lowP{0.65} & 0.61 & 1.00 & \highP{0.76} & 0.97 & 0.97 & \bestM{0.97} & 0.72 & 0.97 & 0.83 & 0.88 & 0.97 & \secondM{0.93} \\
 & Extraction & 0.66 & 0.78 & \highP{0.70} & 0.67 & 0.78 & 0.69 & 0.59 & 0.64 & 0.61 & 0.67 & 0.77 & \secondM{0.70} & 0.69 & 0.76 & \bestM{0.71} \\
 & Average & 0.67 & 0.88 & 0.74 & 0.72 & 0.90 & 0.78 & 0.82 & 0.84 & \secondM{0.83} & 0.76 & 0.89 & 0.81 & 0.83 & 0.89 & \bestM{0.85} \\
\midrule
\multirow{4}{*}{\textbf{Overall}} & Flagging & 0.90 & 0.91 & 0.91 & 0.90 & 0.91 & 0.90 & 0.89 & 0.85 & 0.87 & 0.91 & 0.93 & \secondM{0.92} & 0.93 & 0.93 & \bestM{0.93} \\
 & Counts & 0.52 & 0.99 & 0.68 & 0.56 & 1.00 & 0.72 & 0.87 & 0.99 & \secondM{0.92} & 0.69 & 0.99 & 0.81 & 0.88 & 0.99 & \bestM{0.93} \\
 & Extraction & 0.63 & 0.74 & 0.67 & 0.64 & 0.76 & 0.67 & 0.63 & 0.68 & 0.65 & 0.65 & 0.75 & \bestM{0.68} & 0.66 & 0.73 & \secondM{0.68} \\
 & Average & 0.68 & 0.88 & 0.75 & 0.70 & 0.89 & 0.77 & 0.80 & 0.84 & \secondM{0.81} & 0.75 & 0.89 & 0.81 & 0.82 & 0.88 & \bestM{0.85} \\
\bottomrule
\end{tabular}
}
\end{table*}

\subsubsection*{Outbreak Extraction}
Outbreak extraction results, evaluated across Lassa and Zika, are shown in Table~\ref{tab:outbreak_extraction_model_ablation}. GPT-5.2 achieves the highest overall average $F_1$ ($0.77$), followed closely by \texttt{Kimi-K2.5} ($0.76$), \texttt{DeepSeek-V3.2} ($0.73$), \texttt{GLM-4.7} ($0.72$), and \texttt{gpt-oss-120b} ($0.70$). Results diverge sharply between pathogens. Lassa Counts are strong across all models, ranging from $F_1$ $0.77$ (GPT-5.2) to $0.95$ (\texttt{Kimi-K2.5}), and field-level Extraction is uniformly high ($0.76$ to $0.83$). Zika Counts performance falls substantially for \texttt{gpt-oss-120b} ($F_1$ $0.47$) and \texttt{GLM-4.7} ($0.52$), whilst GPT-5.2 remains comparatively strong ($0.83$). A notable divergence in Flagging is also observed. \texttt{DeepSeek-V3.2} performs weakest on Lassa Flagging ($F_1$ $0.62$) whilst achieving the second-best result on Zika ($0.68$). These per-pathogen contrasts are consistent with the field-level analysis of suspected cases and epidemiological context fields reported below.

\clearpage

\begin{table*}[t!] \centering \caption{\textbf{Outbreak extraction metrics across LLMs.} \textbf{Average} denotes means across sub-tasks; \textbf{Overall} denotes means across pathogens. \highP{Green} and \lowP{Red} denote the best- and worst-performing pathogens for each model (in terms of $F_1$ score). \bestM{Bold} indicates the best-performing model for each pathogen, and \secondM{Underline} indicates the second-best. $P=\text{precision}$; $R=\text{recall}$; $F_1=\text{F1-Score}$.} \label{tab:outbreak_extraction_model_ablation} 
{\footnotesize
\setlength{\tabcolsep}{3pt} \renewcommand{\arraystretch}{1.1} 
\begin{tabular}{ll|ccc|ccc|ccc|ccc|ccc}
\toprule
\textbf{Pathogen} & \textbf{Type} & \multicolumn{3}{c|}{\textbf{\texttt{gpt-oss-120b}}} & \multicolumn{3}{c|}{\texttt{GPT-5.2}} & \multicolumn{3}{c|}{\textbf{\texttt{DeepSeek-V3.2}}} & \multicolumn{3}{c|}{\textbf{\texttt{Kimi-K2.5}}} & \multicolumn{3}{c}{\textbf{\texttt{GLM-4.7}}} \\
& & $P$ & $R$ & $F1$ & $P$ & $R$ & $F1$ & $P$ & $R$ & $F1$ & $P$ & $R$ & $F1$ & $P$ & $R$ & $F1$ \\
\midrule
\multirow{4}{*}{\textbf{Lassa}} & Flagging & 0.69 & 0.82 & \highP{\secondM{0.70}} & 0.72 & 0.84 & \highP{\bestM{0.74}} & 0.61 & 0.65 & \lowP{0.62} & 0.67 & 0.77 & \highP{0.69} & 0.65 & 0.68 & \lowP{0.66} \\
 & Counts & 0.83 & 1.00 & \highP{0.91} & 0.62 & 1.00 & \lowP{0.77} & 0.83 & 1.00 & \highP{0.91} & 0.90 & 1.00 & \highP{\bestM{0.95}} & 1.00 & 0.86 & \highP{\secondM{0.92}} \\
 & Extraction & 0.85 & 0.73 & \lowP{0.77} & 0.84 & 0.79 & \lowP{\secondM{0.81}} & 0.75 & 0.78 & \highP{0.76} & 0.84 & 0.83 & \highP{\bestM{0.83}} & 0.83 & 0.76 & \highP{0.78} \\
 & Average & 0.79 & 0.85 & \secondM{0.80} & 0.73 & 0.88 & 0.78 & 0.73 & 0.81 & 0.77 & 0.80 & 0.87 & \bestM{0.82} & 0.83 & 0.76 & 0.79 \\
\midrule
\multirow{4}{*}{\textbf{Zika}} & Flagging & 0.58 & 0.71 & \lowP{0.53} & 0.59 & 0.75 & \lowP{0.57} & 0.65 & 0.82 & \highP{\secondM{0.68}} & 0.61 & 0.80 & \lowP{0.59} & 0.67 & 0.87 & \highP{\bestM{0.70}} \\
 & Counts & 0.49 & 0.45 & \lowP{0.47} & 0.76 & 0.92 & \highP{\bestM{0.83}} & 0.68 & 0.61 & \lowP{0.64} & 0.72 & 0.88 & \lowP{\secondM{0.79}} & 0.84 & 0.38 & \lowP{0.52} \\
 & Extraction & 0.84 & 0.78 & \highP{\secondM{0.80}} & 0.88 & 0.87 & \highP{\bestM{0.87}} & 0.69 & 0.81 & \lowP{0.73} & 0.71 & 0.78 & \lowP{0.73} & 0.74 & 0.78 & \lowP{0.76} \\
 & Average & 0.64 & 0.65 & 0.60 & 0.75 & 0.84 & \bestM{0.76} & 0.68 & 0.74 & 0.69 & 0.68 & 0.82 & \secondM{0.70} & 0.75 & 0.68 & 0.66 \\
\midrule
\multirow{4}{*}{\textbf{Overall}} & Flagging & 0.63 & 0.76 & 0.61 & 0.66 & 0.79 & \secondM{0.66} & 0.63 & 0.73 & 0.65 & 0.64 & 0.78 & 0.64 & 0.66 & 0.77 & \bestM{0.68} \\
 & Counts & 0.66 & 0.72 & 0.69 & 0.69 & 0.96 & \secondM{0.80} & 0.76 & 0.80 & 0.78 & 0.81 & 0.94 & \bestM{0.87} & 0.92 & 0.62 & 0.72 \\
 & Extraction & 0.85 & 0.76 & \secondM{0.79} & 0.86 & 0.83 & \bestM{0.84} & 0.72 & 0.79 & 0.75 & 0.78 & 0.81 & 0.78 & 0.78 & 0.77 & 0.77 \\
 & Average & 0.71 & 0.75 & 0.70 & 0.74 & 0.86 & \bestM{0.77} & 0.70 & 0.78 & 0.73 & 0.74 & 0.84 & \secondM{0.76} & 0.79 & 0.72 & 0.72 \\
\bottomrule
\end{tabular}
}
\end{table*}

\subsection{Field-Level Results Possible with \name{}}\label{app:data_extraction_extended}
The main text reports the aggregated data extraction metrics for each model. However, the \name{} harness allows us to use field-level values for understanding localised LLM errors on the dataset. This section gives the disaggregated extraction results possible with the harness for \texttt{gpt-oss-120b}. Results are reported by pathogen, data type and field where applicable. 

The human reference labelled datasets are provided open source by the Pathogen Epidemiology Review Group, available through the R \texttt{epireview} package or on GitHub at \href{https://github.com/mrc-ide/epireview/tree/main/inst/extdata}{https://github.com/mrc-ide/epireview/tree/main/inst/extdata}. As of March 2026, owing to PERG's continual progress through SLRs on nine priority pathogens, human reference extraction data is available in a standardised format for four pathogens: Lassa, Ebola, SARS, and Zika. For each pathogen, we evaluate classification measures for each of the \textit{Flagging}, \textit{Count}, and \textit{Extraction} metrics defined formally in \Cref{app:data_extraction_extended_methods}.

\subsubsection*{Parameters}
\Cref{tab:parameters_extended_flagging_count} presents the results for parameter extraction Flagging and Count metrics. These results are used to produce the aggregate data presented in the main body text (\Cref{tab:data_extraction_f1} in \Cref{sec:pipeline_experimental_results}). For flagging relevant parameters, \name{} performs consistently across all pathogens with high recall ($0.92$ average), though precision is lower and more variable ($0.51$ average). The results suggest that while the model identifies nearly all relevant extractions, this coverage comes at the cost of many false positive flags that may propagate errors to later sub-stages. In terms of overall parameter extraction counts, the performance flips in favour of precision ($0.83$) now at the expense of lower recall ($0.47$). The discrepancy with parameter flagging performance is understandable, as the structured extraction step allows flagged parameters to be discarded through tool calls. When the model produces a final extraction, this extraction is likely correct. However, it often fails to produce all required extractions according to human reference data. An article may have multiple extractions of the same parameter class, and in these cases \texttt{gpt-oss-120b} can underestimate the number of extractions required.

\begin{table}[h]

\centering
\caption{\textbf{Flagging and Count classification metrics for parameter extraction with \name{} harness (\texttt{gpt-oss-120b}).} $P=\text{precision}$; $R=\text{recall}$; $F_1=\text{F1-Score}$.}
\label{tab:parameters_extended_flagging_count}
\renewcommand{\arraystretch}{1.1}
{\footnotesize
\begin{tabular}{l|cccccccccccc}
\toprule
\bf Metric & \multicolumn{3}{c}{\textbf{Lassa}} & \multicolumn{3}{c}{\textbf{Ebola}} & \multicolumn{3}{c}{\textbf{SARS}} & \multicolumn{3}{c}{\textbf{Zika}} \\
 & $P$ & $R$ & $F_1$ & $P$ & $R$ & $F_1$ & $P$ & $R$ & $F_1$ & $P$ & $R$ & $F_1$ \\
\midrule
\bf Flagging & 0.56 &
0.98 &
0.71 &
0.60 &
0.92 &
0.72 &
0.50 &
0.81 &
0.62 &
0.40 &
0.96 &
0.57 \\
\midrule
\bf Count & 1.00 &
0.35 &
0.51 &
0.79 &
0.47 &
0.59 &
0.80 &
0.61 &
0.69 &
0.72 &
0.47 &
0.57 \\
\bottomrule
\end{tabular}
}
\end{table}

\begin{table}[h]
\setlength{\tabcolsep}{3pt}
\centering
\caption{\textbf{Field-level precision, recall, and $F_1$ for Extraction on parameters with \name{} harness (\texttt{gpt-oss-120b}).} \textit{Group} corresponds to the sub-stage of parameter extraction where the field is collected. The final row shows averages across all fields. $P=\text{precision}$; $R=\text{recall}$; $F_1=\text{F1-Score}$.}
\label{tab:parameters_detailed_metrics}
\renewcommand{\arraystretch}{1.1}
{\footnotesize
\begin{tabular}{ll|ccc|ccc|ccc|ccc}
\toprule
\bf Group & \bf Field & \multicolumn{3}{c}{\textbf{Lassa}} & \multicolumn{3}{c}{\textbf{Ebola}} & \multicolumn{3}{c}{\textbf{SARS}} & \multicolumn{3}{c}{\textbf{Zika}} \\
 & & $P$ & $R$ & $F_1$ & $P$ & $R$ & $F_1$ & $P$ & $R$ & $F_1$ & $P$ & $R$ & $F_1$ \\
\midrule
\multirow{4}{*}{\textbf{Value}} & value & 0.22 & 0.22 & 0.22 & 0.20 & 0.20 & 0.20 & 0.23 & 0.23 & 0.23 & 0.14 & 0.14 & 0.14 \\
 & unit & 0.50 & 0.43 & 0.46 & 0.62 & 0.35 & 0.44 & 0.69 & 0.61 & 0.65 & 0.65 & 0.43 & 0.52 \\
 & method & 1.00 & 0.89 & 0.94 & 0.48 & 0.78 & 0.59 & 0.76 & 0.83 & 0.79 & 0.86 & 0.80 & 0.83 \\
\cmidrule{2-14}
 & Average & 0.57 & 0.51 & 0.54 & 0.44 & 0.44 & 0.41 & 0.56 & 0.56 & 0.56 & 0.55 & 0.46 & 0.50 \\
\midrule
\multirow{5}{*}{\textbf{Uncertainty}} & value type & 0.38 & 0.43 & 0.40 & 0.30 & 0.33 & 0.32 & 0.35 & 0.57 & 0.43 & 0.12 & 0.22 & 0.16 \\
 & statistical approach & -- & -- & -- & -- & -- & -- & -- & -- & -- & 0.44 & 0.66 & 0.53 \\
 & single type uncertainty & 1.00 & 1.00 & 1.00 & 0.98 & 0.94 & 0.96 & 0.81 & 0.95 & 0.88 & 0.97 & 0.99 & 0.98 \\
 & paired uncertainty & 0.25 & 0.40 & 0.31 & 0.59 & 0.72 & 0.65 & 0.39 & 0.88 & 0.54 & 0.46 & 0.90 & 0.61 \\
\cmidrule{2-14}
 & Average & 0.54 & 0.61 & 0.57 & 0.62 & 0.67 & 0.64 & 0.52 & 0.80 & 0.62 & 0.50 & 0.69 & 0.57 \\
\midrule
\multirow{4}{*}{\textbf{Population}} & population sex & 0.86 & 0.67 & 0.75 & 0.62 & 0.79 & 0.69 & 0.59 & 0.75 & 0.66 & 0.59 & 0.70 & 0.64 \\
 & population group & 0.14 & 0.11 & 0.12 & 0.24 & 0.33 & 0.28 & 0.23 & 0.25 & 0.24 & 0.54 & 0.54 & 0.54 \\
 & population sample type & 0.86 & 0.67 & 0.75 & 0.32 & 0.41 & 0.36 & 0.58 & 0.63 & 0.60 & 0.37 & 0.36 & 0.37 \\
\cmidrule{2-14}
 & Average & 0.62 & 0.48 & 0.54 & 0.40 & 0.51 & 0.44 & 0.47 & 0.54 & 0.50 & 0.50 & 0.54 & 0.52 \\
\midrule
\textbf{Overall} & & 0.58 & 0.53 & 0.55 & 0.49 & 0.54 & 0.50 & 0.51 & 0.63 & 0.56 & 0.51 & 0.57 & 0.53 \\
\bottomrule
\end{tabular}
}
\end{table}

For Extraction, complete field-level results are presented in \Cref{tab:parameters_detailed_metrics}. The aggregate results in the main text show parameter extractions to have moderate quality across all pathogens, with little variation among them. Analysing the results at the field level reveals patterns in the model's handling of different data modalities and different types of epidemiological context. The system performs worst on value fields, with population fields also showing relatively weak performance compared to other groups. We suspect the difficulty with population context arises from the large numbers of valid options for many fields, notably population group and population sample type. Many of these options have precise interpretations in epidemiological literature. Without fine-tuning, \texttt{gpt-oss-120b} may struggle to apply these interpretations in a complex tool-calling environment. Classification is near perfect for single type uncertainty and generally strong for method, with the exception of Ebola. Fields with unrestricted domains, like value, are much harder to classify correctly. Expert validation in \Cref{app:expert_validaation_data} suggests that at least some of this difficulty may stem from our exact-match criteria being overly punitive of equivalent numbers in different formats.

\subsubsection*{Transmission Models}
\Cref{tab:models_detailed_metrics} presents the complete results for Flagging, Count, and Extraction evaluations of transmission models across the four priority pathogens. Screening performance was strong across all pathogens for article flagging, with recall ranging from $0.86$ to $0.99$ and precision from $0.86$ to $0.96$, indicating reliable identification of modelling studies. Model count extraction achieved consistently high recall ($0.97$--$1.00$) but notably lower precision ($0.48$--$0.60$), suggesting a systematic tendency to overestimate the number of models reported per article rather than failing to identify them.

Field-level extraction showed a clear gradient in task difficulty. Core structural characteristics were extracted with high accuracy. Model type classification and the theoretical versus data-fitted distinction achieved balanced precision and recall between $0.62$ and $0.89$ across pathogens. Single-value fields such as stochastic versus deterministic modelling and code availability frequently exceeded $0.75$ and reached perfect scores for some pathogens. In contrast, more complex or multi-value fields exhibited substantially lower performance. Transmission route extraction was particularly challenging for Ebola and SARS, while assumptions and interventions showed modest precision and recall across all pathogens. Overall, across screening and extraction tasks, precision ranged from $0.61$ to $0.70$ and recall from $0.75$ to $0.81$. This indicates that the model reliably captures core model characteristics, with remaining limitations concentrated in the extraction of nuanced descriptive details.

\begin{table}[t!]
\setlength{\tabcolsep}{3pt}
\centering
\caption{\textbf{Precision, recall, and $F_1$ metrics for transmission model screening and extraction across four pathogens with \name{} harness (\texttt{gpt-oss-120b}).} \textit{Screening} includes article flagging and model count accuracy. \textit{Extraction} evaluates field-level accuracy for matched model pairs, covering core structural characteristics (model type, stochastic vs deterministic, theoretical vs data-fitted, code availability) and more complex multi-value fields (transmission routes, assumptions, interventions). Strong performance is observed for core model characteristics, while extraction of assumptions, interventions, and transmission routes remains more challenging. $P=\text{precision}$; $R=\text{recall}$; $F_1=\text{F1-Score}$.}
\label{tab:models_detailed_metrics}
\renewcommand{\arraystretch}{1.1}
{\footnotesize
\begin{tabular}{l|cccccccccccc}
\toprule
 & \multicolumn{3}{c}{\textbf{Lassa}} & \multicolumn{3}{c}{\textbf{Ebola}} & \multicolumn{3}{c}{\textbf{SARS}} & \multicolumn{3}{c}{\textbf{Zika}} \\
 & $P$ & $R$ & $F1$ & $P$ & $R$ & $F1$ & $P$ & $R$ & $F1$ & $P$ & $R$ & $F1$ \\
\midrule
\multicolumn{13}{l}{\textbf{Flagging}} \\
Article Flagging
& 0.95 & 0.99 & 0.97
& 0.92 & 0.92 & 0.92
& 0.86 & 0.86 & 0.86
& 0.87 & 0.89 & 0.88 \\
\midrule
\multicolumn{13}{l}{\textbf{Counts}} \\
Model Count
& 0.60 & 1.00 & 0.75
& 0.50 & 1.00 & 0.67
& 0.49 & 0.97 & 0.65
& 0.48 & 0.98 & 0.65 \\
\midrule
\multicolumn{13}{l}{\textbf{Extraction}} \\
Model Type
& 0.89 & 0.89 & 0.89
& 0.89 & 0.89 & 0.89
& 0.77 & 0.77 & 0.77
& 0.88 & 0.88 & 0.88 \\
\midrule
Compartmental Type
& 0.00 & 0.00 & 0.00
& --- & --- & ---
& 0.80 & 0.80 & 0.80
& 0.83 & 0.83 & 0.83 \\
\midrule
Stochastic vs Deterministic
& 1.00 & 1.00 & 1.00
& 0.75 & 0.85 & 0.80
& 0.76 & 0.78 & 0.77
& 0.82 & 0.79 & 0.81 \\
\midrule
Theoretical vs Data-Fitted
& 0.78 & 0.78 & 0.78
& 0.88 & 0.88 & 0.88
& 0.62 & 0.62 & 0.62
& 0.81 & 0.81 & 0.81 \\
\midrule
Code Available
& 1.00 & 0.89 & 0.94
& 0.85 & 0.84 & 0.84
& 1.00 & 1.00 & 1.00
& 0.82 & 0.76 & 0.79 \\
\midrule
Transmission Routes
& 1.00 & 0.94 & 0.97
& 0.13 & 0.15 & 0.14
& 0.26 & 0.32 & 0.29
& 0.68 & 0.74 & 0.71 \\
\midrule
Assumptions
& 0.29 & 0.46 & 0.36
& 0.27 & 0.46 & 0.34
& 0.21 & 0.39 & 0.28
& 0.31 & 0.52 & 0.39 \\
\midrule
Interventions
& 0.54 & 0.64 & 0.58
& 0.48 & 0.69 & 0.56
& 0.46 & 0.79 & 0.58
& 0.32 & 0.69 & 0.43 \\
\midrule
\textbf{Overall}
& 0.70 & 0.78 & 0.73
& 0.62 & 0.77 & 0.67
& 0.61 & 0.75 & 0.66
& 0.66 & 0.81 & 0.71 \\
\bottomrule
\end{tabular}
}
\end{table}

\subsubsection*{Outbreaks}
\begin{table}[b!]
\centering
\caption{\textbf{Outbreak screening and extraction with \name{} harness (\texttt{gpt-oss-120b}) across feature categories.} \textit{Screening} measured article flagging (identifying papers containing outbreaks) and outbreak count accuracy (extracting the correct number of outbreaks per paper). \textit{Extraction} evaluated field-level accuracy for matched outbreak pairs across four epidemiological categories: temporal features (start/end dates), geographic features (country, specific location), case burden (confirmed cases, deaths), and epidemiological context (detection mode, pre-outbreak status, ongoing status, asymptomatic transmission). Overall metrics represent the average across all extraction fields. $P=\text{precision}$; $R=\text{recall}$; $F_1=\text{F1-Score}$.}
\renewcommand{\arraystretch}{1.1}
{\footnotesize
\label{tab:outbreak_aggregated_metrics}
\begin{tabular}{ll|ccc|ccc}
\toprule
 \multirow{2}{*}{\bf Metric} & \multirow{2}{*}{\bf Field}
 & \multicolumn{3}{c}{\textbf{Lassa}}
 & \multicolumn{3}{c}{\textbf{Zika}} \\
 & & $P$ & $R$ & $F1$ & $P$ & $R$ & $F1$ \\
\midrule
\bf Flagging & Article Flagging & 0.69 & 0.82 & 0.75 & 0.58 & 0.71 & 0.64 \\
\midrule
\bf Counts & Outbreak Counts & 0.83 & 1.00 & 0.91 & 0.49 & 0.45 & 0.47 \\
\midrule
\multirow{4}{*}{\textbf{Extraction}} & Temporal Features & 0.83 & 0.74 & 0.78 & 0.85 & 0.82 & 0.83 \\
 & Geographic and Spatial Features & 0.75 & 0.78 & 0.76 & 0.75 & 0.75 & 0.75 \\
 & Case Burden & 0.82 & 0.75 & 0.79 & 0.93 & 0.93 & 0.93 \\
 & Epidemiological Context and Metadata & 0.93 & 0.70 & 0.80 & 0.84 & 0.67 & 0.75 \\
\midrule
\textbf{Overall} &  & 0.85 & 0.73 & 0.79 & 0.84 & 0.78 & 0.81 \\
\bottomrule
\end{tabular}
}
\end{table}

Applying the evaluation framework described in Appendix~\ref{app:additional_evaluation_details}, we analysed outbreak extraction performance across two priority pathogens, Lassa and Zika (Table~\ref{tab:outbreak_aggregated_metrics}). Screening performance differed between pathogens and screening subtasks. For Lassa, article flagging achieved moderate precision ($0.69$) and strong recall ($0.82$), indicating that most outbreak-containing papers were identified, although a non-trivial fraction of flagged papers were false positives. For Zika, article flagging was more balanced (precision $0.58$, recall $0.71$), suggesting improved sensitivity relative to precision, but still leaving missed outbreak descriptions and over-inclusion of non-outbreak papers. Outbreak counting remained strong for Lassa (precision $0.83$, recall $1.00$), while Zika outbreak counting was substantially lower (precision $0.49$, recall $0.45$), consistent with continued difficulty in reliably enumerating outbreak events in the Zika corpus.

\begin{table}[t!]
\centering
\caption{\textbf{Outbreak feature extraction by category for \name{} harness (\texttt{gpt-oss-120b}).} Each row shows field-level performance within the four major epidemiological categories. Temporal and case burden features showed consistently high performance, while location-specific fields and epidemiological context features showed greater variability. $P=\text{precision}$; $R=\text{recall}$; $F_1=\text{F1-Score}$.}
\label{tab:outbreak_detailed_metrics}
\renewcommand{\arraystretch}{1.1}
{\footnotesize
\begin{tabular}{l|cccccc}
\toprule
 & \multicolumn{3}{c}{\textbf{Lassa}} & \multicolumn{3}{c}{\textbf{Zika}} \\
 & $P$ & $R$ & $F1$ & $P$ & $R$ & $F1$ \\
\midrule
\multicolumn{7}{l}{\textbf{Temporal Features}} \\
Start Year & 0.89 & 0.80 & 0.84 & 0.90 & 0.82 & 0.86 \\
Start Month & 0.78 & 0.78 & 0.78 & 0.80 & 0.80 & 0.80 \\
Start Day & 0.86 & 0.67 & 0.75 & 0.95 & 0.95 & 0.95 \\
End Month & 0.78 & 0.78 & 0.78 & 0.65 & 0.65 & 0.65 \\
End Day & 0.86 & 0.67 & 0.75 & 0.95 & 0.86 & 0.90 \\
\textbf{Average} & 0.83 & 0.74 & 0.78 & 0.85 & 0.82 & 0.83 \\
\midrule
\multicolumn{7}{l}{\textbf{Geographic and Spatial Features}} \\
Outbreak Country & 1.00 & 1.00 & 1.00 & 1.00 & 1.00 & 1.00 \\
Location & 0.50 & 0.56 & 0.53 & 0.50 & 0.50 & 0.50 \\
\textbf{Average} & 0.75 & 0.78 & 0.77 & 0.75 & 0.75 & 0.75 \\
\midrule
\multicolumn{7}{l}{\textbf{Case Burden}} \\
Confirmed Cases & 0.75 & 0.60 & 0.67 & 0.86 & 0.86 & 0.86 \\
Deaths & 0.90 & 0.90 & 0.90 & 1.00 & 1.00 & 1.00 \\
\textbf{Average} & 0.83 & 0.75 & 0.79 & 0.93 & 0.93 & 0.93 \\
\midrule
\multicolumn{7}{l}{\textbf{Epidemiological Context and Metadata}} \\
Mode of Detection & 0.71 & 0.50 & 0.59 & 0.50 & 0.41 & 0.45 \\
Pre-outbreak Status & 1.00 & 0.30 & 0.46 & 1.00 & 0.41 & 0.58 \\
Ongoing Status & 1.00 & 1.00 & 1.00 & 0.86 & 0.86 & 0.86 \\
Asymptomatic Transmission & 1.00 & 1.00 & 1.00 & 1.00 & 1.00 & 1.00 \\
\textbf{Average} & 0.93 & 0.70 & 0.76 & 0.84 & 0.67 & 0.72 \\
\midrule
\textbf{Overall} & 0.85 & 0.73 & 0.79 & 0.84 & 0.78 & 0.81 \\
\bottomrule
\end{tabular}
}
\end{table}

Field-level extraction performance, grouped by epidemiological feature categories, revealed consistent strengths and persistent weaknesses (Table~\ref{tab:outbreak_detailed_metrics}). Temporal features remained robust across both pathogens, with precision $0.83$ and recall $0.74$ for Lassa, and precision $0.85$ and recall $0.82$ for Zika. Case burden metrics also showed strong extraction, with precision $0.83$ and recall $0.75$ for Lassa, and precision $0.93$ and recall $0.93$ for Zika. Death extraction was especially accurate, with $0.90$ precision and recall for Lassa and $1.00$ precision and recall for Zika.

Geographic extraction continued to show a split between coarse and fine granularity. Outbreak country identification was perfect for both pathogens ($1.00$ precision and recall), while specific location extraction was notably weaker for Lassa (precision $0.50$, recall $0.56$) and Zika (precision $0.50$, recall $0.50$), consistent with variability in how places are described in scientific text. Epidemiological context fields showed the greatest variability. For Lassa, mode of detection was moderate ($0.71$ precision, $0.50$ recall) and pre-outbreak status exhibited high precision but low recall ($1.00$/$0.30$), indicating frequent omission of this attribute. Ongoing status was extracted perfectly for Lassa ($1.00$/$1.00$) but showed moderate performance for Zika ($0.86$/$0.86$). For Zika, mode of detection ($0.50$/$0.41$) and pre-outbreak status ($1.00$/$0.41$) remained challenging, although asymptomatic transmission was extracted perfectly for both pathogens ($1.00$/$1.00$). The overall extraction performance averaged $0.85$ precision and $0.73$ recall for Lassa, and $0.84$ precision and $0.78$ recall for Zika, suggesting reliable field-level accuracy with remaining gaps concentrated in context-dependent and location-specific attributes.

\subsection{Article Screening Strategy Ablations}
\label{app:extended_results_article_screening}
This section gives the precision, recall and $F_1$ values for the article screening ablation strategies summarised in \Cref{fig:fulltext_recall_barchart}.

\subsubsection*{Title and Abstract Screening}
Table~\ref{tab:abstract_screening_metrics} summarises title-and-abstract screening performance across seven pathogens, showing moderate overall recall ($0.72$) alongside high precision ($0.79$), for an overall $F_1$ of $0.74$. This pattern suggests the abstract-stage screening is tuned toward specificity, prioritising the rejection of irrelevant studies at the cost of more false negatives. Performance is broadly consistent across pathogens, with the strongest balance for MERS ($F_1$ of $0.78$) and the weakest for Marburg ($F_1$ of $0.69$).

\begin{table}[h]
\centering
\caption{\textbf{Precision, recall, and $F_1$ for title-and-abstract screening across seven pathogens with \name{} harness (\texttt{gpt-oss-120b}).} Metrics summarise how well the abstract-stage classifier retained studies judged relevant under PERG screening criteria, reported for each pathogen and overall. $P=\text{precision}$; $R=\text{recall}$; $F_1=\text{F1-Score}$.}
\label{tab:abstract_screening_metrics}
\renewcommand{\arraystretch}{1.1}
\setlength{\tabcolsep}{8pt}
{\footnotesize
\begin{tabular}{l|ccc}
\toprule
\textbf{Pathogen} & $P$ & $R$ & $F_1$ \\
\midrule
Marburg & 0.80 & 0.64 & 0.69 \\
Ebola & 0.74 & 0.75 & 0.75 \\
Lassa & 0.82 & 0.72 & 0.75 \\
SARS & 0.78 & 0.76 & 0.77 \\
Zika & 0.73 & 0.77 & 0.75 \\
MERS & 0.83 & 0.74 & 0.78 \\
Nipah & 0.84 & 0.66 & 0.70 \\
\midrule
Overall & 0.79 & 0.72 & 0.74 \\
\bottomrule
\end{tabular}
}
\end{table}

\subsubsection*{Full-text Screening}
Table~\ref{tab:fulltext_screening_metrics} compares three full-text screening strategies and highlights a clear precision-recall trade-off. Human abstract $\rightarrow$ AI full-text achieves the strongest overall performance (precision $0.83$, recall $0.92$), while the two-stage strategy (AI abstract $\rightarrow$ AI full-text) shows lower recall ($0.81$), consistent with error propagation from abstract gating. Direct AI full-text screening improves recall ($0.89$) but reduces precision ($0.68$), reflecting a recall-maximising approach when abstracts are treated as an information bottleneck.

\begin{table}[h!]
\centering
\caption{\textbf{Full-text screening performance on \name{} harness (\texttt{gpt-oss-120b}) under three operational strategies for identifying relevant articles.} Metrics compare a two-stage AI pipeline (AI abstract$\rightarrow$AI full-text), a mixed workflow (human abstract$\rightarrow$AI full-text), and direct AI full-text screening, reported for each pathogen and overall. $R=\text{recall}$; $F_1=\text{F1-Score}$.}
\footnotesize
\label{tab:fulltext_screening_metrics}
\renewcommand{\arraystretch}{1.1}
\setlength{\tabcolsep}{6pt}
{\footnotesize
\begin{tabular}{l|ccc|ccc|ccc}
\toprule
\multirow{2}{*}{\textbf{Pathogen}}
& \multicolumn{3}{c|}{\makecell{\textbf{AI Screen (Abstract)}\\$\rightarrow$\\\textbf{AI Screen (Full-text)}}}
& \multicolumn{3}{c|}{\makecell{\textbf{Human Screen (Abstract)}\\$\rightarrow$\\\textbf{AI Screen (Full-text)}}}
& \multicolumn{3}{c}{\makecell{~\\\textbf{AI Screen (Direct Full-text)}\\~}} \\
\cline{2-10}
\rule{0pt}{9pt}
& $P$ & $R$ & $F_1$
& $P$ & $R$ & $F_1$
& $P$ & $R$ & $F_1$ \\
\midrule
Marburg & 0.75 & 0.76 & 0.75 & 0.77 & 0.83 & 0.80 & 0.64 & 0.82 & 0.69 \\
Ebola & 0.73 & 0.84 & 0.77 & 0.86 & 0.97 & 0.91 & 0.67 & 0.93 & 0.72 \\
Lassa & 0.79 & 0.78 & 0.78 & 0.83 & 0.94 & 0.88 & 0.71 & 0.91 & 0.77 \\
SARS & 0.71 & 0.85 & 0.76 & 0.80 & 0.95 & 0.86 & 0.64 & 0.91 & 0.68 \\
Zika & 0.66 & 0.79 & 0.69 & 0.81 & 0.91 & 0.85 & 0.64 & 0.85 & 0.67 \\
MERS & 0.76 & 0.83 & 0.79 & 0.83 & 0.96 & 0.88 & 0.69 & 0.95 & 0.76 \\
Nipah & 0.87 & 0.84 & 0.85 & 0.89 & 0.90 & 0.90 & 0.74 & 0.88 & 0.79 \\
\midrule
Overall & 0.75 & 0.81 & 0.77 & 0.83 & 0.92 & 0.87 & 0.68 & 0.89 & 0.73 \\
\bottomrule
\end{tabular}
}
\end{table}

\clearpage
\section{Extended Expert Validation Results}\label{app:expert_validaation_data}
Six epidemiology researchers contributed to our validation survey. We collected $62$ submissions for parameters, $50$ for models, and $31$ for outbreaks. \Cref{tab:expert_validation_results} reports all metrics collected from the survey. The validation survey was designed as an expert audit of \name{} harness outputs, not as a duplicate annotation study of the PERG reference set. Each item contained the source article and one LLM extraction. Experts judged whether the extraction was relevant and whether populated fields were correct. The sampling was not designed to provide systematic overlap of the same PERG fields across multiple raters. We therefore do not report Cohen's $\kappa$ for PERG annotations. This limits our ability to separate model errors from residual reference annotation noise. To reduce schema-level noise, we filter reference entries with invalid enum values before evaluation and report the filtering rates in Appendix~\ref{app:data_extraction_extended_methods}. We interpret the PERG data as peer-reviewed reference annotations rather than as an errorless target.

{\footnotesize
\begin{longtable}{>{\raggedright\arraybackslash}p{10.8cm} S[table-format=1.2, round-mode=places, round-precision=2]}
\caption{\textbf{Extended expert validation results.} Results are reported as expert-rated flagging precision and expert-rated extraction accuracy. Within each section, rows are ordered from overall scores to subgroup scores and then field-level scores.}
\label{tab:expert_validation_results} \\
\toprule
\textbf{Item} & {\textbf{Score}} \\
\midrule
\endfirsthead

\multicolumn{2}{l}{\textit{Table \thetable\ continued from previous page}} \\
\toprule
\textbf{Item} & {\textbf{Score}} \\
\midrule
\endhead

\midrule
\multicolumn{2}{r}{\textit{Continued on next page}} \\
\endfoot

\bottomrule
\endlastfoot

\multicolumn{2}{l}{\textbf{Parameters}} \\
\addlinespace[0.2em]
Overall --- Flagging precision & 0.66 \\
Overall --- Extraction accuracy & 0.77 \\

\addlinespace[0.35em]
\textit{Precision by class} & {} \\
\quad Attack rate & 0.25 \\
\quad Growth rate & 1.00 \\
\quad Human delay & 0.62 \\
\quad Reproduction number & 1.00 \\
\quad Seroprevalence & 0.50 \\
\quad Severity & 0.57 \\

\addlinespace[0.35em]
\textit{Accuracy by group} & {} \\
\quad Value & 0.89 \\
\quad Uncertainty & 0.76 \\
\quad Population & 0.59 \\
\quad Aggregation & 0.83 \\

\addlinespace[0.35em]
\textit{Value fields} & {} \\
\quad Value & 0.81 \\
\quad Unit & 0.96 \\
\quad Type & 0.88 \\
\quad Bounds & 0.79 \\
\quad Value type & 0.90 \\
\quad Statistical approach & 0.97 \\

\addlinespace[0.35em]
\textit{Uncertainty fields} & {} \\
\quad Single-type uncertainty & 0.88 \\
\quad Paired uncertainty & 0.84 \\
\quad Distribution type & 0.57 \\

\addlinespace[0.35em]
\textit{Population fields} & {} \\
\quad Sample type & 0.74 \\
\quad Population group & 0.49 \\
\quad Sample size & 0.66 \\
\quad Sex & 0.50 \\
\quad Age range & 0.58 \\
\quad Countries & 0.82 \\
\quad Locations & 0.71 \\
\quad Method moment value & 0.23 \\

\addlinespace[0.35em]
\textit{Aggregation fields} & {} \\
\quad Aggregation & 0.83 \\

\addlinespace[0.6em]
\multicolumn{2}{l}{\textbf{Models}} \\
\addlinespace[0.2em]
Overall --- Flagging precision & 0.40 \\
Overall --- Extraction accuracy & 0.83 \\

\addlinespace[0.35em]
\textit{Field accuracy} & {} \\
\quad Model type & 0.89 \\
\quad Compartmental type & 0.89 \\
\quad Stochastic or deterministic & 0.70 \\
\quad Theoretical model & 0.84 \\

\addlinespace[0.6em]
\multicolumn{2}{l}{\textbf{Outbreaks}} \\
\addlinespace[0.2em]
Overall --- Flagging precision & 0.61 \\
Overall --- Extraction accuracy & 0.80 \\

\addlinespace[0.35em]
\textit{Accuracy by group} & {} \\
\quad Temporal & 0.62 \\
\quad Geographical & 0.87 \\
\quad Case burden & 0.85 \\
\quad Epidemiological & 0.85 \\

\addlinespace[0.35em]
\textit{Temporal fields} & {} \\
\quad Start year & 0.84 \\
\quad Start month & 0.70 \\
\quad Start day & 0.62 \\
\quad End year & 0.50 \\
\quad End month & 0.60 \\
\quad End day & 0.56 \\
\quad Duration in months & 0.50 \\

\addlinespace[0.35em]
\textit{Geographical fields} & {} \\
\quad Country & 0.95 \\
\quad Location & 0.80 \\

\addlinespace[0.35em]
\textit{Case burden fields} & {} \\
\quad Confirmed cases & 0.88 \\
\quad Suspected cases & 0.64 \\
\quad Asymptomatic cases & 1.00 \\
\quad Severe cases & 1.00 \\
\quad Deaths & 0.71 \\

\addlinespace[0.35em]
\textit{Epidemiological fields} & {} \\
\quad Mode of detection & 0.82 \\
\quad Pre-outbreak status & 0.82 \\
\quad Asymptomatic transmission described & 0.89 \\

\end{longtable}
}

For flagging (sub-task) precision, models and outbreaks are reported only at the overall level (as in the main text). For parameters, precision is averaged over flagging decisions made for each parameter class. The random subsample of articles assigned gave six relevant parameter classes: attack rate, growth rate, human delay, reproduction number, seroprevalence, and severity. The remaining two parameter classes, mutation rate and relative contribution, were absent from the sample. Since there is a flagging decision made for each parameter class on each article, each parameter class-level precision is calculated over the same sample size ($N=62$).

For extraction accuracy, the aggregate statistics are normalised over groups of similar fields. For example, outbreaks have clusters of fields related to temporal features (start date, end date, and duration), geographical features (country and location), case burden (case counts and fatalities) and epidemiological factors (mode of detection, status pre-outbreak, and asymptomatic transmission). We normalise at the group level to treat each aspect of the extraction as equally important, in order to avoid overemphasising groups with larger numbers of metadata fields. We omit group-level normalisation for models owing to the smaller number of validated fields.

The disaggregated statistics reveal findings that are masked in the average statistics. For example, among parameter classes, \name{} workflow (with \texttt{gpt-oss-120b}) performs worst on flagging attack rate (experts reported several instances where the system confused attack rate with seroprevalence information). At the field level, the LLM struggles the most with understanding parameter population context and with the temporal outbreak features (group accuracy $0.62$). Parameter population fields are multiple-choice selections with many options (see \Cref{tab:population_tool_call}), and these designations often have specific interpretations in epidemiology. For example, ``persons under investigation'' is a population group of patients exhibiting clinical and epidemiological risk factors, a definition that an LLM may struggle to apply consistently in different article contexts.

\clearpage
\section{Article Search and Retrieval}\label{app:search_queries}

This section details the search query construction, database-specific adaptations, and PDF retrieval strategy used for article acquisition across priority pathogens in the \name~workflow. Following the Pathogen Epidemiology Review Group (PERG) methodology\footnote{\url{https://github.com/mrc-ide/priority-pathogens/wiki/Search-terms}}, we developed a standardised base query structure that captures core epidemiological domains including transmission dynamics, disease severity, temporal parameters, transmission heterogeneity, and evolutionary characteristics.

Different bibliographic databases support different search capabilities, requiring tailored query implementations. We maintain two versions of each pathogen query: one for PubMed and Europe PMC (which support wildcard truncation operators using \texttt{*}), and another for OpenAlex (which requires fully expanded term variants).

\subsection{Base Search Query (PubMed and Europe PMC)}

The base query for PubMed and Europe PMC uses Boolean operators with truncation symbols to capture morphological term variations:

\begin{lstlisting}[basicstyle=\small\ttfamily, breaklines=true, backgroundcolor=\color{gray!10}]
[PATHOGEN_IDENTIFIER] AND (
    (transmissi* OR epidemiolog*) OR 
    (model* NOT imag*) OR 
    (severity OR "case fatality ratio*" OR CFR OR "case fatality rate*" 
        OR "mortality rate*" OR "attack rate*") OR 
    ("infectious period*" OR "serial interval*" OR "incubation period*" 
        OR "generation time*" OR "generation interval*" OR "latent period*" 
        OR latency) OR 
    (heterogeneit* OR superspread* OR "super spread*" OR super-spread* 
        OR overdispersion OR overdispersed OR over-dispersion OR over-dispersed 
        OR "over dispersion" OR "over dispersed") OR 
    (infectivity OR infectiousness OR "growth rate*" OR "reproduction number*" 
        OR "reproductive number*" OR R0 OR "reproduction ratio*" 
        OR "reproductive rate*") OR 
    ("pre-existing immunity" OR serological OR serology OR serosurvey*) OR 
    (evolution* OR mutation* OR substitution*) OR 
    (outbreak* OR cluster* OR epidemic*) OR 
    ("risk factor*")
    [ADDITIONAL_TERMS]
) [EXCLUSION_CRITERIA]
\end{lstlisting}

\subsection{OpenAlex Adapted Queries}

Because the OpenAlex API does not support wildcard operators\footnote{\url{https://docs.openalex.org/how-to-use-the-api/get-lists-of-entities/search-entities}} and strips these characters during query processing, we expanded all truncated terms into their common morphological variants:

\begin{lstlisting}[basicstyle=\small\ttfamily, breaklines=true, backgroundcolor=\color{gray!10}]
[PATHOGEN_IDENTIFIER] AND (
    (transmission OR transmissibility OR transmissible OR transmitted 
        OR transmitting OR transmit OR epidemiology OR epidemiological 
        OR epidemiologic) OR 
    (model OR models OR modeling OR modelling OR modeled OR modelled 
        NOT (image OR images OR imaging)) OR 
    (severity OR "case fatality ratio" OR "case fatality ratios" OR CFR 
        OR "case fatality rate" OR "case fatality rates" OR "mortality rate" 
        OR "mortality rates" OR "attack rate" OR "attack rates") OR 
    ("infectious period" OR "infectious periods" OR "serial interval" 
        OR "serial intervals" OR "incubation period" OR "incubation periods" 
        OR "generation time" OR "generation interval" OR "generation intervals" 
        OR "latent period" OR "latent periods" OR latency) OR 
    (heterogeneity OR heterogeneous OR superspread OR superspreader 
        OR superspreaders OR superspreading OR "super spread" 
        OR "super spreader" OR "super spreaders" OR "super spreading" 
        OR overdispersion OR overdispersed OR "over dispersion" 
        OR "over dispersed") OR 
    (infectivity OR infectiousness OR "growth rate" OR "growth rates" 
        OR "reproduction number" OR "reproduction numbers" 
        OR "reproductive number" OR "reproductive numbers" OR R0 
        OR "reproduction ratio" OR "reproduction ratios" 
        OR "reproductive rate" OR "reproductive rates" 
        OR "basic reproduction number") OR 
    ("pre-existing immunity" OR serological OR serology OR serosurvey 
        OR serosurveys OR seroprevalence OR serosurveillance) OR 
    (evolution OR evolutionary OR evolving OR evolved OR mutation 
        OR mutations OR mutant OR mutants OR mutate OR mutated 
        OR substitution OR substitutions) OR 
    (outbreak OR outbreaks OR cluster OR clusters OR clustering 
        OR epidemic OR epidemics OR pandemic OR pandemics) OR 
    ("risk factor" OR "risk factors")
    [ADDITIONAL_TERMS]
) [EXCLUSION_CRITERIA]
\end{lstlisting}

\subsection{Pathogen-Specific Query Modifications}

Table~\ref{tab:search_queries} summarises the pathogen-specific modifications applied across all database implementations. Most pathogens require only customised identifiers to ensure relevant literature retrieval. However, the queries for SARS explicitly exclude COVID-19 literature to prevent cross-contamination with SARS-CoV-2 studies. Similarly, queries for Zika include vector-specific epidemiological parameters (extrinsic incubation period, vector competence) that are essential for capturing mosquito-borne transmission dynamics. For Rift Valley fever, Crimean-Congo hemorrhagic fever (CCHF) and MERS, we incorporated additional virus-specific identifiers and spelling variants to enhance retrieval comprehensiveness. Despite these modifications, all databases share consistent pathogen identifiers and exclusion criteria, differing only in their use of wildcard forms (PubMed/Europe PMC) versus expanded term variants (OpenAlex).

\begin{table*}[h]
\footnotesize
\caption{\textbf{Pathogen-specific modifications to the standardised search query}. All databases share consistent pathogen identifiers and exclusion criteria; PubMed/Europe PMC use wildcard forms while OpenAlex uses expanded variants.}
\label{tab:search_queries}
\begin{center}
\begin{tabular}{lp{5cm}p{3.2cm}p{3.5cm}}
\toprule
\textbf{Pathogen} & \textbf{\texttt{PATHOGEN\_IDENTIFIER}} & \textbf{\texttt{ADDITIONAL\_TERMS}} & \textbf{\texttt{EXCLUSION\_CRITERIA}} \\
\midrule
Marburg virus & Marburg virus & --- & --- \\
Ebola virus & Ebola & --- & --- \\
Lassa virus & Lassa & --- & --- \\
SARS-CoV-1 & SARS OR SARS-CoV-1 OR ``Severe acute respiratory syndrome" & --- & NOT (COVID-19 OR SARS-CoV-2) \\
Zika virus & zika & OR (``extrinsic incubation period" OR ``EIP" OR ``vector competence" OR ``vectorial capacity")\textsuperscript{†} & --- \\
Nipah virus & Nipah & --- & --- \\
MERS-CoV & MERS OR MERS-CoV OR ``Middle East respiratory syndrome" OR ``Middle East Respiratory Syndrome Coronavirus"\textsuperscript{‡} & --- & --- \\
Rift Valley fever virus & ``Rift valley fever" OR RVF OR ``Rift Valley Fever Virus" OR RVFV\textsuperscript{‡} & --- & --- \\
CCHF virus & ``Crimean Congo haemorrhagic fever" OR ``Crimean-Congo hemorrhagic fever" OR CCHF OR ``CCHF virus" OR CCHFV\textsuperscript{‡} & --- & --- \\
\bottomrule
\end{tabular}
\end{center}
\vspace{-2mm}
{\small \textsuperscript{†}Vector-specific terms capture mosquito transmission parameters unique to arboviral epidemiology.}\\
{\small \textsuperscript{‡}Expanded identifiers include alternative spellings (American/British English), virus-specific nomenclature, and common abbreviations for comprehensive coverage.}
\end{table*}

\subsection{Metadata Extraction and Deduplication}

We extract bibliographic metadata from each database as summarised in Table~\ref{tab:metadata_fields}. OpenAlex provides direct PDF URLs and internal work identifiers, PubMed supplies standardised medical literature identifiers (PMID: PubMed ID; PMCID: PubMed Central ID), and Europe PMC offers full-text availability metadata. The Digital Object Identifier (DOI) serves as a persistent identifier across databases.

We implement a hierarchical five-level deduplication strategy:

\begin{enumerate}
\item \textbf{DOI-based}: Normalised DOI strings (case-insensitive, URL prefixes stripped);
\item \textbf{PMID-based}: Numeric PMID extraction and normalisation;
\item \textbf{PMCID-based}: Normalised PMC identifiers (uppercase, ``PMC'' prefix standardised);
\item \textbf{OpenAlex ID-based}: Internal OpenAlex work identifiers;
\item \textbf{Title-year combination}: Normalised title strings (lowercase, alphanumeric only) paired with publication year.
\end{enumerate}

When duplicate records are detected, identifier fields (DOI, PMID, PMCID, OpenAlex ID, URLs) preserve all non-null values while narrative fields (title, abstract, journal) retain the first non-null value. Source provenance is marked as ``Both" when records appear in multiple databases.

\begin{table}[h]
\centering
\footnotesize
\begin{minipage}{0.48\textwidth}
    \centering
    \caption{\textbf{Metadata fields extracted during article search.} PMID: PubMed ID; PMCID: PubMed Central ID; DOI: Digital Object Identifier.}
    \label{tab:metadata_fields}
    \begin{footnotesize}
    \begin{tabular}{ll}
    \toprule
    \textbf{Field} & \textbf{Description} \\
    \midrule
    article\_id & Generated unique identifier \\
    source & Database origin \\
    pmid & PubMed Identifier \\
    pmcid & PubMed Central Identifier \\
    doi & Digital Object Identifier \\
    title & Article title \\
    authors & Semicolon-delimited author list \\
    journal & Publication venue \\
    year & Publication year \\
    abstract & Article abstract \\
    url & Canonical article URL \\
    openalex\_id & OpenAlex work identifier \\
    openalex\_pdf\_url & Direct PDF link from OpenAlex \\
    pathogen & Target pathogen \\
    query & Search query used \\
    harvested\_at & ISO 8601 timestamp \\
    \bottomrule
    \end{tabular}
    \end{footnotesize}
\end{minipage}
\hfill
\begin{minipage}{0.48\textwidth}
    \centering
    \caption{\textbf{Additional fields populated during PDF retrieval attempts.}}
    \label{tab:download_fields}
    \begin{footnotesize}
    \begin{tabular}{ll}
    \toprule
    \textbf{Field} & \textbf{Description} \\
    \midrule
    downloaded & Boolean success flag \\
    downloaded\_path & Filesystem path to PDF \\
    download\_source & Source that provided PDF \\
    download\_attempted\_at & ISO 8601 timestamp \\
    download\_error & Error messages from attempts \\
    \bottomrule
    \end{tabular}
    \end{footnotesize}
\end{minipage}
\end{table}

\subsection{PDF Retrieval}\label{app:article_download}

We attempt PDF downloads through multiple open access sources using a cascading retrieval strategy. Before attempting downloads, available identifiers (PMID, PMCID, DOI) are cross-referenced using NCBI's PMC ID Converter API\footnote{\url{https://www.ncbi.nlm.nih.gov/pmc/tools/id-converter-api/}} to maximise source compatibility. The system then attempts downloads from up to four sources in priority order (Table~\ref{tab:download_sources}), stopping at the first successful retrieval.

\subsubsection{Implementation Details}

Downloads employ HTTP streaming to temporary files with 64~KB chunks and validate each file through two stages: (1) magic byte verification (\texttt{\%PDF} header), and (2) content inspection for HTML access denial pages. Files exceeding 500~MB or failing validation are immediately discarded. Thread-pool parallelism with 16 workers processes downloads concurrently while respecting per-source rate limits. In-memory caches keyed by normalised identifiers store both successful PDF URLs and negative markers to eliminate redundant API calls. Progress is checkpointed every 50 records for crash recovery.

Successfully validated PDFs are saved with standardised filenames following identifier priority (PMID $\rightarrow$ PMCID $\rightarrow$ DOI hash $\rightarrow$ title hash). Metadata is augmented with download provenance including source, timestamp, and error diagnostics.

\begin{table}[t]
\centering
\caption{\textbf{PDF retrieval sources in cascading priority order.} Sources are queried sequentially until success or exhaustion. Identifier cross-referencing via NCBI PMC ID Converter API precedes all download attempts (10 req/s, cached).}
\label{tab:download_sources}
\begin{small}
\begin{tabular}{clcc}
\toprule
\textbf{Priority} & \textbf{Source \& Endpoint} & \textbf{Rate Limit} & \textbf{Cached} \\
\midrule
1 & \textbf{OpenAlex Direct PDF URL} & 30 req/s & No \\
  & \quad Metadata field \texttt{openalex\_pdf\_url} & & \\
\midrule
2 & \textbf{Europe PMC Fulltext API} & 20 req/s & Yes \\
  & \quad \texttt{ebi.ac.uk/europepmc/webservices/rest/search} & & \\
\midrule
3 & \textbf{Unpaywall API} & 50 req/s & Yes \\
  & \quad \texttt{api.unpaywall.org/v2/\{DOI\}?email=\{EMAIL\}} & & \\
\midrule
4 & \textbf{OpenAlex DOI Lookup} & 30 req/s & Yes \\
  & \quad \texttt{api.openalex.org/works/https://doi.org/\{DOI\}} & & \\
\bottomrule
\end{tabular}
\end{small}
\end{table}

\subsection{Final Quality Control}

After retrieval, we applied deduplication and quality filtering that removes: records lacking abstracts, duplicate article IDs, duplicate DOIs (retaining first occurrence) and records with file validation failures.

\clearpage
\section{Article Screening Criteria and Prompts}
\label{app:study_objectives}
\label{app:article_screening_prompts_criteri_fulltext}

\label{app:article_screening_prompts_criteria}
Following article search and retrieval, the articles are screened for relevance to the study. The screening is conducted on abstracts, and then on full-text articles. We present the study objectives, inclusion and exclusion criteria, along with the detailed prompts used to screen for relevant priority pathogen articles. Through prompt sensitivity tests with the open-source models under evaluation, similar to \citet{alhetelahahmad2026measuring} but replacing yes/no with include/exclude, we settle on using a modified version of the prompt from ScreenPrompt \citep{ottoSRPromptPaper}. The prompts follow a structured format: basic instruction, study objectives, inclusion/exclusion criteria, article content, and chain-of-thought screening instructions with parsable output request.


\begin{tcolorbox}[
    enhanced,
    colback=orange!5,
    colframe=orange!20,
    boxrule=0pt,
    leftrule=3pt,
    arc=0pt,
    left=6pt, right=6pt, top=2pt, bottom=2pt
]
{\small
{\textbf{\textcolor{orange!70!black}{Study Objectives}}}

\smallskip

This systematic review aims to collate transmission and modelling parameters for \texttt{\{pathogen\_name\}}. The review seeks to:

\begin{enumerate}[leftmargin=*, itemsep=-3pt, topsep=2pt]
    \item Provide estimates of key infectious disease metrics (reproduction number, CFR, generation time, serial interval, incubation period, etc.)
    \item Document historical outbreak characteristics (size, location, duration, deaths)
    \item Identify mathematical/statistical models of transmission
    \item Collate risk factors for infection, severe disease, and death
    \item Summarize seroprevalence data
    \item Support infectious disease modelling and outbreak response efforts
\end{enumerate}
}

\vspace{2pt}
{\small
This information enables effective outbreak preparedness, resource targeting, and mathematical modelling for nowcasting and forecasting of \texttt{\{pathogen\_name\}}.}
\end{tcolorbox}


\begin{tcolorbox}[
    enhanced,
    colback=green!5,
    colframe=green!20,
    boxrule=0pt,
    leftrule=3pt,
    arc=0pt,
    left=6pt, right=6pt, top=4pt, bottom=4pt
]

{\small
{\textbf{\textcolor{green!60!black}{Inclusion Criteria}}}

{ALL must be met:}

\begin{enumerate}[leftmargin=*, itemsep=-3pt, topsep=2pt]
    \item {Pathogen:} Must be about \texttt{\{pathogen\_name\}}
    
    \item {Language:} English only
    
    \item {Study type:} Peer-reviewed, original research (note systematic reviews/meta-analyses for special consideration)
    
    \item {Population:} Human subjects (animal studies acceptable if reporting EITHER: (a) transmission parameters: $R_0$, $R_t$, $R_e$, $r$, growth rate, mutation rate, OR (b) vector parameters: extrinsic incubation period, vector reproduction numbers, vector competence, mosquito delays)
    
    \item {Content:} Must contain AT LEAST ONE of:
    \begin{enumerate}[label=(\alph*), leftmargin=*, itemsep=-1pt]
        \item Quantitative details of concluded/ongoing human outbreak (size, year, location, duration, spatial scale)
        \item Mathematical or statistical model of disease transmission
        \item Measures/estimates of transmission parameters: $R$, $R_0$, $R_t$, $r$, $R_e$, growth rate, doubling time
        \item Measures/estimates of timing parameters: generation time, serial interval, incubation period, latent period, infectious period
        \item Measures/estimates of severity: CFR, IFR, hospitalization rate, mortality rate, attack rate
        \item Measures/estimates of genetic evolution: mutation rate, substitution rate, evolutionary rate
        \item Measures of overdispersion or superspreading ($k$ parameter, transmission heterogeneity)
        \item Seroprevalence data or serological surveys
        \item Risk factors for infection, severe disease, death, or hospitalization (with statistical measures)
        \item Measures/estimates of vector parameters: extrinsic incubation period (EIP), mosquito reproduction numbers, vector competence, mosquito delays, or relative transmission contributions (human-to-human vs vector-borne/zoonotic)
    \end{enumerate}

\dashtext{Full-text only}
    
    \item
    {Data Extraction Requirement:} Must contain extractable mathematical models, transmission models, or quantitative parameter estimates (with values or ranges) for disease modeling. This includes: reproduction numbers, transmission rates, incubation periods, case fatality ratios, model structures, intervention effects, or other modeling parameters. Articles without extractable quantitative parameters or models should be excluded.
\end{enumerate}}
\end{tcolorbox}





\vspace{8pt}

\newpage

\begin{tcolorbox}[
    enhanced,
    breakable,
    colback=white,
    colframe=blue!30,
    boxrule=0.8pt,
    arc=2pt,
    left=0pt, right=0pt, top=0pt, bottom=0pt,
    title={\small\textbf{Title \& Abstract Screening Prompt}},
    fonttitle=\sffamily,
    coltitle=blue!70!black,
    colbacktitle=blue!15,
    attach boxed title to top left={yshift=-2mm, xshift=4mm},
    boxed title style={boxrule=0.5pt, arc=1pt}
]

\begin{tcolorbox}[
    enhanced,
    colback=gray!5,
    colframe=gray!20,
    boxrule=0pt,
    leftrule=3pt,
    arc=0pt,
    left=6pt, right=6pt, top=4pt, bottom=4pt
]
    
    \vspace{4pt}
    {\small{You are an expert epidemiologist screening full-text articles for a systematic review on the target pathogen.}}

\end{tcolorbox}

\begin{tcolorbox}[
    enhanced,
    colback=orange!5,
    colframe=orange!20,
    boxrule=0pt,
    leftrule=3pt,
    arc=0pt,
    left=6pt, right=6pt, top=4pt, bottom=4pt
]
{\small\textbf{\textcolor{orange!70!black}{Study Objectives}}}

\vspace{4pt}
{\small\textit{[See Study Objectives above]}}
\end{tcolorbox}

\begin{tcolorbox}[
    enhanced,
    colback=gray!5,
    colframe=gray!20,
    boxrule=0pt,
    leftrule=3pt,
    arc=0pt,
    left=6pt, right=6pt, top=4pt, bottom=4pt
]
{\small\textbf{\textcolor{gray!70!black}{Screening Criteria}}}

{\small
\vspace{4pt}
The following is an excerpt of 2 sets of criteria. A study is considered included if it meets ALL inclusion criteria. If a study meets ANY exclusion criteria, it should be excluded. Here are the 2 sets of criteria:}

\begin{tcolorbox}[
    enhanced,
    colback=green!5,
    colframe=green!20,
    boxrule=0pt,
    leftrule=3pt,
    arc=0pt,
    left=6pt, right=6pt, top=4pt, bottom=4pt
]

{\small\textbf{\textcolor{green!60!black}{Inclusion Criteria}}}

\vspace{2pt}
{\small\textit{[See Inclusion Criteria 1--5 above]}}
\end{tcolorbox}

\begin{tcolorbox}[
    enhanced,
    colback=red!5,
    colframe=red!20,
    boxrule=0pt,
    leftrule=3pt,
    arc=0pt,
    left=6pt, right=6pt, top=4pt, bottom=4pt
]
{\small\textbf{\textcolor{red!60!black}{Exclusion Criteria}}}

\vspace{2pt}
{\small
{Exclude if ANY apply:}
\begin{enumerate}
    [leftmargin=*, itemsep=0pt, topsep=4pt]
    \item Pathogen: Not about \texttt{\{pathogen\_name\}} (excludes studies on other pathogens)
    \item Language: Non-English
    \item Publication type: Conference proceedings, abstract-only, posters, correspondence
    \item Study design: \textit{In-vitro} studies only (no human or animal component)
    \item Study design: Solely animal studies AND animal studies that do not report transmission parameters ($R_0$, $R_t$, $R_e$, $r$, growth rate, mutation rate)
    \item {Outbreak type:} Accidental laboratory outbreaks (not natural disease transmission)
\end{enumerate}
}
\end{tcolorbox}

\end{tcolorbox}

\begin{tcolorbox}[
    enhanced,
    colback=gray!5,
    colframe=gray!20,
    boxrule=0pt,
    leftrule=3pt,
    arc=0pt,
    left=6pt, right=6pt, top=4pt, bottom=4pt
]
{\small\textbf{\textcolor{gray!70!black}{Abstract (To Screen)}}}

\vspace{4pt}
{\small\ttfamily
Title: \{\{title\}\}\\[2pt]
Abstract: \{\{abstract\}\}
}
\end{tcolorbox}

\begin{tcolorbox}[
    enhanced,
    colback=gray!5,
    colframe=gray!20,
    boxrule=0pt,
    leftrule=3pt,
    arc=0pt,
    left=6pt, right=6pt, top=4pt, bottom=4pt
]
{\small\textbf{\textcolor{gray!70!black}{Screening Instructions}}}

\vspace{4pt}
{\small
We now assess whether the paper should be included in the systematic review by evaluating it against each and every predefined inclusion and exclusion criterion. First, we will reflect on how we will decide whether a paper should be included or excluded. Then, we will think step by step for each criterion, giving reasons for why they are met or not met.

\vspace{4pt}
Studies that may not fully align with the primary focus of our inclusion criteria but provide data or insights potentially relevant to our review deserve thoughtful consideration. Given the nature of abstracts as concise summaries of comprehensive research, some degree of interpretation is necessary.

\vspace{4pt}
Our aim should be to inclusively screen abstracts, ensuring broad coverage of pertinent studies while filtering out those that are clearly irrelevant.

\vspace{4pt}
We will conclude by outputting (on the very last line) \texttt{<decision>EXCLUDE</decision>} if the paper warrants exclusion, or \texttt{<decision>INCLUDE</decision>} if inclusion is advised or uncertainty persists.
}
\end{tcolorbox}

\end{tcolorbox}


\newpage
Finally, the articles that pass the abstract screening have their full text screened as follows.

\begin{tcolorbox}[
    enhanced,
    breakable,
    colback=white,
    colframe=blue!30,
    boxrule=0.8pt,
    arc=2pt,
    left=0pt, right=0pt, top=0pt, bottom=0pt,
    title={\small\textbf{Full-Text Screening Prompt}},
    fonttitle=\sffamily,
    coltitle=blue!70!black,
    colbacktitle=blue!15,
    attach boxed title to top left={yshift=-2mm, xshift=4mm},
    boxed title style={boxrule=0.5pt, arc=1pt}
]

\begin{tcolorbox}[
    enhanced,
    colback=gray!5,
    colframe=gray!20,
    boxrule=0pt,
    leftrule=3pt,
    arc=0pt,
    left=6pt, right=6pt, top=4pt, bottom=4pt
]
    
    \vspace{4pt}
    {\small{You are an expert epidemiologist screening abstracts for a systematic review on the target pathogen.}}

\end{tcolorbox}

\begin{tcolorbox}[
    enhanced,
    colback=orange!5,
    colframe=orange!20,
    boxrule=0pt,
    leftrule=3pt,
    arc=0pt,
    left=6pt, right=6pt, top=4pt, bottom=4pt
]
{\small\textbf{\textcolor{orange!70!black}{Study Objectives}}}

\vspace{4pt}
{\small\textit{[See Study Objectives above]}}
\end{tcolorbox}

\begin{tcolorbox}[
    enhanced,
    colback=gray!5,
    colframe=gray!20,
    boxrule=0pt,
    leftrule=3pt,
    arc=0pt,
    left=6pt, right=6pt, top=4pt, bottom=4pt
]
{\small\textbf{\textcolor{gray!70!black}{Screening Criteria}}}

{\small
\vspace{4pt}
The following is an excerpt of 2 sets of criteria. A study is considered included if it meets ALL inclusion criteria. If a study meets ANY exclusion criteria, it should be excluded. Here are the 2 sets of criteria:}

\begin{tcolorbox}[
    enhanced,
    colback=green!5,
    colframe=green!20,
    boxrule=0pt,
    leftrule=3pt,
    arc=0pt,
    left=6pt, right=6pt, top=4pt, bottom=4pt
]
{\small\textbf{\textcolor{green!60!black}{Inclusion Criteria}}}

\vspace{2pt}
{\small\textit{[See Inclusion Criteria 1--6 above, including full-text criterion]}}
\end{tcolorbox}

\begin{tcolorbox}[
    enhanced,
    colback=red!5,
    colframe=red!20,
    boxrule=0pt,
    leftrule=3pt,
    arc=0pt,
    left=6pt, right=6pt, top=4pt, bottom=4pt
]
{\small\textbf{\textcolor{red!60!black}{Exclusion Criteria}} 

{Exclude if ANY apply:}

\vspace{2pt}
{\small
\begin{enumerate}
[leftmargin=*, itemsep=0pt, topsep=4pt]
    \item Not about \texttt{\{pathogen\_name\}} (excludes other pathogens)
    \item Non-English language
    \item Conference proceedings, abstract-only, posters, correspondence, Literature reviews, meta-analyses
    \item \textit{In-vitro} studies only (no human or animal component)
    \item Animal studies without transmission parameters ($R_0$, $R_t$, $R_e$, $r$, growth rate, mutation rate) or solely animal studies.
\item{\parbox{0.88\linewidth}{Case studies/reports with $<$10 human cases}}
    \item Accidental laboratory outbreaks
\end{enumerate}
}}
\end{tcolorbox}
\end{tcolorbox}

\begin{tcolorbox}[
    enhanced,
    colback=gray!5,
    colframe=gray!20,
    boxrule=0pt,
    leftrule=3pt,
    arc=0pt,
    left=6pt, right=6pt, top=4pt, bottom=4pt
]
{\small\textbf{\textcolor{gray!70!black}{Full-Text Article (To Screen)}}}

\vspace{4pt}
{\small\ttfamily
Title: \{\{title\}\}\\[2pt]
Full Text: \{\{fulltext\}\}
}
\end{tcolorbox}

\begin{tcolorbox}[
    enhanced,
    colback=gray!5,
    colframe=gray!20,
    boxrule=0pt,
    leftrule=3pt,
    arc=0pt,
    left=6pt, right=6pt, top=4pt, bottom=4pt
]
{\small\textbf{\textcolor{gray!70!black}{Screening Instructions}}}

\vspace{4pt}
{\small
We now assess whether the paper should be included in the systematic review by evaluating it against each and every predefined inclusion and exclusion criterion. First, we will reflect on how we will decide whether a paper should be included or excluded. Then, we will think step by step for each criterion, giving reasons for why they are met or not met.

\vspace{4pt}
\textbf{Critically evaluate:} Does this paper contain extractable quantitative data, models, or parameters relevant to disease transmission and outbreak response? This is essential for inclusion.

\vspace{4pt}
We will conclude by outputting (on the very last line) \texttt{<decision>EXCLUDE</decision>} if the paper warrants exclusion, or \texttt{<decision>INCLUDE</decision>} if inclusion is advised or uncertainty persists.
}
\end{tcolorbox}

\end{tcolorbox}

\newpage

\section{Data Extraction Process}\label{app:data_extraction_details}

After screening, the finalised pool of relevant articles underwent rigorous data extraction. This extraction stage evaluates the use of structured tool-calling to extract three categories of data: epidemiological parameters, transmission models and outbreak data from full-text articles. Each category followed a multi-stage workflow with validation on each tool output.




\subsection{Parameters}\label{app:parameters_extended_extraction_process}

\paragraph{Valid Epidemiological Parameters for Extraction}
Epidemiological parameters are quantitative summaries of how an infection behaves in a population, such as its rate of spread, the delays between key stages of infection, the infection and fatality rates, and risk factors across demographic groups. We used PERG's data entry tool, a REDCap survey, as the reference list of epidemiological quantities that human reviewers would extract from the literature.\footnote{\url{https://redcap.imperial.ac.uk/surveys/?s=CEX3YKW8W47NMFA4}} This gave a fixed catalogue of 47 \textit{parameter types} that cover mutation processes, transmission intensity, delay distributions in humans and mosquitoes, severity, seroprevalence, and risk factors. These higher-order groupings are labelled \textit{parameter classes}, and \name{} defines data extraction criteria at the parameter class-level. \Cref{tab:perg_parameter_types} lists all parameter types targeted by our pipeline, together with brief definitions that match the guidance given to human experts.

\begin{footnotesize}
\begin{longtable}
{p{0.32\textwidth}p{0.18\textwidth}p{.45\textwidth}}
    \caption{\textbf{Valid parameters for extraction, according to PERG's process.}}
    \label{tab:perg_parameter_types}\\
        \toprule
        \bf Parameter type & \bf Parameter class & \bf Description \\\midrule
        Attack rate & Attack rate & Proportion of a population that becomes infected during an outbreak. \\
        \addlinespace
        Secondary attack rate & Attack rate & Proportion of contacts of a primary case who become infected. \\
        \addlinespace
        Doubling time & Doubling time & Time required for the number of infections to double. \\
        \addlinespace
        Growth rate & Growth rate & Exponential rate at which new infections increase over time. \\
        \addlinespace
        Evolutionary rate & Mutations & Rate of genetic change in a population over time, typically substitutions per site per year. \\
        \addlinespace
        Mutation rate & Mutations & Frequency at which new genetic mutations arise per site per replication cycle. \\
        \addlinespace
        Substitution rate & Mutations & Speed at which mutations become fixed in a population’s genome. \\
        \addlinespace
        Generation time & Human delay & Average interval between infection in a case and infection in a secondary case. \\
        \addlinespace
        Serial interval & Human delay & Time between symptom onset in a primary and secondary case. \\
        \addlinespace
        Latent period & Human delay & Time from infection to becoming infectious. \\
        \addlinespace
        Incubation period & Human delay & Time from infection to symptom onset. \\
        \addlinespace
        Infectious period & Human delay & Duration during which an infected person can transmit the pathogen. \\
        \addlinespace
        Time in care & Human delay & Average duration of hospitalisation or clinical care. \\
        \addlinespace
        Symptom onset $\rightarrow$ admission to care & Human delay & Time from symptom onset to hospital or clinical admission. \\
        \addlinespace
        Symptom onset $\rightarrow$ discharge / recovery & Human delay & Time from symptom onset to recovery or discharge. \\
        \addlinespace
        Symptom onset $\rightarrow$ death & Human delay & Time from symptom onset to death. \\
        \addlinespace
        Admission $\rightarrow$ discharge / recovery & Human delay & Time from hospital admission to recovery or discharge. \\
        \addlinespace
        Admission $\rightarrow$ death & Human delay & Time from hospital admission to death. \\
        \addlinespace
        Other human delay & Human delay & Other reported delays related to human infection or response. \\
        \addlinespace
        Overdispersion & Overdispersion & Measure of variation in the distribution of individual infectiousness. \\
        \addlinespace
        Human-to-human & Relative contribution & Proportion of total transmission attributable to human-to-human spread. \\
        \addlinespace
        Zoonotic-to-human & Relative contribution & Proportion of total transmission from animal or vector sources to humans. \\
        \addlinespace
        Basic (R0) & Reproduction number & Average number of secondary cases from one case in a fully susceptible population. \\
        \addlinespace
        Effective (Re) & Reproduction number & Average number of secondary cases in a population with partial immunity or interventions. \\
        \addlinespace
        Case fatality rate (CFR) & Severity & Proportion of diagnosed cases that result in death. \\
        \addlinespace
        Infection fatality rate (IFR) & Severity & Proportion of all infections (symptomatic and asymptomatic) that result in death. \\
        \addlinespace
        Proportion of symptomatic cases & Severity & Proportion of infections that develop symptoms. \\
        \addlinespace
        IgM & Seroprevalence & Proportion of individuals with detectable IgM antibodies, indicating recent infection. \\
        \addlinespace
        IgG & Seroprevalence & Proportion of individuals with IgG antibodies, indicating past infection or immunity. \\
        \addlinespace
        PRNT & Seroprevalence & Proportion with neutralising antibodies detected by plaque reduction neutralization test. \\
        \addlinespace
        HAI/HI & Seroprevalence & Proportion with antibodies detected by hemagglutination inhibition assay. \\
        \addlinespace
        IFA & Seroprevalence & Proportion with antibodies detected by immunofluorescence assay. \\
        \addlinespace
        Unspecified & Seroprevalence & Seroprevalence reported without specifying assay type. \\
        \addlinespace
        Risk factors & Risk factors & Host, environmental, or behavioural characteristics associated with infection risk. \\
    \bottomrule
\end{longtable}
\end{footnotesize}


\paragraph{Multi-Stage Parameter Extraction Pipeline}
Parameter extraction utilises a five-step workflow that mirrors how a careful human reader would process scientific articles. Starting from full-text contents, we identify relevant estimates in the text, extract them into a standardised format, and collect relevant metadata about population context and parameter uncertainty.

For our first step, we ask a reasoning language model with tool calling to ``screen'' each article for each parameter class. The reasoning model is provided with a tool to extract (potentially discontiguous) quotations from the source text that relate to the parameter class. We provide specific details for each parameter class as displayed in \cref{tab:parameter_class_screening_details}, which are copied quotations from the parameter extraction documentation from the \texttt{priority-pathogens} codebase \cite{nashPrioritypathogens2026}, accessible at \href{https://github.com/mrc-ide/priority-pathogens/wiki/Parameter-Data}{https://github.com/mrc-ide/priority-pathogens/wiki/Parameter-Data}.


\begin{footnotesize}
\begin{longtable}{p{0.2\textwidth}p{0.75\textwidth}}
    \caption{\textbf{Screening details for each parameter class.} This is inputted into the {``Parameter Class Screening Details''} section of the \textcolor{blue!70!black}{\textbf{\sffamily{Parameter Screening Prompt}}} below.} \\
        \toprule
        \bf Parameter Class & \bf Screening Details \\
        \midrule
        Attack rate & The attack rate is the proportion of an at-risk population contracting the disease during a specified time interval. It is often reported as a percentage or rate, e.g. 52 people per 10,000 people. \\
        \addlinespace

        Growth rate & The epidemic growth rate is a key metric that reflects how quickly the number of infections is changing day by day in a population. It is a time-dependent measure, usually expressed as a percentage or a rate per unit of time (e.g. per day), and is crucial for monitoring the speed and trajectory of an outbreak. \\
        \addlinespace

        Human delay & These parameters all refer to time intervals in the natural history of infection of the host. \\
        \addlinespace

        Mutation rate & Mutation rates, like substitution rate or evolutionary rate, describe the speed at which genetic changes accumulate in a population. \\
        \addlinespace

        Relative contribution & This parameter is intended for pathogens (e.g. MERS) where there is both human to human (h2h) and animal to human (a2h) transmission, and aims to capture the relative magnitude of these two routes of infection in humans. We expect these to be proportions or percentages. E.g. a study might estimate 60\% of infections in humans to be from h2h infection. \\
        \addlinespace

        Reproduction number & We are extracting either the basic reproduction number R0 or the effective reproduction number Re. \\
        \addlinespace

        Risk factors & We are extracting general information about risk factors in the included papers. We are extracting both univariate (naive) and multivariate (adjusted) risk factors, even if they are both available. \\
        \addlinespace

        Seroprevalence & These parameters refer to estimations of seroprevalence in the paper. This may also be referred to as antibody prevalence. These parameters will all be expressed in a proportion or percentage of the population. \\
        \addlinespace

        Severity & Severity refers to either the case fatality ratio or the infection fatality ratio. The case fatality ratio is the proportion of cases who end up dying of the disease. Note this depends on the case definition used, as the denominator is people identified as ``cases''. The infection fatality ratio is the proportion of infections who end up dying of the disease. \\
        \bottomrule
    \label{tab:parameter_class_screening_details}
\end{longtable}
\end{footnotesize}

The model is also provided with the study objectives from \cref{app:study_objectives} and instructed to only extract parameters ``estimated from or fitted to actual data''. If no relevant information is found, the model is told to refrain from calling the tool. The full prompt for this step is templatised as follows:

\begin{tcolorbox}[
    enhanced,
    breakable,
    colback=white,
    colframe=blue!30,
    boxrule=0.8pt,
    arc=2pt,
    left=0pt, right=0pt, top=0pt, bottom=0pt,
    title={\small\textbf{Parameter Screening Prompt}},
    fonttitle=\sffamily,
    coltitle=blue!70!black,
    colbacktitle=blue!15,
    attach boxed title to top left={yshift=-2mm, xshift=4mm},
    boxed title style={boxrule=0.5pt, arc=1pt}
]

\begin{tcolorbox}[
    enhanced,
    colback=gray!5,
    colframe=gray!20,
    boxrule=0pt,
    leftrule=3pt,
    arc=0pt,
    left=6pt, right=6pt, top=4pt, bottom=4pt
]

\vspace{2pt}
{\small{You are an expert epidemiologist extracting epidemiological parameters from scientific articles. You will be provided with the processed text of a scientific article. Your task is to extract information about epidemiological parameters according to the provided schema.}}
\end{tcolorbox}

\begin{tcolorbox}[
    enhanced,
    colback=orange!5,
    colframe=orange!20,
    boxrule=0pt,
    leftrule=3pt,
    arc=0pt,
    left=6pt, right=6pt, top=4pt, bottom=4pt
]
{\small\textbf{\textcolor{orange!60!black}{Study Objectives}}}

\vspace{2pt}
{\small

\textit{See study objectives in \cref{app:study_objectives}.}}
\end{tcolorbox}

\begin{tcolorbox}[
    enhanced,
    colback=gray!5,
    colframe=gray!20,
    boxrule=0pt,
    leftrule=3pt,
    arc=0pt,
    left=6pt, right=6pt, top=4pt, bottom=4pt
]
{\small\textbf{\textcolor{gray!60!black}{Summary Extraction Task Definition}}}

\vspace{2pt}
{
\small

For your first task, you will be provided with the full text of a scientific article and a specific type of parameter. We are only extracting parameters that are estimated from or fitted to actual data. For transmission models, if it is only a theoretical model and they have just chosen parameters from other studies/randomly, then please don’t extract these.

\vspace{1em}

Your task is to scan the provided text and determine whether this article estimates any parameters of the provided type. If it does, you must use the provided tool to extract relevant summaries from the text about this parameter. If the article makes no mention of the parameter, simply do not call the tool.

\vspace{1em}

If there are multiple pieces of information about the same parameter, return them as separate list items. You will need to call the tool multiple times if there are multiple separate parameter estimates of the provided type.

\vspace{1em}

In future steps, we will be using the provided summaries to extract structured information about the parameter, including:
\begin{enumerate}[label=(\alph*), leftmargin=*, itemsep=-3pt]
    \item The estimated value
    \item Uncertainty intervals
    \item Sample study population
\end{enumerate}

Please make sure your summaries provide all of this information if it is provided. Please be thorough: err on the side of extracting more information rather than less.

}
\end{tcolorbox}

\begin{tcolorbox}[
    enhanced,
    colback=gray!5,
    colframe=gray!20,
    boxrule=0pt,
    leftrule=3pt,
    arc=0pt,
    left=6pt, right=6pt, top=4pt, bottom=4pt
]
{\small\textbf{\textcolor{gray!70!black}{Parameter Class Screening Details}}}

\vspace{2pt}
{\small\textit{
See the details provided for each parameter class in \cref{tab:parameter_class_screening_details}.
}}
\end{tcolorbox}

\begin{tcolorbox}[
    enhanced,
    colback=gray!5,
    colframe=gray!20,
    boxrule=0pt,
    leftrule=3pt,
    arc=0pt,
    left=6pt, right=6pt, top=4pt, bottom=4pt
]
{\small\textbf{\textcolor{gray!70!black}{Full Text}}}

\vspace{2pt}
{\small\ttfamily
Title: \{\{title\}\}\\[2pt]
Full Text: \{\{fulltext\}\}
}
\end{tcolorbox}

\end{tcolorbox}



Our next steps are executed for each value of the \texttt{summaries} array returned by the model's tool call. We prompt the model in a new context, omitting the full text, to focus the model on the relevant text snippets from \texttt{summaries} and to save both inference time and API cost. If no relevant parameters are identified for a given article, \texttt{summaries} will be empty, and the extraction process will terminate.

Otherwise, we move to our second step, value extraction. At this step, the model utilises the \texttt{value\_info} of the parameter to extract structured information about its value and uncertainty bounds. As before, we provide instructions for using the tool for each parameter class. These are listed below:


\begin{tcolorbox}[
    enhanced,
    breakable,
    enforce breakable,
    colback=violet!5,
    colframe=violet!20,
    boxrule=0pt,
    leftrule=3pt,
    arc=0pt,
    left=6pt, right=6pt, top=4pt, bottom=4pt
]
{\small\textbf{\textcolor{violet!60!black}{Value Extraction Details for Attack rate}}}

\vspace{2pt}
{\small
If the attack rate is reported as a percentage, extract the percentage in the \texttt{value} field and set \texttt{unit} to \texttt{percentage}. If the attack rate is reported as a rate, extract the numerator in the \texttt{value} field and set \texttt{rate\_denominator} to the denominator of the rate.

Please extract attack rates as written in the paper.
}
\end{tcolorbox}

\begin{tcolorbox}[
    enhanced,
    breakable,
    enforce breakable,
    colback=violet!5,
    colframe=violet!20,
    boxrule=0pt,
    leftrule=3pt,
    arc=0pt,
    left=6pt, right=6pt, top=4pt, bottom=4pt
]
{\small\textbf{\textcolor{violet!60!black}{Value Extraction Details for Growth rate}}}

\vspace{2pt}
{\small
Please extract growth rates from the paper. Populate the \texttt{value} field with a numerical value as it is specified in the paper. If the paper provides a percentage value like $33\%$, record this value as \texttt{0.33}. Populate the \texttt{unit} field with one of the provided units according to the tool schema.
}
\end{tcolorbox}

\begin{tcolorbox}[
    enhanced,
    breakable,
    enforce breakable,
    colback=violet!5,
    colframe=violet!20,
    boxrule=0pt,
    leftrule=3pt,
    arc=0pt,
    left=6pt, right=6pt, top=4pt, bottom=4pt
]
{\small\textbf{\textcolor{violet!60!black}{Value Extraction Details for Human delay}}}

\vspace{2pt}
{\small
\textbf{Delay type}

The \texttt{delay\_type} field records the specific type of time interval. It can take one of the following values:
\begin{itemize}[leftmargin=*, itemsep=-3pt]
    \item \texttt{generation\_time}: The generation time is the time interval between infector exposure to infection and infectee exposure to infection. It may be used in reproduction number estimation, but given the difficulties in its observation, it may be replaced by the serial interval (see below).
    \item \texttt{serial\_interval}: The serial interval is the time interval between infector symptom onset and infectee symptom onset. It is frequently used in reproduction number estimation, as a substitute for the generation time.
    \item \texttt{latent\_period}: The latent period is the time interval between exposure to infection and becoming infectious. It is sometimes used interchangeably with the incubation period (see below). It may also be referred to as the latency period or the pre-infectious period.
    \item \texttt{incubation\_period}: The incubation period is the time interval between exposure to infection and symptom onset. It often coincides with the latent period, but may be shorter (symptom onset before infectiousness, e.g. SARS) or longer (infectiousness before symptom onset, e.g. Covid-19). It may also be referred to as the intrinsic incubation period (in the context of vector-borne diseases) or a subclinical infection.
    \item \texttt{infectious\_period}: The infectious period is the time interval during which the host remains infectious. It directly follows the latent period (see above). It may also be referred to as the infective period, the contagious period, the transmission period or the communicability period.
    \item \texttt{time\_in\_care}: The time in care is the time interval between admission to care and discharge from care or death. Unless there is a delay in receiving care, it directly follows the time from symptom to careseeking. It may vary according to health outcome and is typically highly skewed. It may also be referred to as the length of stay (LOS).
\end{itemize}

Human delays other than the six listed above may also be reported, for example the time from symptom onset to recovery, symptom onset to death, time from seeking care to admission to care etc. We allow \texttt{delay\_type} to take on one of these other time interval values:
\begin{itemize}[leftmargin=*, itemsep=-3pt]
    \item \texttt{admission\_\_to\_\_death}
    \item \texttt{admission\_\_to\_\_discharge\_or\_recovery}
    \item \texttt{symptom\_onset\_\_to\_\_admission}
    \item \texttt{symptom\_onset\_\_to\_\_death}
    \item \texttt{symptom\_onset\_\_to\_\_discharge\_or\_recovery}
\end{itemize}

In the case that \textit{none} of the above values apply to a human delay parameter you have found, set \texttt{delay\_type = 'other'} and record the type of delay in the \texttt{delay\_type\_note} field.

\vspace{1em}

\textbf{Value and unit}

Use the \texttt{value} and \texttt{unit} fields to record the parameter estimate (e.g. $x$ hours, days, weeks, or other).
}
\end{tcolorbox}

\begin{tcolorbox}[
    enhanced,
    breakable,
    enforce breakable,
    colback=violet!5,
    colframe=violet!20,
    boxrule=0pt,
    leftrule=3pt,
    arc=0pt,
    left=6pt, right=6pt, top=4pt, bottom=4pt
]
{\small\textbf{\textcolor{violet!60!black}{Value Extraction Details for Mutation rate}}}

\vspace{2pt}
{\small
For this task, we extract parameters estimated from pathogen genetic sequences. If no parameters were derived from genetic sequences, then this section can be skipped \textit{even if sequencing was performed and reported}.

\vspace{1em}

\texttt{substitution\_rate}, \texttt{evolutionary\_rate}, and \texttt{mutation\_rate} are different \texttt{parameter\_type} values for describing the speed at which genetic changes accumulate in a population. When selecting the \texttt{parameter\_type}, choose the value type and units based on the wording used by the authors in the article. If there are multiple terms used for the same measure (e.g. substitution rate is used in the text, evolutionary rate is used in the table), choose either the most frequently used term or default to \texttt{substitution\_rate} (if the units are substitutions per site per year). These values are often in the supplemental material. So if genetic sequences or phylogenetic analyses are presented, check the supplement. We are not extracting parameters associated with selection pressure or synonymous/nonsynonymous mutations, unless based on data or methodological limitations they have only been able to calculate substitution rate from nonsynonymous mutations (in that case specify this in the `Gene' field, similar to \textit{in vitro} experiments - see next bullet point). If substitution rates are calculated for subgroups (e.g. `clades,' `strains,' `branches', etc), report the global estimate and indicate disaggregated data is available in the Parameter Disaggregation section.

\vspace{1em}

As always, the \texttt{unit} value is very important for these parameters. The most common unit is \texttt{substitutions\_per\_site\_per\_year}. If units are not clear or they do not match the available options in the drop-down menu, set to \texttt{unspecified}.

\vspace{1em}

Fill the \texttt{genome\_site} field with the portion of the pathogen's genome used to estimate any extracted parameters (e.g. reproduction number, growth rate, substitution rate). This can be a gene, a gene segment, a codon position, or a more generic description (e.g. `whole genome' or `intergenic positions'). If parameter values are independently estimated for different portions of the genome, please enter each on a separate parameter value form. If a mutation rate is estimated by \textit{in vitro} experiments of recombinant variants (for example, measuring the rate of mutation in an inserted gene, such as green fluorescent protein [GFP]), enter the name of the inserted gene used, even though this gene might not be naturally occurring in the virus's genome. In addition, they may measure different types of mutations (SNPs vs indels) during \textit{in vitro} experiments. If this is the case, enter the type of mutation used to calculate the rate (ex. GFP-SNP, to signify that SNP mutations in the GFP gene were used to calculate the mutation rate).
}
\end{tcolorbox}

\begin{tcolorbox}[
    enhanced,
    breakable,
    enforce breakable,
    colback=violet!5,
    colframe=violet!20,
    boxrule=0pt,
    leftrule=3pt,
    arc=0pt,
    left=6pt, right=6pt, top=4pt, bottom=4pt
]
{\small\textbf{\textcolor{violet!60!black}{Value Extraction Details for Severity}}}

\vspace{2pt}
{\small
\begin{itemize}
    \item \texttt{parameter\_type} -- we extract case fatality ratios (CFR), infection fatality ratios (IFR), and the proportion of cases that are symptomatic and asymptomatic.
    \begin{itemize}
        \item Case fatality ratio (\texttt{CFR}) -- the proportion of cases who end up dying of the disease. Note this depends on the case definition used, as the denominator is people identified as ``cases''. All CFRs should be extracted, even when a subset of the population is selected (e.g. severe cases); make sure to describe the population denominator in the context and notes.
        \item Infection fatality ratio (\texttt{IFR}) -- the proportion of infections who end up dying of the disease (harder to calculate but less context dependent).
        \item Symptomatic proportion of infections -- the proportion of total infections that are symptomatic.
        \item Asymptomatic proportion of infections -- the proportion of total infections that are asymptomatic.
    \end{itemize}
    \item Parameter value -- we don't do any calculation ourselves i.e. if a paper quotes number of deaths and number of cases, but not a CFR, we don't calculate the CFR.
    \item Ratio/prevalence values -- please extract the \texttt{numerator} and \texttt{denominator} that generate the severity ratio. In line with the rule of 3, only extract the numerator and denominator of the central CFR value, even if disaggregated numerators and denominators are available. If there is no central value, do not extract any numerator or denominator. If the numerator and denominator are presented, but the percentage severity is not, extract the numerator, denominator and context, but leave the central value blank.
    \item \texttt{method} -- we extract information about the method used to calculate CFR (or IFR), mainly whether it is:
    \begin{itemize}
        \item a ``\texttt{naive}'' method, i.e. percentage mortality which computes total deaths divided by total cases (or infections); this is wrong because there may be many cases or infections who do not have final status information, so the naive estimate is typically an underestimate of true CFR (or IFR).
        \item an \texttt{adjusted} method, which somehow accounts for infections or cases with unknown final status (e.g. calculates deaths / (deaths + recoveries) or does something more fancy).
        \item an \texttt{unknown} method.
    \end{itemize}
    \item \texttt{value\_type}: mean, median, shape, etc. Please note that it may be the case that multiple measures of central tendency are provided, especially when the entire distribution of a parameter is presented. To avoid extracting multiple measures of centrality for the same parameter and to avoid bias, only one parameter \texttt{value\_type} can be extracted. Central parameter types are prioritised based on the available uncertainty types in the following way:
    \begin{itemize}
        \item When SD/variance/CIs are available: extract \texttt{mean}.
        \item Else when only IQR/CrIs are available: extract \texttt{median}.
        \item If \texttt{mode} is presented, this should be prioritised \textit{after} the \texttt{mean} or \texttt{median}.
        \item If Weibull distribution parameters are presented: prioritise extraction of the \texttt{shape} rather than mean/CIs or median/CrIs. We can get mean/CIs from shape/scale analytically but can only get shape/scale from mean/CIs numerically.
    \end{itemize}
    \item \texttt{statistical\_approach} -- if the central parameter estimates are summarised directly from empirical data, select \texttt{observed\_sample\_statistic}. If the central parameter is estimated using a transmission model, select \texttt{estimated\_model\_parameter}. Due to limited data sources, the Oropouche systematic review \textit{only} was extended to include \texttt{case\_study} data.
\end{itemize}
}
\end{tcolorbox}


The full prompt for the value extraction step is templatised below, incorporating text from both the parameter class screening details and the value extraction details.

\begin{tcolorbox}[
    enhanced,
    breakable,
    colback=white,
    colframe=blue!30,
    boxrule=0.8pt,
    arc=2pt,
    left=0pt, right=0pt, top=0pt, bottom=0pt,
    title={\small\textbf{Value Extraction Prompt}},
    fonttitle=\sffamily,
    coltitle=blue!70!black,
    colbacktitle=blue!15,
    attach boxed title to top left={yshift=-2mm, xshift=4mm},
    boxed title style={boxrule=0.5pt, arc=1pt}
]

\begin{tcolorbox}[
    enhanced,
    colback=gray!5,
    colframe=gray!20,
    boxrule=0pt,
    leftrule=3pt,
    arc=0pt,
    left=6pt, right=6pt, top=4pt, bottom=4pt
]

\vspace{2pt}
{\small{You are an expert epidemiologist extracting epidemiological parameters from scientific articles. You will be provided with the processed text of a scientific article. Your task is to extract information about epidemiological parameters according to the provided schema.}}
\end{tcolorbox}

\begin{tcolorbox}[
    enhanced,
    colback=orange!5,
    colframe=orange!20,
    boxrule=0pt,
    leftrule=3pt,
    arc=0pt,
    left=6pt, right=6pt, top=4pt, bottom=4pt
]
{\small\textbf{\textcolor{orange!60!black}{Study Objectives}}}

\vspace{2pt}
{\small

\textit{See study objectives in \cref{app:study_objectives}.}}
\end{tcolorbox}

\begin{tcolorbox}[
    enhanced,
    colback=gray!5,
    colframe=gray!20,
    boxrule=0pt,
    leftrule=3pt,
    arc=0pt,
    left=6pt, right=6pt, top=4pt, bottom=4pt
]
{\small\textbf{\textcolor{gray!60!black}{Value Extraction Task Definition}}}

\vspace{2pt}
{
\small
\textbf{Value extraction task}

For your next task, you will be provided with excerpts from a scientific article and a specific type of parameter. We are only extracting parameters that are estimated from or fitted to actual data. For transmission models, if it is only a theoretical model and they have just chosen parameters from other studies/randomly, then please don’t extract these.

\vspace{1em}

Scan the provided text and for the requested parameter and return all estimated parameter values using the provided tool. You will need to call the tool multiple times if there are multiple separate estimates.

}
\end{tcolorbox}

\begin{tcolorbox}[
    enhanced,
    colback=gray!5,
    colframe=gray!20,
    boxrule=0pt,
    leftrule=3pt,
    arc=0pt,
    left=6pt, right=6pt, top=4pt, bottom=4pt
]
{\small\textbf{\textcolor{gray!70!black}{Parameter Class Screening Details}}}

\vspace{2pt}
{\small
\textbf{\{\{\texttt{parameter\_class}\}\}: parameter value extraction}

\vspace{1em}

\{\{\textit{Screening details from \cref{tab:parameter_class_screening_details}}\}\}

\vspace{1em}

\begin{tcolorbox}[
    enhanced,
    breakable,
    enforce breakable,
    colback=violet!5,
    colframe=violet!20,
    boxrule=0pt,
    leftrule=3pt,
    arc=0pt,
    left=6pt, right=6pt, top=4pt, bottom=4pt
]
{\small\textbf{\textcolor{violet!60!black}{Value Extraction Details for \{\texttt{parameter\_class}\}}}

\textit{See the specific details of value extraction above.}
}
\end{tcolorbox}

}
\end{tcolorbox}

\begin{tcolorbox}[
    enhanced,
    colback=gray!5,
    colframe=gray!20,
    boxrule=0pt,
    leftrule=3pt,
    arc=0pt,
    left=6pt, right=6pt, top=4pt, bottom=4pt
]
{\small\textbf{\textcolor{gray!70!black}{Value Excerpts}}}

\vspace{2pt}
{\small
The following are excerpts from the scientific article about parameter value:

\vspace{1em}

\{\{\texttt{value\_info}\}\}

}
\end{tcolorbox}

\end{tcolorbox}

The tool provided to the language model is distinct per parameter class. In \cref{tab:value_extraction_schemas}, we specify the schemas utilised for these tool calls.

\begin{footnotesize}
\begin{longtable}{p{0.15\textwidth}p{0.15\textwidth}p{0.08\textwidth}p{0.23\textwidth}p{0.25\textwidth}}
    \caption{\textbf{Schemas used for value extraction tool calls for each parameter class.} Here ``--'' means that any values of the correct type are allowed.}\label{tab:value_extraction_schemas} \\
        \toprule
        \bf Parameter class & \bf Variable & \bf Type & \bf Allowed values & \bf Description \\
        \midrule
        Attack rate & value & Float & -- & The value of the attack rate. \\
        & unit & Enum & percentage; rate & The unit of the provided attack rate. \\
        & type & Enum & primary; secondary & Whether primary or secondary attack rate. \\
        & rate denominator & Integer; Null & -- & The denominator of the value, if the parameter is provided as a rate. \\
        \addlinespace
        Doubling time & value & Float & -- & The value of the doubling time, in days. \\
        \addlinespace
        Growth rate & value & Float & -- & The value of the growth rate. \\
        & unit & Enum & per hour; per day; per week; per month; per year; other; unspecified & The unit of the provided growth rate. \\
        \addlinespace
        Human delay & value & Float & -- & The value of the human delay parameter. \\
        & delay type & Enum & admission to death; admission to discharge or recovery; generation time; incubation period; infectious period; serial interval; symptom onset to admission; symptom onset to death; symptom onset to discharge or recovery; time in care; other & The specific delay parameter reported. \\
        \addlinespace
        Mutation rate & value & Float & -- & The value of the mutation rate parameter. \\
        & type & Enum & evolutionary rate; mutation rate; substitution rate & The specific mutation rate parameter reported. \\
        & unit & Enum & substitutions per site per year; mutations per genome per generation; percentage; other; unspecified & The unit of the mutation rate parameter value. \\
        & genome site & String & -- & The specific genome site or region associated with the mutation rate value. \\
        \addlinespace
        Overdispersion & value & Float & -- & The value of the overdispersion parameter \\
        & unit & Enum & no units; max number of cases superspreading & The unit of the overdispersion parameter \\
        \addlinespace
        Relative contribution & value & Float & -- & The value of the relative contribution parameter. \\
        & type & Enum & human-to-human; zoonotic-to-human & The type of relative contribution reported. \\

        \addlinespace
        Reproduction number & value & Float & -- & The value of the reproduction number parameter. \\
        & type & Enum & basic R0; effective Re & The type of reproduction number reported. \\
        & transmission & Enum & human; mosquito; unspecified; other & The type of transmission for this reproduction number estimate. \\
        & method & Enum & branching process; growth rate; compartmental model; next generation matrix; empirical; genomic; other & The method used to obtain the reproduction number estimate. \\
        \addlinespace
        Risk factors & name & List[Enum] & age; close contact; breastfeeding; comorbidity; contact with animal; environmental; funeral; hospitalisation; household contact; humidity; non-household contact; occupation; prior immunity to arboviruses; rainfall; sex; social gathering; temperature; other & The name of the risk factor. \\
        & outcome & List[Enum] & death in general population; Guillain Barre Syndrome; infection; low birthweight; microcephaly; miscarriage or stillbirth; other neurological symptoms in general population; premature birth; serology; severe disease in general population; spillover risk; recovery; Zika congenital syndrome or other birth defects; other & The outcome for which the risk factor was evaluated. \\
        & occupation & List[Enum] & abattoir services; correctional facilities; education; funeral and burial services; healthcare; laboratory; livestock and animal herders; public transport; quarantine facilities; veterinary; other; unspecified & If \texttt{name} is set to `occupation', the occupation(s) that correspond(s) most closely to that described in the paper. \\
        & significant & Enum & significant; not significant; unspecified & Whether the risk factor is significant or not. \\
        & adjusted & Enum & adjusted; not adjusted; unspecified & Whether the estimates of the risk factors are adjusted or unadjusted. \\
        \addlinespace
        Seroprevalence & value & Float & -- & The seroprevalence value as a proportion between 0.0 and 1.0. \\
        & parameter type & Enum & IgG; IgM; PRNT; HAI; IFA; unspecified & The type of seroprevalence parameter. \\
        & numerator & Integer; Null & The numerator used to calculate the seroprevalence value. If not provided, set to \texttt{Null}. \\
        & denominator & Integer; Null & The denominator used to calculate the seroprevalence value. If not provided, set to \texttt{Null}. \\

        \addlinespace
        Severity & value & Float & -- & The value of the severity parameter as a proportion between 0.0 and 1.0. \\
        & numerator & Integer; Null & -- & The numerator of the CFR or IFR parameter, if provided. \\
        & denominator & Integer; Null & -- & The denominator of the CFR or IFR parameter, if provided. \\
        & parameter type & Enum & CFR; IFR; proportion of symptomatic cases; proportion of asymptomatic cases & The type of severity parameter reported. \\
        & method & Enum; Null & naive; adjusted; unknown & The method used to calculate the CFR or IFR. \\
        \bottomrule
\end{longtable}
\end{footnotesize}

Following value extraction, all parameters move to our third step: population context extraction. We extract population context with the same prompt and tool for all parameter classes (see below).  

\begin{tcolorbox}[
    enhanced,
    breakable,
    colback=white,
    colframe=blue!30,
    boxrule=0.8pt,
    arc=2pt,
    left=0pt, right=0pt, top=0pt, bottom=0pt,
    title={\small\textbf{Population Extraction Prompt}},
    fonttitle=\sffamily,
    coltitle=blue!70!black,
    colbacktitle=blue!15,
    attach boxed title to top left={yshift=-2mm, xshift=4mm},
    boxed title style={boxrule=0.5pt, arc=1pt}
]

\begin{tcolorbox}[
    enhanced,
    colback=gray!5,
    colframe=gray!20,
    boxrule=0pt,
    leftrule=3pt,
    arc=0pt,
    left=6pt, right=6pt, top=4pt, bottom=4pt
]

\vspace{2pt}
{\small{You are an expert epidemiologist extracting epidemiological parameters from scientific articles. You will be provided with the processed text of a scientific article. Your task is to extract information about epidemiological parameters according to the provided schema.}}
\end{tcolorbox}

\begin{tcolorbox}[
    enhanced,
    colback=orange!5,
    colframe=orange!20,
    boxrule=0pt,
    leftrule=3pt,
    arc=0pt,
    left=6pt, right=6pt, top=4pt, bottom=4pt
]
{\small\textbf{\textcolor{orange!60!black}{Study Objectives}}}

\vspace{2pt}
{\small

\textit{See study objectives in \cref{app:study_objectives}.}}
\end{tcolorbox}

\begin{tcolorbox}[
    enhanced,
    colback=gray!5,
    colframe=gray!20,
    boxrule=0pt,
    leftrule=3pt,
    arc=0pt,
    left=6pt, right=6pt, top=4pt, bottom=4pt
]
{\small\textbf{\textcolor{gray!60!black}{Population Extraction Task Definition}}}

\vspace{2pt}
{
\small

For your next task, you will be provided with excerpts from a scientific article and an estimated parameter that has been extracted from that article. Your task is to scan the provided text and extract relevant sample population information for the given parameter. You will use the provided tool, which sets the schema you should follow when returning population information.
}
\end{tcolorbox}

\begin{tcolorbox}[
    enhanced,
    colback=gray!5,
    colframe=gray!20,
    boxrule=0pt,
    leftrule=3pt,
    arc=0pt,
    left=6pt, right=6pt, top=4pt, bottom=4pt
]
{\small\textbf{\textcolor{gray!70!black}{Population Excerpts}}}

\vspace{2pt}
{\small
The following are excerpts from the scientific article about parameter population context:

\vspace{1em}

\{\{\texttt{population\_info}\}\}

}
\end{tcolorbox}

\end{tcolorbox}

\newpage
The population tool call is schematised as follows:

\begin{footnotesize}
\begin{longtable}
    {p{0.22\textwidth}p{0.08\textwidth}p{0.3\textwidth}p{0.28\textwidth}}
    \caption{\textbf{The schema for the population context extraction tool call.}}
    \label{tab:population_tool_call} \\
        \toprule
        \bf Variable & \bf Type & \bf Allowed values & \bf Description \\
        \midrule
        population sex & Enum & female; male; both; unspecified & The sex composition of your study population. If you have 99 men and 1 woman you would still put both in this option. \\
        \addlinespace
        
        population sample type & Enum & community based; hospital based; household based; housing estate based; population based; school based; travel based; trade or business based; contact based; mixed settings; other; unspecified & How was the study conducted? \\
        \addlinespace
        population group & Enum & healthcare workers; farmers; outdoor workers; animal workers; butchers; abattoir workers; pregnant women; children; sex workers; people who inject drugs; household contacts of survivors; persons under investigation; general population; persons with symptoms; mixed settings; unspecified; other & Demographic i.e. who was sampled? \\
        \addlinespace
        population sample size & Integer; Null & -- & Number of participants/samples tested etc. \\
        \addlinespace
        population age min & Integer; Null & -- & These must be number fields. If your sample is people over 18 you would put age min = 18 and leave age max blank. \\
        \addlinespace

        population age max & Integer; Null & -- & These must be number fields. If your sample is people over 18 you would put age min = 18 and leave age max blank. \\
        \addlinespace

        population countries & List[String] & -- & Where was the study undertaken? \\
        \addlinespace

        population location & String & -- & Location reported i.e. Kerry Town Ebola Treatment Centre. \\
        \addlinespace

        method moment value & Enum & start outbreak; mid outbreak; end outbreak; post outbreak; endemic; unspecified & When in the outbreak was this study undertaken? \\
        \bottomrule
\end{longtable}
\end{footnotesize}

\clearpage

For our final step, if we have multiple extractions of the same class for an article, we ask the language model to aggregate parameters that should be reported as ranges over population disaggregations. Our aggregation logic follows PERG's \textit{rule of three}, which specifies certain pathogen-specific conditions for when aggregated reporting is appropriate. These are detailed to the language model in the instruction prompt below, which is provided identically for all parameter classes.

\bigskip

\begin{tcolorbox}[
    enhanced,
    breakable,
    colback=white,
    colframe=blue!30,
    boxrule=0.8pt,
    arc=2pt,
    left=0pt, right=0pt, top=0pt, bottom=0pt,
    title={\small\textbf{Aggregation Prompt}},
    fonttitle=\sffamily,
    coltitle=blue!70!black,
    colbacktitle=blue!15,
    attach boxed title to top left={yshift=-2mm, xshift=4mm},
    boxed title style={boxrule=0.5pt, arc=1pt}
]

\begin{tcolorbox}[
    enhanced,
    colback=gray!5,
    colframe=gray!20,
    boxrule=0pt,
    leftrule=3pt,
    arc=0pt,
    left=6pt, right=6pt, top=4pt, bottom=4pt
]

\vspace{2pt}
{\small{You are an expert epidemiologist extracting epidemiological parameters from scientific articles. You will be provided with the processed text of a scientific article. Your task is to extract information about epidemiological parameters according to the provided schema.}}
\end{tcolorbox}

\begin{tcolorbox}[
    enhanced,
    colback=orange!5,
    colframe=orange!20,
    boxrule=0pt,
    leftrule=3pt,
    arc=0pt,
    left=6pt, right=6pt, top=4pt, bottom=4pt
]
{\small\textbf{\textcolor{orange!60!black}{Study Objectives}}}

\vspace{2pt}
{\small

\textit{See study objectives in \cref{app:study_objectives}.}}
\end{tcolorbox}

\begin{tcolorbox}[
    enhanced,
    colback=gray!5,
    colframe=gray!20,
    boxrule=0pt,
    leftrule=3pt,
    arc=0pt,
    left=6pt, right=6pt, top=4pt, bottom=4pt
]
{\small\textbf{\textcolor{gray!60!black}{Aggregation Task Definition}}}

\vspace{2pt}
{
\small
\textbf{Aggregation task}

For your next task, you will be provided with a list of parameters already extracted from an epidemiological study. Your task is to provide \textit{aggregations} of these parameter values when suitable.

\vspace{1em}

\textbf{Aggregation context}

Some epidemiological papers have a huge level of parameter disaggregation (e.g. age, sex, location) and so we have established different rules to ease our meta-review process. For non-location-related disaggregations, please remember the \textbf{rule of three}. If there are three or more disaggregations for a parameter, e.g. Rt values for three or more age groups, extract these as a \textbf{range} and specify that disaggregated data is available and what the parameter is disaggregated by.

Each pathogen has different rules on location, which we state here:
\begin{itemize}
    \item marburg; ebola; MERS: Location is included within the rule of three.
    \item lassa; SARS; zika; nipah: Please \textit{do not aggregate} values if the disaggregation is by location as much as possible and do not apply the rule of three for geographic regions down to admin level 2 (sub-regions) of a country. However, please respect the rule of three for estimates by neighborhood for example.
\end{itemize}

If the provided parameters do not contain adequate population information to perform an aggregation, then do not return any aggregated values.

If you decide that an aggregation is necessary, use the provided tool to submit aggregated values according to the tool's schema. Provide the \texttt{lower\_bound} and \texttt{upper\_bound} of the parameter values, and list the types of population disaggregation (like ``age'', ``sex'', etc.) in the \texttt{disaggregated\_by} field. Fill the \texttt{aggregated\_ids} list with all of the \texttt{id}s from the parameters you aggregated.}
\end{tcolorbox}

\begin{tcolorbox}[
    enhanced,
    colback=gray!5,
    colframe=gray!20,
    boxrule=0pt,
    leftrule=3pt,
    arc=0pt,
    left=6pt, right=6pt, top=4pt, bottom=4pt
]
{\small\textbf{\textcolor{gray!70!black}{Extracted parameters}}}

\vspace{2pt}
{\small
\textbf{Extracted parameters}: \{\{\texttt{parameters}\}\}
}
\end{tcolorbox}

\end{tcolorbox}

\clearpage

\subsection{Models}

\paragraph{Valid transmission models for extraction}
Epidemiological transmission models are mathematical frameworks that simulate how infectious diseases spread through populations by mechanistically describing the interactions between infected and susceptible individuals. We extract models that mechanistically represent disease transmission dynamics, excluding purely statistical analyses, regression-based forecasting without transmission mechanisms, and risk factor studies. Table~\ref{tab:model_characteristics} defines the categories of model characteristics extracted in this study, organised into structural properties, epidemiological features, assumptions, intervention categories, and reproducibility indicators.

\begin{footnotesize}
\begin{table}[H]
    \centering
    \caption{\textbf{Model characteristic categories targeted by the extraction pipeline.}}
    \label{tab:model_characteristics}
    \begin{tabular}{p{0.25\textwidth}p{0.65\textwidth}}
        \toprule
        \bf Category & \bf Description \\
        \midrule
        Structural Properties & 
        Model type (compartmental, agent-based, branching process) and compartmental architecture (SIS, SIR, SEIR, etc.). Whether the model is stochastic or deterministic. \\
        
        Epidemiological Features & 
        Primary transmission routes (airborne, direct contact, vector-borne, sexual). Spatial heterogeneity and spillover dynamics from animal reservoirs. \\
        
        Behavioural Assumptions & 
        Mixing patterns (homogeneous or heterogeneous), age-dependent susceptibility, cross-immunity between pathogens, and temporal variation in transmission rates. \\
        
        Theoretical vs. Fitted & 
        Whether the model was fitted to actual data or uses parameters from literature or arbitrary values. \\
        
        Intervention Categories & 
        Control measures evaluated including vaccination, quarantine, vector control, treatment, contact tracing, behaviour changes, and various vector management strategies. \\
        
        Reproducibility Indicators & 
        Code availability, programming language used, data sharing status, and presence of documentation (README). \\
        \bottomrule
    \end{tabular}
\end{table}
\end{footnotesize}

\paragraph{Model extraction schema}
Table~\ref{tab:model_extraction_schema} defines the complete extraction schema with field specifications, data types, allowed values, and descriptions. The schema uses controlled vocabularies to ensure consistency and enable structured analysis of modelling approaches across the literature.

\begin{footnotesize}
\renewcommand{\arraystretch}{1.3}
\begin{longtable}{>{\raggedright\small}p{.19\textwidth}>{\raggedright\small}p{.09\textwidth}>{\raggedright\small}p{.30\textwidth}>{\raggedright\arraybackslash\small}p{.32\textwidth}}
    \caption{\textbf{Model extraction schema with field specifications, data types, allowed values, and descriptions.}}\label{tab:model_extraction_schema} \\
    \toprule
    \bf Field Name & \bf Type & \bf Allowed Values & \bf Description \\
    \midrule
    
    
    
    
    model\_type & Enum & Compartmental; Branching process; Agent/Individual based; Other; Unspecified & Primary modeling framework used for transmission dynamics. \\
    \addlinespace
    
    compartmental\_type & Enum & SIS; SIR; SEIR; SEIR-SEI; SAIR-SEI; Not compartmental; Other compartmental & Specific compartmental model architecture if applicable. Use ``Not compartmental'' for non-compartmental models. \\
    \addlinespace
    
    stoch\_deter & Enum; Null & Stochastic; Deterministic & Whether the model is stochastic or deterministic. Only extract if explicitly stated. Null if not specified. \\
    \addlinespace
    
    transmission\_route & List[Enum] & Airborne or close contact; Human to human (direct contact); Human to human (direct non-sexual contact); Vector/Animal to human; Sexual; Unspecified & Primary pathway(s) through which pathogen transmission occurs. Multiple routes can be selected. \\
    \addlinespace
    
    uncertainty\_was\_considered & Boolean; Null & True; False & Whether uncertainty was considered through stochasticity, multiple models, or parameter variation (e.g. sensitivity analyses, Bayesian analysis). Null if not specified. \\
    \addlinespace
    
    spatial\_model & Boolean; Null & True; False & Whether the model explicitly incorporated spatial or geographic heterogeneity. Null if not specified. \\
    \addlinespace
    
    spillover\_included & Boolean; Null & True; False & Whether the model explicitly modelled spillover (e.g. animal reservoir component, contribution to force of infection from zoonosis). Null if not specified. \\
    \addlinespace
    
    assumptions & List[Enum] & Homogeneous mixing; Latent period is same as incubation period; Heterogeneity in transmission rates (between human groups; between groups; between human and vector; over time); Age dependent susceptibility; Cross-immunity between Zika and dengue; Other; Unspecified & Key structural and behavioural assumptions. Should be explicitly described in the paper or clear from model equations. Multiple assumptions can be selected. \\
    \addlinespace
    
    theoretical\_model & Boolean & True; False & Whether the model was NOT fitted to data (parameters from literature or arbitrary values). True if not fitted; False if fitted to actual data. \\
    \addlinespace
    
    interventions\_type & List[Enum] & Vaccination; Quarantine; Vector/Animal control; Treatment; Contact tracing; Hospitals; Treatment centres; Safe burials; Behaviour changes; Wolbachia replacement/suppression; Genetically modified mosquitoes; Mechanical removal of breeding sites; Pesticides/larvicides; Insecticide-treated nets; Indoor residual spraying; Other; Unspecified & Categories of control measures evaluated or incorporated in the model(s). Multiple interventions can be selected. \\
    \addlinespace
    
    code\_available & Boolean & True; False & Whether model implementation code was made publicly available. \\
    \addlinespace
    
    coding\_language & Enum; Null & R; Python; Matlab; Julia; C++; Other & Programming language(s) used for model implementation if code is available. Null if not specified. \\
    \addlinespace
    
    is\_data\_used\_available & Enum; Null & Yes (as an attachment; with a DOI; on Github; on another platform); Not available; Unspecified & Whether input data used for the model was shared and how it was shared. Null if not specified. \\
    \addlinespace
    
    readme\_included & Boolean; Null & True; False & Whether a README or detailed documentation accompanied the code repository. Null if not applicable. \\
    \addlinespace
    
    notes & String; Null & -- & Additional context or notes about the extracted model. Free text field. \\
    \bottomrule
\end{longtable}
\renewcommand{\arraystretch}{1.0}
\end{footnotesize}

\paragraph{Multi-stage model extraction pipeline}
Model extraction employs a two-stage agentic workflow operating on full-text article content. Unlike parameter extraction, which requires fine-grained text excerpting and value parsing, model extraction focuses on identifying the presence of dynamic transmission models and characterising their structural properties using controlled vocabularies.

In the first stage, a binary screening step identifies articles containing dynamic transmission models while excluding purely statistical analyses, regression-based forecasting, and risk factor studies without transmission dynamics (see the ``Model Screening Prompt'' below). The language model returns a simple ``True'' or ``False'' response indicating whether the article contains models suitable for extraction.

For articles passing this screen, the extraction stage deploys a structured tool-calling approach where the language model iteratively invokes an \texttt{extract\_model\_data} function once per distinct model identified in the article (see the ``Model Extraction Prompt'' below). Each tool call populates the standardised schema defined in Table~\ref{tab:model_extraction_schema}.

The schema enforces controlled vocabularies for all fields through strict JSON validation. Multiple-select fields (\texttt{transmission\_route}, \texttt{assumptions}, \texttt{interventions\_type}) accept arrays of values from predefined enumerations, while single-select fields enforce unique values or null for optional characteristics. Validation logic rejects outputs violating vocabulary constraints or logical rules. For example, a non-compartmental \texttt{model\_type} must have \texttt{compartmental\_type} set to ``Not compartmental'', this prompts the model to correct errors before proceeding.

The complete extraction workflow is coordinated by the \texttt{ModelExtractionRunner} class, which loads full-text data, applies screening decisions, manages iterative tool calls with validation feedback, and logs all outputs to structured CSV files.

\begin{tcolorbox}[
    enhanced,
    breakable,
    colback=white,
    colframe=blue!30,
    boxrule=0.8pt,
    arc=2pt,
    left=0pt, right=0pt, top=0pt, bottom=0pt,
    title={\small\textbf{Model Screening Prompt}},
    fonttitle=\sffamily,
    coltitle=blue!70!black,
    colbacktitle=blue!15,
    attach boxed title to top left={yshift=-2mm, xshift=4mm},
    boxed title style={boxrule=0.5pt, arc=1pt}
]

\begin{tcolorbox}[
    enhanced,
    colback=gray!5,
    colframe=gray!20,
    boxrule=0pt,
    leftrule=3pt,
    arc=0pt,
    left=6pt, right=6pt, top=4pt, bottom=4pt
]

\vspace{2pt}
{\small{You are an epidemiologist specializing in infectious disease modeling. Determine if this article contains dynamic transmission models for infectious disease.}}
\end{tcolorbox}

\begin{tcolorbox}[
    enhanced,
    colback=gray!5,
    colframe=gray!20,
    boxrule=0pt,
    leftrule=3pt,
    arc=0pt,
    left=6pt, right=6pt, top=4pt, bottom=4pt
]
{\small\textbf{\textcolor{gray!60!black}{Screening Task Definition}}}

\vspace{2pt}
{\small
\textbf{Include (respond ``True''):}
\begin{itemize}[leftmargin=*, itemsep=-3pt]
\item Compartmental models (SIR, SEIR, etc.)
\item Agent-based or individual-based models
\item Branching process models
\item Network transmission models
\end{itemize}

\vspace{4pt}
\textbf{Exclude (respond ``False''):}
\begin{itemize}[leftmargin=*, itemsep=-3pt]
\item Pure statistical/regression analyses
\item Time series forecasting without mechanistic transmission
\item Risk factor analyses without transmission dynamics
\item Seroprevalence studies without modeling
\end{itemize}

\vspace{4pt}
Respond with only ``True'' or ``False''.
}
\end{tcolorbox}

\begin{tcolorbox}[
    enhanced,
    colback=gray!5,
    colframe=gray!20,
    boxrule=0pt,
    leftrule=3pt,
    arc=0pt,
    left=6pt, right=6pt, top=4pt, bottom=4pt
]
{\small\textbf{\textcolor{gray!70!black}{Full Text}}}

\vspace{2pt}
{\small\ttfamily
Title: \{\{title\}\}\\[2pt]
Full Text: \{\{fulltext\}\}
}
\end{tcolorbox}

\end{tcolorbox}

\vspace{-3pt}

\clearpage

\begin{tcolorbox}[
    enhanced,
    breakable,
    colback=white,
    colframe=blue!30,
    boxrule=0.8pt,
    arc=2pt,
    left=0pt, right=0pt, top=0pt, bottom=0pt,
    title={\small\textbf{Model Extraction Prompt}},
    fonttitle=\sffamily,
    coltitle=blue!70!black,
    colbacktitle=blue!15,
    attach boxed title to top left={yshift=-2mm, xshift=4mm},
    boxed title style={boxrule=0.5pt, arc=1pt}
]

\begin{tcolorbox}[
    enhanced,
    colback=gray!5,
    colframe=gray!20,
    boxrule=0pt,
    leftrule=3pt,
    arc=0pt,
    left=6pt, right=6pt, top=4pt, bottom=4pt
]

\vspace{2pt}
{\small{You are an epidemiologist specializing in infectious disease modeling. Extract information about transmission models from scientific articles.}}
\end{tcolorbox}

\begin{tcolorbox}[
    enhanced,
    colback=gray!5,
    colframe=gray!20,
    boxrule=0pt,
    leftrule=3pt,
    arc=0pt,
    left=6pt, right=6pt, top=4pt, bottom=4pt
]
{\small\textbf{\textcolor{gray}{Study Objectives}}}

\vspace{2pt}
{\small
\textbf{Study Objectives}

This systematic review collates transmission models, outbreaks and parameters for \{\{pathogen\}\}.
}
\end{tcolorbox}

\begin{tcolorbox}[
    enhanced,
    colback=gray!5,
    colframe=gray!20,
    boxrule=0pt,
    leftrule=3pt,
    arc=0pt,
    left=6pt, right=6pt, top=4pt, bottom=4pt
]
{\small\textbf{\textcolor{gray!60!black}{Extraction Task Definition}}}

\vspace{2pt}
{\small
\textbf{Model extraction task}

Extract \textbf{ALL dynamic transmission models} described in the article that were actually used or implemented.

\vspace{6pt}
Do \textbf{not} extract:
\vspace{-4pt}
\begin{itemize}[leftmargin=*, itemsep=-3pt]
\item Models only mentioned as possible alternatives without implementation
\item Regression-only analyses
\item Purely statistical forecasting
\end{itemize}

\vspace{4pt}
\textbf{Tool Calling:}
\vspace{-4pt}
\begin{itemize}[leftmargin=*, itemsep=-3pt]
\item Call \texttt{extract\_model\_data} \textbf{once per model} identified in the article
\item After extracting all model/s, stop calling the tool (no completion call needed)
\end{itemize}

\vspace{4pt}
\textbf{Schema Requirements:}
\vspace{-4pt}
\begin{itemize}[leftmargin=*, itemsep=-3pt]
\item \texttt{transmission\_route}, \texttt{assumptions}, \texttt{interventions\_type} are \textbf{arrays} (multiple-select)
\item All other fields are \textbf{single values} (single-select)
\item Use \texttt{null} for optional single-select fields when not stated
\item Use \texttt{["Unspecified"]} for required arrays when not stated
\end{itemize}
}
\end{tcolorbox}

\begin{tcolorbox}[
    enhanced,
    colback=gray!5,
    colframe=gray!20,
    boxrule=0pt,
    leftrule=3pt,
    arc=0pt,
    left=6pt, right=6pt, top=4pt, bottom=4pt
]
{\small\textbf{\textcolor{gray!70!black}{Full Text}}}

\vspace{2pt}
{\small\ttfamily
Title: \{\{title\}\}\\[2pt]
Full Text: \{\{fulltext\}\}
}
\end{tcolorbox}

\end{tcolorbox}

The language model uses the \texttt{extract\_model\_data()} tool (provided to it) to populate the schema defined in Table~\ref{tab:model_extraction_schema}. The tool enforces strict JSON validation with controlled vocabularies for all fields, rejecting invalid outputs and prompting corrections. The complete tool specification follows standard OpenAI function calling conventions with enum constraints for single-select fields and array validation for multiple-select fields.

\clearpage
\subsection{Outbreaks}

\paragraph{Valid outbreak data for extraction}
Outbreak data capture the epidemiological characteristics of concluded epidemic events, including temporal bounds, geographic scope, transmission sources, case counts stratified by confirmation status, and demographic breakdowns. We extracted outbreak information as stated in articles, without performing additional calculations or inferring missing values.

Following extraction guidelines suggested by PERG,\footnote{\url{https://github.com/mrc-ide/priority-pathogens/wiki/Outbreak-data}} outbreak inclusion criteria varied by pathogen based on reporting completeness and literature volume. For Marburg and Lassa, all reported outbreaks were captured regardless of size. For Zika, only outbreaks with at least 10 confirmed, probable, or suspected cases were extracted, reflecting the assumption that smaller events may not be systematically documented and contribute minimally to population-level immunity estimates. Table~\ref{tab:outbreak_field_descriptions} defines the outbreak characteristics and their meanings in natural language.

\begin{footnotesize}
\begin{longtable}{p{.35\textwidth}p{.65\textwidth}}
\caption{\textbf{Outbreak field descriptions and meanings.}}
\label{tab:outbreak_field_descriptions} \\
\toprule
\bf Outbreak Characteristic & \bf Description \\
\midrule
\endfirsthead
\multicolumn{2}{c}%
{{\bfseries Table \thetable\ continued from previous page}} \\
\toprule
\bf Outbreak Characteristic & \bf Description \\
\midrule
\endhead
\midrule
\multicolumn{2}{r}{{Continued on next page}} \\
\endfoot
\bottomrule
\endlastfoot

Outbreak start day & Day of outbreak onset (1--31). Extracted as stated in paper. \\
Outbreak start month & Month of outbreak onset. Extracted as stated in paper. \\
Outbreak start year & Year of outbreak onset. Extracted as stated in paper. \\
Outbreak end day & Day of outbreak conclusion (1--31). Extracted as stated in paper. \\
Outbreak end month & Month of outbreak conclusion. Extracted as stated in paper. \\
Outbreak end year & Year of outbreak conclusion. Extracted as stated in paper. \\
Outbreak duration (months) & Duration of outbreak in months. ONLY extracted if explicitly stated in paper; not calculated. \\
Outbreak is currently ongoing & Whether outbreak was concluded or ongoing at time of publication. \\
Outbreak country & Country where outbreak occurred, using WHO standard country names. Refers to reporting country rather than infection source for imported cases. \\
Outbreak location & Specific geographic location within country (city, district, province) as written in paper. Multiple locations separated by semicolons. \\
Outbreak location type & Administrative or geographic unit type of outbreak location. \\
Outbreak source & Known or suspected source of outbreak introduction. \\
Mode of detection & Primary method(s) used to identify and confirm cases. \\
Method of case definition & Criteria used to classify cases. Extracted as described in paper. \\
Pre-outbreak baseline & Baseline disease status in affected area prior to outbreak. Rarely reported. \\
Cases confirmed & Number of laboratory-confirmed cases (e.g. via molecular testing). \\
Cases probable & Number of probable cases as defined in paper. Definition may vary across studies. \\
Cases suspected & Number of suspected cases as defined in paper. Definition may vary across studies. \\
Cases unspecified & Number of cases where confirmation status was not specified. \\
Cases asymptomatic & Number of asymptomatic infections as defined in paper. \\
Cases severe & Number of severe cases or hospitalizations as stated in paper. \\
Deaths & Number of deaths attributed to outbreak. \\
Asymptomatic transmission described & Whether article explicitly described or discussed asymptomatic transmission. \\
Population size (geographical area) & Total population of affected geographic area. Rarely reported. \\
Type of cases (sex disaggregation) & Case type for which sex disaggregation was reported. \\
Male cases & Number of cases in males for specified case type. \\
Proportion male cases & Proportion (0.0--1.0) or percentage (0--100) of cases in males. \\
Female cases & Number of cases in females for specified case type. \\
Proportion female cases & Proportion (0.0--1.0) or percentage (0--100) of cases in females. \\
Notes & Additional context or clarifications about outbreak characteristics. \\
\end{longtable}
 
\end{footnotesize}

\paragraph{Outbreak extraction schema}
Table~\ref{tab:outbreak_extraction_schema} defines the complete extraction schema with field specifications, data types, allowed values, and descriptions. The schema uses controlled vocabularies to ensure consistency and enable structured analysis of outbreak characteristics across the literature.

\renewcommand{\arraystretch}{1.3}
\begin{longtable}{>{\raggedright\small}p{.30\textwidth}>{\raggedright\small}p{.12\textwidth}>{\raggedright\small}p{.25\textwidth}>{\raggedright\arraybackslash\small}p{.25\textwidth}}
    \caption{\textbf{Outbreak extraction schema with field specifications, data types, allowed values, and descriptions.}}\label{tab:outbreak_extraction_schema} \\
    \toprule
    \bf Field Name & \bf Type & \bf Allowed Values & \bf Description \\
    \midrule
    \endfirsthead
    
    \multicolumn{4}{c}%
    {\small{\bfseries Table \thetable\ continued from previous page}} \\
    \toprule
    \bf Field Name & \bf Type & \bf Allowed Values & \bf Description \\
    \midrule
    \endhead
    
    \midrule
    \multicolumn{4}{r}{{Continued on next page}} \\
    \endfoot
    
    \bottomrule
    \endlastfoot
    
    outbreak\_start\_day & Integer; Null & 1-31 & Day of outbreak onset. Null if not provided. \\
    \addlinespace
    
    outbreak\_start\_month & String (Enum); Null & Jan, Feb, Mar, Apr, May, Jun, Jul, Aug, Sep, Oct, Nov, Dec & Month of outbreak onset. Null if not provided. \\
    \addlinespace
    
    outbreak\_start\_year & Integer; Null & Integer year & Year of outbreak onset. Null if not provided. \\
    \addlinespace
    
    outbreak\_end\_day & Integer; Null & 1-31 & Day of outbreak conclusion. Null if not provided. \\
    \addlinespace
    
    outbreak\_end\_month & String (Enum); Null & Jan, Feb, Mar, Apr, May, Jun, Jul, Aug, Sep, Oct, Nov, Dec & Month of outbreak conclusion. Null if not provided. \\
    \addlinespace
    
    outbreak\_end\_year & Integer; Null & Integer year & Year of outbreak conclusion. Null if not provided. \\
    \addlinespace
    
    outbreak\_duration\_months & Float; Null & Numeric value & Duration in months. ONLY if explicitly stated; not calculated. Null if not stated. \\
    \addlinespace
    
    outbreak\_is\_currently\_ongoing & Boolean & True; False & Whether outbreak was concluded (False) or ongoing (True) at publication. \\
    \addlinespace
    
    outbreak\_country & String (Enum) & WHO standard country names (195 member states) & Country where outbreak occurred. MUST match WHO standard names. \\
    \addlinespace
    
    outbreak\_location & String; Null & Free text & Specific location within country. Multiple locations separated by semicolons. Null if not provided. \\
    \addlinespace
    
    outbreak\_location\_type & String; Null & Free text (e.g. district, province, county, state, hospital) & Type of administrative or geographic unit. Null if not specified. \\
    \addlinespace
    
    outbreak\_source & String (Enum); Null & Domestic animal; Wild animal; Date palm sap; Unknown; Other & Known or suspected source of outbreak introduction. Null if not provided. \\
    \addlinespace
    
    mode\_of\_detection & String (Enum); Null & Molecular (PCR etc); Symptoms; Confirmed + Suspected; Unspecified & Primary method used to identify and confirm cases. Null if not provided. \\
    \addlinespace
    
    method\_of\_case\_definition & String; Null & Free text & Criteria used to classify cases as described in paper. Null if not provided. \\
    \addlinespace
    
    pre\_outbreak & String (Enum); Null & Disease-free baseline; Endemic equilibrium; Unspecified; Probable & Baseline disease status prior to outbreak. Null if not provided. \\
    \addlinespace
    
    cases\_confirmed & Float; Null & Non-negative numeric & Number of laboratory-confirmed cases. Null if not provided. \\
    \addlinespace
    
    cases\_probable & Float; Null & Non-negative numeric & Number of probable cases. Null if not provided. \\
    \addlinespace
    
    cases\_suspected & Float; Null & Non-negative numeric & Number of suspected cases. Null if not provided. \\
    \addlinespace
    
    cases\_unspecified & Float; Null & Non-negative numeric & Number of cases with unspecified confirmation status. Null if not provided. \\
    \addlinespace
    
    cases\_asymptomatic & Float; Null & Non-negative numeric & Number of asymptomatic infections. Null if not provided. \\
    \addlinespace
    
    cases\_severe & Float; Null & Non-negative numeric & Number of severe cases or hospitalizations. Null if not provided. \\
    \addlinespace
    
    deaths & Float; Null & Non-negative numeric & Number of deaths attributed to outbreak. Null if not provided. \\
    \addlinespace
    
    asymptomatic\_transmission\_described & Boolean & True; False & Whether article explicitly described or discussed asymptomatic transmission. \\
    \addlinespace
    
    population\_size\_geographical\_area & Float; Null & Non-negative numeric & Total population of affected geographic area. Null if not provided. \\
    \addlinespace
    
    type\_cases\_sex\_disagg & String (Enum); Null & Confirmed; Suspected; Other; Unspecified & Case type for which sex disaggregation was reported. Null if not provided. \\
    \addlinespace
    
    male\_cases & Float; Null & Non-negative numeric & Number of male cases for specified case type. Null if not provided. \\
    \addlinespace
    
    prop\_male\_cases & Float; Null & Numeric (0.0-1.0 or 0-100) & Proportion or percentage of cases in males. Null if not provided. \\
    \addlinespace
    
    female\_cases & Float; Null & Non-negative numeric & Number of female cases for specified case type. Null if not provided. \\
    \addlinespace
    
    prop\_female\_cases & Float; Null & Numeric (0.0-1.0 or 0-100) & Proportion or percentage of cases in females. Null if not provided. \\
    \addlinespace
    
    notes & String; Null & Free text & Additional context or clarifications about outbreak characteristics. Null if not needed. \\
    
\end{longtable}
\renewcommand{\arraystretch}{1.0}

\paragraph{Multi-stage outbreak extraction pipeline}
Outbreak extraction employs a two-stage workflow operating on full-text article content. The first stage applies binary screening to identify articles reporting concluded, real-world outbreak events with defined case counts, excluding ongoing outbreaks, modelled scenarios, routine surveillance, and single case reports (see the ``Outbreak Screening Prompt'' below). The language model returns a simple ``True'' or ``False'' response indicating whether the article contains outbreaks suitable for extraction.

For articles passing this screen, the extraction stage deploys a structured tool-calling approach where the language model iteratively invokes an \texttt{extract\_outbreak\_data} function once per distinct outbreak identified in the article (see the ``Outbreak Extraction Prompt'' below). Outbreaks are considered distinct if they differ by location, time period, or both. Each tool call populates the standardised schema defined in Table~\ref{tab:outbreak_extraction_schema}.

The schema enforces controlled vocabularies for categorical fields through strict JSON validation. The required fields must be provided (\texttt{outbreak\_is\_currently\_ongoing}, \texttt{outbreak\_country}, \texttt{asymptomatic\_transmission\_described}), while all other fields accept null values when data are not stated in the article. The \texttt{outbreak\_country} field enforces WHO standard country names, and the \texttt{outbreak\_location} field prohibits commas, requiring semicolon separators for multiple locations to avoid parsing ambiguities. Validation logic rejects outputs violating vocabulary constraints or data type rules, prompting the model to correct errors before proceeding.

The complete extraction workflow is coordinated by the \texttt{OutbreakExtractionRunner} class, which loads full-text data, applies screening decisions, manages iterative tool calls with validation feedback, and logs all outputs to structured JSONL files for downstream analysis.

\begin{tcolorbox}[
    enhanced,
    breakable,
    colback=white,
    colframe=blue!30,
    boxrule=0.8pt,
    arc=2pt,
    left=0pt, right=0pt, top=0pt, bottom=0pt,
    title={\small\textbf{Outbreak Screening Prompt}},
    fonttitle=\sffamily,
    coltitle=blue!70!black,
    colbacktitle=blue!15,
    attach boxed title to top left={yshift=-2mm, xshift=4mm},
    boxed title style={boxrule=0.5pt, arc=1pt}
]

\begin{tcolorbox}[
    enhanced,
    colback=gray!5,
    colframe=gray!20,
    boxrule=0pt,
    leftrule=3pt,
    arc=0pt,
    left=6pt, right=6pt, top=4pt, bottom=4pt
]

\vspace{2pt}
{\small{You are an epidemiologist conducting systematic review of infectious disease outbreaks. Determine if this article reports concluded, real-world outbreak events with defined case counts.}}
\end{tcolorbox}

\begin{tcolorbox}[
    enhanced,
    colback=gray!5,
    colframe=gray!20,
    boxrule=0pt,
    leftrule=3pt,
    arc=0pt,
    left=6pt, right=6pt, top=4pt, bottom=4pt
]
{\small\textbf{\textcolor{gray!60!black}{Screening Task Definition}}}

\vspace{2pt}
{\small
\textbf{Include (respond ``True''):}
\vspace{-4pt}
\begin{itemize}[leftmargin=*, itemsep=-3pt]
\item Discrete outbreak events with ALL of: specific location, defined time period, and case counts
\item Outbreak investigations describing a bounded epidemic event
\item Case series (2 or more cases) from a specific outbreak
\end{itemize}

\vspace{4pt}
\textbf{Exclude (respond ``False''):}
\vspace{-4pt}
\begin{itemize}[leftmargin=*, itemsep=-3pt]
\item Ongoing outbreaks at time of publication
\item Modeled, simulated, or forecasted outbreaks
\item Routine surveillance or annual disease burden (e.g., ``X cases per year'')
\item Seroprevalence or risk factor studies without outbreak context
\item Single case reports
\end{itemize}

\vspace{4pt}
\textbf{Key Question:} Can you identify a specific outbreak event (not general disease occurrence) with a start/end period and case count?

\vspace{4pt}
Respond with only ``True'' or ``False''.
}
\end{tcolorbox}

\begin{tcolorbox}[
    enhanced,
    colback=gray!5,
    colframe=gray!20,
    boxrule=0pt,
    leftrule=3pt,
    arc=0pt,
    left=6pt, right=6pt, top=4pt, bottom=4pt
]
{\small\textbf{\textcolor{gray!70!black}{Full Text}}}

\vspace{2pt}
{\small\ttfamily
Title: \{\{title\}\}\\[2pt]
Full Text: \{\{fulltext\}\}
}
\end{tcolorbox}

\end{tcolorbox}



\begin{tcolorbox}[
    enhanced,
    colback=blue!5,
    colframe=blue!20,
    boxrule=0pt,
    leftrule=3pt,
    arc=0pt,
    left=6pt, right=6pt, top=4pt, bottom=4pt,
    breakable
]
{\small\textbf{\textcolor{blue!60!black}{Extraction Task Definition}}}

\vspace{2pt}
{\small
\textbf{Outbreak extraction task}

Extract concluded outbreaks with defined case counts from the article. Call \texttt{extract\_outbreak\_data} once for each distinct outbreak (different location, time, or both).

\vspace{6pt}
\textbf{Important Notes:}

We are extracting everything as presented in the paper, even if you think it is an error by the author(s).

Extract data EXACTLY as stated in the paper. Do NOT perform calculations, convert units, or infer missing values.

DO NOT use commas in any field. If you need to separate items within a field, please use a semicolon.

\vspace{6pt}
\textbf{Tool Calling Rules:}
\vspace{-4pt}
\begin{itemize}[leftmargin=*, itemsep=-3pt]
\item Call \texttt{extract\_outbreak\_data} once per distinct outbreak
\item Outbreaks are distinct if they differ by location, time, or both
\item After extracting all outbreaks, stop calling the tool (no completion call needed)
\end{itemize}

\vspace{4pt}
\textbf{Schema Requirements:} Only three fields are required:
\vspace{-4pt}
\begin{itemize}[leftmargin=*, itemsep=-3pt]
\item \texttt{outbreak\_is\_currently\_ongoing}: true or false
\item \texttt{outbreak\_country}: Must be valid WHO country name
\item \texttt{asymptomatic\_transmission\_described}: true or false
\end{itemize}

All other fields: Use null when not stated in the paper.

\vspace{4pt}
\textbf{Extraction Rules:}
\vspace{-4pt}
\begin{itemize}[leftmargin=*, itemsep=-3pt]
\item Only select values that appear in the allowed lists for categorical fields
\item Extract dates as separate components (day, month, year)
\item Do NOT calculate \texttt{outbreak\_duration\_months}; only extract if explicitly stated
\item If you receive a validation error message, correct the tool call and try again
\end{itemize}

\vspace{4pt}
\textbf{Field-Specific Guidance:}

\textbf{Location:}
\begin{itemize}[leftmargin=*, itemsep=-3pt]
\item \texttt{outbreak\_country}: MUST match WHO standard names exactly (e.g., ``United States of America'' not ``USA'', ``Viet Nam'' not ``Vietnam'')
\item \texttt{outbreak\_location}: Extract as written; use semicolons not commas (e.g., ``Lagos; Abuja'')
\end{itemize}

\textbf{Case Counts:} Extract all categories as reported
\vspace{-4pt}
\begin{itemize}[leftmargin=*, itemsep=-3pt]
\item \texttt{cases\_confirmed}: Laboratory-confirmed cases
\item \texttt{cases\_probable}: Probable cases (clinical diagnosis)
\item \texttt{cases\_suspected}: Suspected cases under investigation
\item \texttt{cases\_unspecified}: Cases without clear classification
\item \texttt{cases\_asymptomatic}: Asymptomatic cases identified
\item \texttt{cases\_severe}: Severe cases OR hospitalizations (note if hospitalizations in notes)
\item \texttt{deaths}: Reported deaths
\end{itemize}

\textbf{Mode of Detection:} Select ONE
\begin{itemize}[leftmargin=*, itemsep=-3pt]
\item ``Molecular (PCR etc)'' Laboratory confirmation (PCR, ELISA, culture, etc.)
\item ``Symptoms'': Clinical/syndromic diagnosis only
\item ``Confirmed + Suspected'': Both lab-confirmed and clinical cases
\item ``Unspecified'': Not clearly stated
\end{itemize}

\textbf{Sex Disaggregation:} When provided, extract:
\begin{itemize}[leftmargin=*, itemsep=-3pt]
\item \texttt{male\_cases} / \texttt{female\_cases}: Counts
\item \texttt{prop\_male\_cases} / \texttt{prop\_female\_cases}: Proportion/percentage as reported
\item \texttt{type\_cases\_sex\_disagg}: Which case type is disaggregated (Confirmed/Suspected/Other/Unspecified)
\end{itemize}

\textbf{Pre-Outbreak Baseline:}
\vspace{-4pt}
\begin{itemize}[leftmargin=*, itemsep=-3pt]
\item ``Disease-free baseline'': No previous cases
\item ``Endemic equilibrium'': Disease was endemic
\item ``Probable'': Suggested but not definitive
\item ``Unspecified'': Not discussed
\end{itemize}

\textbf{Dates:} Provide as separate components (day, month, year). Partial dates are acceptable (e.g., only month and year).

\textbf{Duration:} ONLY extract if paper explicitly states duration. Do NOT calculate from dates.

\textbf{Notes:} Use this field for important context, data quality issues, or special circumstances.

\vspace{4pt}
\textbf{Pathogen-Specific Rules:}
\vspace{-4pt}
\begin{itemize}[leftmargin=*, itemsep=-3pt]
\item Zika, RVF: Only extract outbreaks with 10 or more cases
\item Marburg, Lassa, Nipah: Extract all outbreaks
\item OROV: Include even single case reports
\end{itemize}
}
\end{tcolorbox}

\clearpage

\begin{tcolorbox}[
    enhanced,
    breakable,
    colback=white,
    colframe=blue!30,
    boxrule=0.8pt,
    arc=2pt,
    left=0pt, right=0pt, top=0pt, bottom=0pt,
    title={\small\textbf{Outbreak Extraction Prompt}},
    fonttitle=\sffamily,
    coltitle=blue!70!black,
    colbacktitle=blue!15,
    attach boxed title to top left={yshift=-2mm, xshift=4mm},
    boxed title style={boxrule=0.5pt, arc=1pt}
]

\begin{tcolorbox}[
    enhanced,
    colback=gray!5,
    colframe=gray!20,
    boxrule=0pt,
    leftrule=3pt,
    arc=0pt,
    left=6pt, right=6pt, top=4pt, bottom=4pt
]

\vspace{2pt}
{\small{You are an epidemiologist conducting systematic review of infectious disease outbreaks. Extract structured data about concluded outbreak events from scientific articles.}}
\end{tcolorbox}

\begin{tcolorbox}[
    enhanced,
    colback=gray!5,
    colframe=gray!20,
    boxrule=0pt,
    leftrule=3pt,
    arc=0pt,
    left=6pt, right=6pt, top=4pt, bottom=4pt
]
{\small\textbf{\textcolor{gray!60!black}{Study Objectives}}}

{\small

This systematic review collates transmission models, outbreaks and parameters for \{\{pathogen\}\}.
}
\end{tcolorbox}

\begin{tcolorbox}[
    enhanced,
    colback=blue!5,
    colframe=blue!20,
    boxrule=0pt,
    leftrule=3pt,
    arc=0pt,
    left=6pt, right=6pt, top=4pt, bottom=4pt,
    breakable
]
{\small\textbf{\textcolor{blue!60!black}{Extraction Task Definition}}
\textit{See Extraction Task Definition details above.}
}
\end{tcolorbox}

\begin{tcolorbox}[
    enhanced,
    colback=gray!5,
    colframe=gray!20,
    boxrule=0pt,
    leftrule=3pt,
    arc=0pt,
    left=6pt, right=6pt, top=4pt, bottom=4pt
]

{\small\textbf{\textcolor{gray!70!black}{Full Text}}}

\vspace{2pt}
{\small\ttfamily
Title: \{\{title\}\}\\[2pt]
Full Text: \{\{fulltext\}\}
}
\end{tcolorbox}

\end{tcolorbox}

The language model uses the \texttt{extract\_outbreak\_data()} tool (provided to it) to populate the schema defined in Table~\ref{tab:outbreak_extraction_schema}. The tool enforces strict JSON validation with controlled vocabularies for categorical fields, rejecting invalid outputs and prompting corrections. The complete tool specification follows standard OpenAI function calling conventions with enum constraints for single-select fields and null acceptance for optional fields.

\paragraph{Provenance extraction} 
Following successful extraction of parameters, models, and outbreaks, a provenance stage systematically mapped each extracted value to supporting textual excerpts from the article, ensuring complete traceability and grounding of all characteristics in source material. For each extracted record (parameter estimate, model descriptor, or outbreak summary), the provenance extraction invoked a dedicated tool (\texttt{extract\_parameter\_provenance}, \texttt{extract\_model\_provenance}, or \texttt{extract\_outbreak\_provenance}) that received the complete set of previously extracted characteristics and identified verbatim quotes, equation references, or table citations justifying each value selection. For multi-select fields (e.g. transmission routes, assumptions, interventions in models; multiple locations in outbreaks), each selected option required independent textual support. This additional stage enabled potential validation of extraction quality, provided transparency for subsequent data synthesis, and formed an audit trail linking structured outputs to primary literature, with all provenance traces logged to structured files for downstream analysis.

\clearpage
\clearpage
\section{Report Generation: Building Systematic Living Reviews}
Report generation is not evaluated within \name{}. The dataset shared by PERG and the open-access subset retrieved by \name{} overlap but are not identical, which precludes the construction of data-matched reference reports necessary for a fair evaluation. Consultation with epidemiologists at PERG further clarified that screening and data extraction constitute the primary time bottleneck in their workflow, and that research questions, output artefacts, and meta-analysis variables are typically determined once an evidence base has already been assembled. Additionally, the risks associated with incorrect public health artefacts are substantial: In our testing, LLM-generated summaries produced in unconstrained settings tend towards over-interpretation. Report generation along with meta-analysis are future work, beyond the scope of our study.

\label{app:report_generation_methods}

\subsection{Deterministic Report Assembly}
Given a pathogen $p$, we generate human-readable reports directly from the extracted, structured datasets for outbreaks.
The report build aggregates extraction records into descriptive summary tables and figures, then compiles a Markdown draft, and finally renders a PDF. The report build is lightweight relative to retrieval, screening, and extraction and is omitted from our main runtime breakdown (typically $<5$ minutes per pathogen).

\subsubsection*{Inputs and derived artefacts}
\paragraph{Inputs}
Let $\mathcal{D}^{\textsc{O}}_p$ be the set of extracted outbreak records for pathogen $p$ (one row per outbreak entity).
Each record is schema-validated at extraction time (Appendix~\ref{app:data_extraction_details}), so report generation treats the datasets as structured inputs.

\paragraph{Content manifest.}
The manifest stores: pathogen identifier, timestamp, summary statistics (e.g. outbreak counts and geographic coverage), the list of narrative sections, and structured metadata for each figure and table (number, title, caption, path, and row or observation counts). This manifest is later used as part of the evidence packet in the LLM refinement stage (next subsection).

\begin{table}[h]
\centering
\caption{\textbf{Artefact inventory for outbreak report generation (per pathogen $p$).} All artefacts are derived from the extracted outbreak dataset $\mathcal{D}^{\textsc{O}}_p$.}
\label{tab:artifact_inventory_outbreaks}
\footnotesize
\begin{tabular}{p{0.27\linewidth}p{0.38\linewidth}p{0.28\linewidth}}
\toprule
\textbf{Artefact} & \textbf{Path (relative to repo root)} & \textbf{Purpose} \\
\midrule
Outbreak report (Markdown) &
\texttt{writeup/}$p$\texttt{/outbreaks\_writeup.md} &
Human-readable draft with embedded figures and tables. \\
Outbreak report (PDF) &
\texttt{writeup/}$p$\texttt{/outbreaks\_writeup.pdf} &
Portable rendering for sharing and archiving. \\
Figures directory &
\texttt{writeup/}$p$\texttt{/figures/} &
Generated plots referenced by Markdown (e.g., temporal distribution, geographic spread, case counts). \\
Summary tables (embedded) &
(in \texttt{outbreaks\_writeup.md}) &
Count and proportion tables computed from $\mathcal{D}^{\textsc{O}}_p$. \\
Content manifest &
\texttt{writeup/}$p$\texttt{/content\_manifest.json} &
Machine-readable inventory of figures, tables, and dataset statistics. \\
\bottomrule
\end{tabular}
\end{table}

\subsubsection*{Evidence packet construction}
\paragraph{Evidence packet}
For pathogen $p$ and outbreak report type $\textsc{O}$, code constructs an evidence packet
\[
E^{\textsc{O}}_p \;=\; \big(\;\textsc{Stats}^{\textsc{O}}_p,\;\textsc{Figs}^{\textsc{O}}_p,\;\textsc{Tables}^{\textsc{O}}_p,\;W^{\textsc{O},(0)}_p\;\big),
\]
where $\textsc{Stats}^{\textsc{O}}_p$ is a concise text summary of dataset counts and geographic breakdowns, $\textsc{Figs}^{\textsc{O}}_p$ is the required figure list (paths and captions), $\textsc{Tables}^{\textsc{O}}_p$ is the set of tables to be included (as Markdown blocks), and $W^{\textsc{O},(0)}_p$ is the programmatic Markdown draft.
The model is instructed to rely only on $E^{\textsc{O}}_p$ and not to introduce external facts.

\subsection{Evidence grounded narrative refinement}
Report writing proceeds by an LLM revision stage that refines $W^{\textsc{O},(0)}_p$ into a narrative synthesis, while enforcing evidence grounding and artefact presence.

\subsubsection*{Self-refinement loop and non-negotiable checks}
\paragraph{Grounding and asset checks (non-negotiable).}
Two constraints are enforced for every refined version:
\begin{enumerate}[leftmargin=*, itemsep=-2pt]
    \item \textbf{Asset presence:} every required figure path from the manifest must appear at least once as a Markdown image line.
    \item \textbf{Table preservation:} every table provided in the evidence packet must be present, with values unchanged (reformatting is allowed).
\end{enumerate}
If either constraint is violated, we deterministically append missing figures or tables verbatim at the end of the Markdown so the final PDF always renders with the full artefact set.

\paragraph{Minimal formalisation}
Let $W^{(0)}$ denote the initial (programmatic) Markdown draft.
Each iteration applies:
\[
\text{critique}(W^{(k-1)}) \rightarrow C^{(k)}, \qquad \text{revise}(W^{(k-1)}, C^{(k)}) \rightarrow W^{(k)}.
\]
This is only a notation convenience: in practice the evidence packet always accompanies both steps, and the critique output is structured JSON used to drive the next revision.

\subsubsection*{Rubric and prompts}
\paragraph{Rubric}
We use an 8-dimension rubric, each scored from 1 (poor) to 5 (excellent). The dimensions are the same for both report types, except for the scope constraint.

\paragraph{Shared dimensions}
\vspace{-4pt}
\begin{enumerate}[leftmargin=*, itemsep=-2pt]
    \item \texttt{data\_fidelity}: descriptive claims match the evidence packet; no invented statistics or outbreak characteristics.
    \item \texttt{figure\_table\_presence}: all required figures and tables appear.
    \item \texttt{traceability}: outside interpretation blocks, claims cite their source as (Figure X), (Table Y), or (Dataset Statistics).
    \item \texttt{clarity}: consistent terminology, clear writing, minimal ambiguity.
    \item \texttt{completeness}: covers the major patterns visible in the available figures and tables.
    \item \texttt{interpretation\_blocks}: interpretation is confined to dedicated blocks and labelled as such.
    \item \texttt{formatting}: valid Markdown and sensible figure layout hints.
\end{enumerate}

\paragraph{Interpretation policy}
Interpretation is allowed only inside blockquotes beginning with \texttt{> AI-Interpretation:}.
Outside those blocks, the narrative must remain descriptive and evidence-linked; no new numbers may be introduced.

\subsection{Report Generation Prompts}
\label{app:report_generation_prompts_outbreaks}

We present the exact prompts used for outbreak report generation and self-refinement, formatted consistently with the model report prompts.
All prompts are instantiated programmatically by filling placeholders (e.g. \texttt{\{EVIDENCE\_PACKET\}}) at runtime.


\subsubsection*{Outbreak report prompts}

\begin{tcolorbox}[
    enhanced,
    breakable,
    colback=white,
    colframe=blue!30,
    boxrule=0.8pt,
    arc=2pt,
    left=0pt, right=0pt, top=0pt, bottom=0pt,
    title={\small\textbf{Outbreak Report: Initial Synthesis Prompt}},
    fonttitle=\sffamily,
    coltitle=blue!70!black,
    colbacktitle=blue!15,
    attach boxed title to top left={yshift=-2mm, xshift=4mm},
    boxed title style={boxrule=0.5pt, arc=1pt}
]

\begin{tcolorbox}[
    enhanced,
    colback=gray!5,
    colframe=gray!20,
    boxrule=0pt,
    leftrule=3pt,
    arc=0pt,
    left=6pt, right=6pt, top=4pt, bottom=4pt
]

\vspace{2pt}
{\small{You are a senior epidemiologist editing a living outbreak surveillance review. You are revising a first draft prepared by a research assistant who summarized extracted outbreak records.}}
\end{tcolorbox}

\begin{tcolorbox}[
    enhanced,
    colback=gray!5,
    colframe=gray!20,
    boxrule=0pt,
    leftrule=3pt,
    arc=0pt,
    left=6pt, right=6pt, top=4pt, bottom=4pt
]
{\small\textbf{\textcolor{gray!60!black}{Method Basis}}}

\vspace{2pt}
{\small
Do not cite external sources; just follow these behaviors:
\begin{itemize}[leftmargin=*, itemsep=-3pt]
    \item Iterative critique→refine loop (Self-Refine).
    \item Rubric-based form-filling evaluation mindset (G-Eval).
    \item Attribution-first revision: every descriptive claim must be attributable to the provided evidence packet (RARR-style editing for attribution).
    \item Living review principles: explicitly describe what is present in the dataset snapshot and what is missing; avoid academic formatting.
\end{itemize}
}
\end{tcolorbox}

\begin{tcolorbox}[
    enhanced,
    colback=gray!5,
    colframe=gray!20,
    boxrule=0pt,
    leftrule=3pt,
    arc=0pt,
    left=6pt, right=6pt, top=4pt, bottom=4pt
]
{\small\textbf{\textcolor{gray!60!black}{Hard Scope Constraint}}}

\vspace{2pt}
{\small
Focus on documented outbreak events and outbreak characteristics. Do not broaden into transmission modelling, pathogen biology, or clinical management beyond what is supported by the outbreak dataset.
}
\end{tcolorbox}

\begin{tcolorbox}[
    enhanced,
    colback=gray!5,
    colframe=gray!20,
    boxrule=0pt,
    leftrule=3pt,
    arc=0pt,
    left=6pt, right=6pt, top=4pt, bottom=4pt
]
{\small\textbf{\textcolor{gray!60!black}{Truthfulness Constraints}}}

\vspace{2pt}
{\small
\begin{itemize}[leftmargin=*, itemsep=-3pt]
    \item Do not invent outbreak characteristics, case counts, geographic locations, or external facts.
    \item Outside of AI-Interpretation blocks, every numeric or categorical claim must be directly supported by the evidence packet and must cite its support as (Figure X), (Table Y), or (Dataset Statistics).
    \item Interpretation is allowed ONLY inside blockquotes starting with: \texttt{> AI-Interpretation:}
    \item Inside AI-Interpretation blocks, you may propose plausible implications for outbreak surveillance and preparedness, but you must label them as hypotheses and you must not introduce new numbers that are not in the evidence packet.
\end{itemize}
}
\end{tcolorbox}

\begin{tcolorbox}[
    enhanced,
    colback=gray!5,
    colframe=gray!20,
    boxrule=0pt,
    leftrule=3pt,
    arc=0pt,
    left=6pt, right=6pt, top=4pt, bottom=4pt
]
{\small\textbf{\textcolor{gray!60!black}{Figures and Tables Constraints}}}

\vspace{2pt}
{\small
\begin{itemize}[leftmargin=*, itemsep=-3pt]
    \item All figures must appear as markdown images using their existing paths (e.g., \texttt{![Alt](figures/fig1\_...png)}). Placement is free.
    \item Tables must all be present. You may reformat tables, but values must remain identical.
\end{itemize}
}
\end{tcolorbox}

\begin{tcolorbox}[
    enhanced,
    colback=gray!5,
    colframe=gray!20,
    boxrule=0pt,
    leftrule=3pt,
    arc=0pt,
    left=6pt, right=6pt, top=4pt, bottom=4pt
]
{\small\textbf{\textcolor{gray!60!black}{Formatting Agency}}}

\vspace{2pt}
{\small
\begin{itemize}[leftmargin=*, itemsep=-3pt]
    \item You may include an OPTIONAL HTML comment immediately after any figure image line to suggest sizing for PDF rendering.
    \item Format: \texttt{<!-- fig-layout: width\_in=5.5 max\_height\_in=7.5 -->}
    \item If absent, defaults will be used.
\end{itemize}
}
\end{tcolorbox}

\begin{tcolorbox}[
    enhanced,
    colback=gray!5,
    colframe=gray!20,
    boxrule=0pt,
    leftrule=3pt,
    arc=0pt,
    left=6pt, right=6pt, top=4pt, bottom=4pt
]
{\small\textbf{\textcolor{gray!60!black}{Output Requirements}}}

\vspace{2pt}
{\small
\begin{itemize}[leftmargin=*, itemsep=-3pt]
    \item Produce a living outbreak surveillance review in Markdown.
    \item Use descriptive, report-like sections rather than academic paper structure.
    \item For each main section, include: (1) Evidence-based description, then (2) one AI-Interpretation blockquote.
\end{itemize}
}
\end{tcolorbox}

\begin{tcolorbox}[
    enhanced,
    colback=gray!5,
    colframe=gray!20,
    boxrule=0pt,
    leftrule=3pt,
    arc=0pt,
    left=6pt, right=6pt, top=4pt, bottom=4pt
]
{\small\textbf{\textcolor{gray!60!black}{Task Definition}}}

\vspace{2pt}
{\small
Task: Produce Version 1 of the living outbreak surveillance review.
Use the evidence packet below. Maintain honesty and verifiability.

\vspace{1em}

\textbf{Required structure} (you may adapt headings, but keep these concepts):
\begin{enumerate}[label=\arabic*), leftmargin=*, itemsep=-3pt]
    \item Snapshot (dataset size, temporal coverage, geographic scope, what this review represents)
    \item Outbreak temporal distribution (outbreak frequency over time, identification of major epidemic periods)
    \item Geographic distribution and spread patterns (countries affected, spatial clustering, cross-border transmission)
    \item Outbreak size and severity (case counts, fatality rates, outbreak durations)
    \item Detection and reporting patterns (modes of detection, case definitions used, reporting delays if mentioned)
    \item Demographic patterns (sex disaggregation, age patterns if available)
    \item Data quality and gaps (completeness of reporting, missing information, asymptomatic transmission documentation)
    \item Evidence-based recommendations (only tied to observed gaps in outbreak surveillance)
    \item Change log stub (for future updates)
\end{enumerate}
}
\end{tcolorbox}

\begin{tcolorbox}[
    enhanced,
    colback=gray!5,
    colframe=gray!20,
    boxrule=0pt,
    leftrule=3pt,
    arc=0pt,
    left=6pt, right=6pt, top=4pt, bottom=4pt
]
{\small\textbf{\textcolor{gray!70!black}{Evidence Packet}}}

\vspace{2pt}
{\small\ttfamily
\{EVIDENCE\_PACKET\}
}
\end{tcolorbox}

\end{tcolorbox}

\newpage
\begin{tcolorbox}[
    enhanced,
    breakable,
    colback=white,
    colframe=blue!30,
    boxrule=0.8pt,
    arc=2pt,
    left=0pt, right=0pt, top=0pt, bottom=0pt,
    title={\small\textbf{Outbreak Report: Critique Prompt}},
    fonttitle=\sffamily,
    coltitle=blue!70!black,
    colbacktitle=blue!15,
    attach boxed title to top left={yshift=-2mm, xshift=4mm},
    boxed title style={boxrule=0.5pt, arc=1pt}
]

\begin{tcolorbox}[
    enhanced,
    colback=gray!5,
    colframe=gray!20,
    boxrule=0pt,
    leftrule=3pt,
    arc=0pt,
    left=6pt, right=6pt, top=4pt, bottom=4pt
]

\vspace{2pt}
{\small{You are a meticulous scientific editor. Return only valid JSON.}}
\end{tcolorbox}

\begin{tcolorbox}[
    enhanced,
    colback=gray!5,
    colframe=gray!20,
    boxrule=0pt,
    leftrule=3pt,
    arc=0pt,
    left=6pt, right=6pt, top=4pt, bottom=4pt
]
{\small\textbf{\textcolor{gray!60!black}{Critique Task Definition}}}

\vspace{2pt}
{\small
You are a scientific editor evaluating a living outbreak surveillance review for faithfulness to the provided evidence packet.
Return STRICT JSON only.
}
\end{tcolorbox}

\begin{tcolorbox}[
    enhanced,
    colback=gray!5,
    colframe=gray!20,
    boxrule=0pt,
    leftrule=3pt,
    arc=0pt,
    left=6pt, right=6pt, top=4pt, bottom=4pt
]
{\small\textbf{\textcolor{gray!60!black}{Evidence Packet Summary}}}

\vspace{2pt}
{\small\ttfamily
\{DATASET\_STATISTICS\}
}
\end{tcolorbox}

\begin{tcolorbox}[
    enhanced,
    colback=gray!5,
    colframe=gray!20,
    boxrule=0pt,
    leftrule=3pt,
    arc=0pt,
    left=6pt, right=6pt, top=4pt, bottom=4pt
]
{\small\textbf{\textcolor{gray!70!black}{Required Figure Paths}}}

\vspace{2pt}
{\small\ttfamily
All of the following must appear at least once:\\[6pt]
\{REQUIRED\_FIGURE\_PATHS\}
}
\end{tcolorbox}

\begin{tcolorbox}[
    enhanced,
    colback=gray!5,
    colframe=gray!20,
    boxrule=0pt,
    leftrule=3pt,
    arc=0pt,
    left=6pt, right=6pt, top=4pt, bottom=4pt
]
{\small\textbf{\textcolor{gray!70!black}{Report to Critique}}}

\vspace{2pt}
{\small\ttfamily
\{CURRENT\_REPORT\}
}
\end{tcolorbox}

\begin{tcolorbox}[
    enhanced,
    colback=gray!5,
    colframe=gray!20,
    boxrule=0pt,
    leftrule=3pt,
    arc=0pt,
    left=6pt, right=6pt, top=4pt, bottom=4pt
]
{\small\textbf{\textcolor{gray!60!black}{Evaluation Dimensions}}}

\vspace{2pt}
{\small
Evaluate dimensions (score 1-5). Provide issues and concrete suggestions.

\vspace{1em}

\textbf{Dimensions:}
\begin{enumerate}[label=\arabic*), leftmargin=*, itemsep=-3pt]
    \item \texttt{data\_fidelity}: descriptive claims supported by evidence packet; no invented outbreak characteristics, case counts, or geographic information.
    \item \texttt{outbreak\_focus}: stays centered on documented outbreak events and outbreak surveillance rather than transmission modelling or pathogen biology.
    \item \texttt{figure\_table\_presence}: all required figures present; all tables present.
    \item \texttt{traceability}: outside AI-Interpretation blocks, claims cite support as (Figure X)/(Table Y)/(Dataset Statistics).
    \item \texttt{clarity}: readable, precise, minimal ambiguity, consistent terminology for outbreak characteristics and surveillance metrics.
    \item \texttt{completeness}: covers major patterns in outbreak temporal distribution, geographic spread, and detection practices described by available figures/tables.
    \item \texttt{interpretation\_blocks}: each main section includes a blockquote starting with \texttt{> AI-Interpretation:} and interpretation stays inside it.
    \item \texttt{formatting}: figure layout directives used sensibly where needed; no broken markdown.
\end{enumerate}
}
\end{tcolorbox}

\begin{tcolorbox}[
    enhanced,
    colback=gray!5,
    colframe=gray!20,
    boxrule=0pt,
    leftrule=3pt,
    arc=0pt,
    left=6pt, right=6pt, top=4pt, bottom=4pt
]
{\small\textbf{\textcolor{gray!70!black}{JSON Response Format}}}

\vspace{2pt}
{\small\ttfamily
Return JSON of the form:\\[6pt]
\{\\
\ \ "dimensions": \{\\
\ \ \ \ "data\_fidelity": \{"score": 1-5, "issues": [...], "suggestions": [...]\},\\
\ \ \ \ "outbreak\_focus": \{"score": 1-5, "issues": [...], "suggestions": [...]\},\\
\ \ \ \ "figure\_table\_presence": \{"score": 1-5, "issues": [...], "suggestions": [...]\},\\
\ \ \ \ "traceability": \{"score": 1-5, "issues": [...], "suggestions": [...]\},\\
\ \ \ \ "clarity": \{"score": 1-5, "issues": [...], "suggestions": [...]\},\\
\ \ \ \ "completeness": \{"score": 1-5, "issues": [...], "suggestions": [...]\},\\
\ \ \ \ "interpretation\_blocks": \{"score": 1-5, "issues": [...], "suggestions": [...]\},\\
\ \ \ \ "formatting": \{"score": 1-5, "issues": [...], "suggestions": [...]\}\\
\ \ \},\\
\ \ "priority\_fixes": [...]\\
\}
}
\end{tcolorbox}

\end{tcolorbox}

\begin{tcolorbox}[
    enhanced,
    breakable,
    colback=white,
    colframe=blue!30,
    boxrule=0.8pt,
    arc=2pt,
    left=0pt, right=0pt, top=0pt, bottom=0pt,
    title={\small\textbf{Outbreak Report: Revision Prompt}},
    fonttitle=\sffamily,
    coltitle=blue!70!black,
    colbacktitle=blue!15,
    attach boxed title to top left={yshift=-2mm, xshift=4mm},
    boxed title style={boxrule=0.5pt, arc=1pt}
]

\begin{tcolorbox}[
    enhanced,
    colback=gray!5,
    colframe=gray!20,
    boxrule=0pt,
    leftrule=3pt,
    arc=0pt,
    left=6pt, right=6pt, top=4pt, bottom=4pt
]

\vspace{2pt}
{\small{You are a senior epidemiologist performing an evidence-grounded revision.}}
\end{tcolorbox}

\begin{tcolorbox}[
    enhanced,
    colback=gray!5,
    colframe=gray!20,
    boxrule=0pt,
    leftrule=3pt,
    arc=0pt,
    left=6pt, right=6pt, top=4pt, bottom=4pt
]
{\small\textbf{\textcolor{gray!60!black}{Revision Constraints}}}

\vspace{2pt}
{\small
\begin{itemize}[leftmargin=*, itemsep=-3pt]
    \item Follow an attribution-first editing approach: outside AI-Interpretation blocks, every claim must be supported by the evidence packet.
    \item Keep the document outbreak-focused.
    \item All figures must appear at least once as markdown images with their existing paths.
    \item All tables must be present; you may reformat, but values must not change.
    \item Interpretation is permitted only within blockquotes beginning with \texttt{> AI-Interpretation:}.
    \item You may add optional figure sizing directives as HTML comments immediately after image lines: \texttt{<!-- fig-layout: width\_in=5.5 max\_height\_in=7.5 -->}
\end{itemize}
}
\end{tcolorbox}

\begin{tcolorbox}[
    enhanced,
    colback=gray!5,
    colframe=gray!20,
    boxrule=0pt,
    leftrule=3pt,
    arc=0pt,
    left=6pt, right=6pt, top=4pt, bottom=4pt
]
{\small\textbf{\textcolor{gray!60!black}{Quality Scores}}}

\vspace{2pt}
{\small\ttfamily
\{DIMENSION\_SCORES\}
}
\end{tcolorbox}

\begin{tcolorbox}[
    enhanced,
    colback=gray!5,
    colframe=gray!20,
    boxrule=0pt,
    leftrule=3pt,
    arc=0pt,
    left=6pt, right=6pt, top=4pt, bottom=4pt
]
{\small\textbf{\textcolor{gray!70!black}{Priority Fixes}}}

\vspace{2pt}
{\small\ttfamily
\{PRIORITY\_FIXES\}
}
\end{tcolorbox}

\begin{tcolorbox}[
    enhanced,
    colback=gray!5,
    colframe=gray!20,
    boxrule=0pt,
    leftrule=3pt,
    arc=0pt,
    left=6pt, right=6pt, top=4pt, bottom=4pt
]
{\small\textbf{\textcolor{gray!70!black}{Evidence Packet}}}

\vspace{2pt}
{\small\ttfamily
\{EVIDENCE\_PACKET\}
}
\end{tcolorbox}

\begin{tcolorbox}[
    enhanced,
    colback=gray!5,
    colframe=gray!20,
    boxrule=0pt,
    leftrule=3pt,
    arc=0pt,
    left=6pt, right=6pt, top=4pt, bottom=4pt
]
{\small\textbf{\textcolor{gray!70!black}{Current Report}}}

\vspace{2pt}
{\small\ttfamily
\{CURRENT\_REPORT\}
}
\end{tcolorbox}

\begin{tcolorbox}[
    enhanced,
    colback=gray!5,
    colframe=gray!20,
    boxrule=0pt,
    leftrule=3pt,
    arc=0pt,
    left=6pt, right=6pt, top=4pt, bottom=4pt
]
{\small\textbf{\textcolor{gray!60!black}{Revision Requirements}}}

\vspace{2pt}
{\small
\begin{itemize}[leftmargin=*, itemsep=-3pt]
    \item Fix all critique issues.
    \item Ensure each main section has (1) evidence-based description with citations (Figure/Table/Dataset Statistics), then (2) \texttt{> AI-Interpretation:} block.
    \item Remove or relabel any statement not supported by the evidence packet.
    \item Ensure outbreak-only framing (documented outbreak events and surveillance patterns).
    \item Keep document a living surveillance review (descriptive, update-ready), not an academic paper.
\end{itemize}

\vspace{1em}

Return the complete revised Markdown.
}
\end{tcolorbox}

\end{tcolorbox}

\clearpage
\section{Living Systematic Reviews with \name{}}\label{app:report_writeup}

Utilising \name{} harness workflow on the data extracted corpus from previous stages (with \texttt{gpt-oss-120b}), we generated living reviews for nine WHO priority pathogens: Marburg virus, Ebola virus, Lassa virus, SARS-CoV-1, Zika virus, MERS-CoV, Nipah virus, Rift Valley fever (RVF) virus, and Crimean Congo haemorrhagic fever (CCHF) virus. Each review comprises two complementary documents (a transmission-modelling review synthesising extracted model characteristics, and an outbreak surveillance review aggregating historical outbreak data) alongside structured datasets and visualisations. While four of these pathogens (Ebola, Lassa, SARS, Zika) have been validated against PERG's expert annotations, the remaining five represent preliminary syntheses for pathogens where PERG's systematic review process has not yet commenced or is in early stages.

\begin{figure}[h]
    \centering
    \begin{subfigure}[t]{0.6\textwidth}
        \centering
        \includegraphics[width=0.6\textwidth]{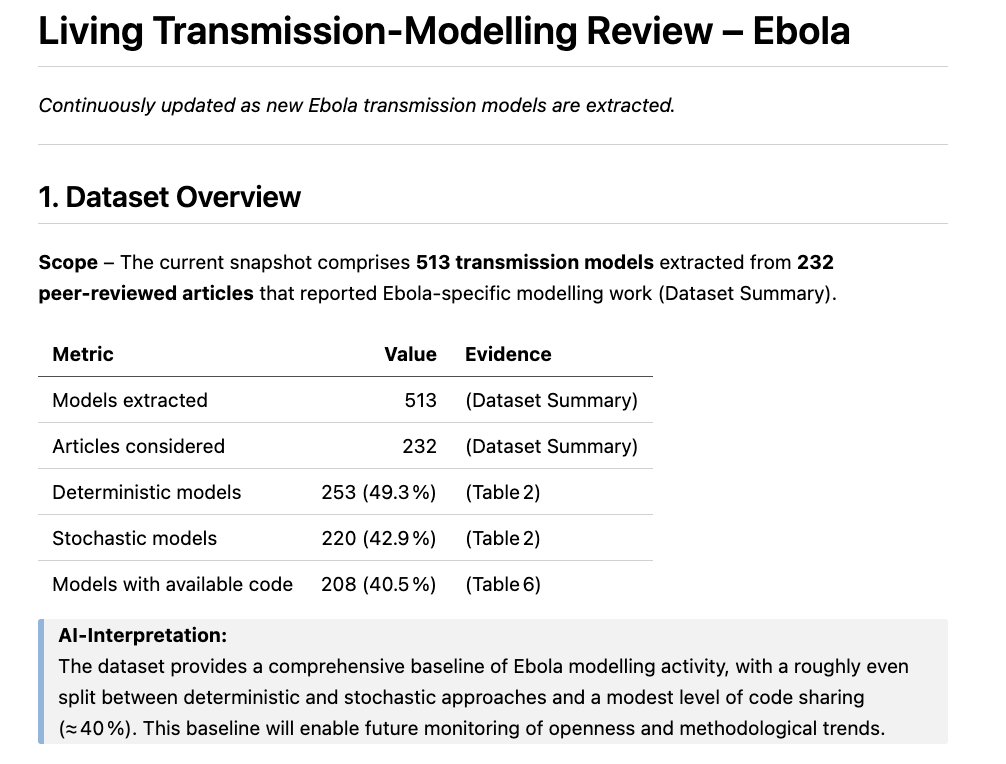}
        \caption{Transmission-Modelling Review excerpt showing dataset scope, model architecture distribution, and reproducibility indicators for $513$ Ebola models extracted from $232$ articles.}
        \label{fig:ebola_models}
    \end{subfigure}

    \vspace{0.5em} 

    \begin{subfigure}[t]{0.6\textwidth}
        \centering
        \includegraphics[width=0.6\textwidth]{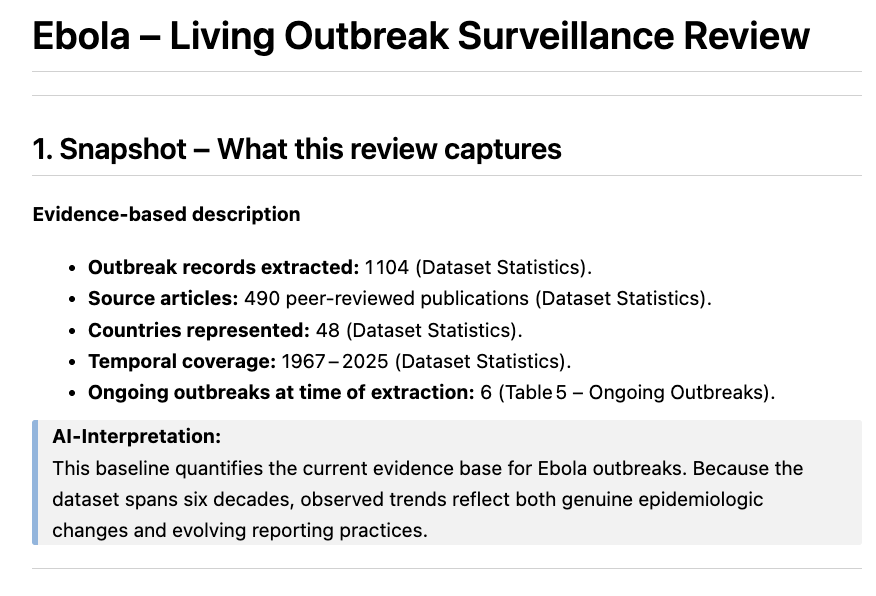}
        \caption{Outbreak Surveillance Review excerpt presenting snapshot statistics for $1,104$ outbreak records from $490$ publications, covering temporal span 1967--2025 and $48$ countries.}
        \label{fig:ebola_outbreaks}
    \end{subfigure}

    \caption{\textbf{Ebola living reviews generated by \name{}.} Both reviews follow a structured format: evidence-based descriptions citing supporting figures and tables, followed by interpretation blocks explicitly labelled as AI-generated synthesis.}
    \label{fig:ebola_reviews}
\end{figure}

Figure~\ref{fig:ebola_reviews} presents excerpts from the Ebola living reviews, illustrating the structure and content of \name{} workflow's outputs for a validated pathogen. The transmission-modelling review (Figure~\ref{fig:ebola_models}) provides a quantitative overview of the $513$ extracted models, including distributions across model architectures, stochasticity classifications, and code availability. The outbreak surveillance review (Figure~\ref{fig:ebola_outbreaks}) synthesises $1{,}104$ outbreak records spanning nearly six decades, with temporal coverage, geographic distribution, and detection methodology patterns presented through evidence-based descriptions paired with interpretive commentary blocks.

\begin{figure}[h]
    \centering
    \begin{subfigure}[t]{0.48\textwidth}
        \centering
        \includegraphics[width=\textwidth]{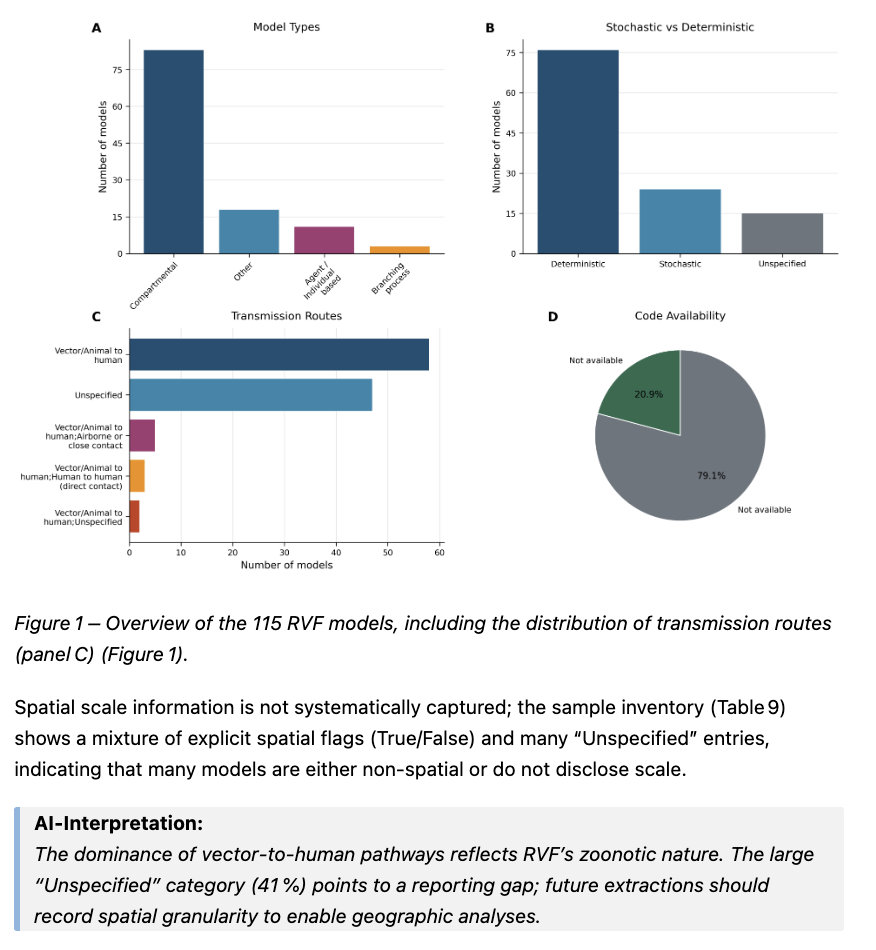}
        \caption{RVF transmission-modelling review showing distribution of $115$ extracted models across architecture types, stochasticity, transmission routes, and code availability.}
        \label{fig:rvf_models}
    \end{subfigure}
    \hfill
    \begin{subfigure}[t]{0.48\textwidth}
        \centering
        \includegraphics[width=\textwidth]{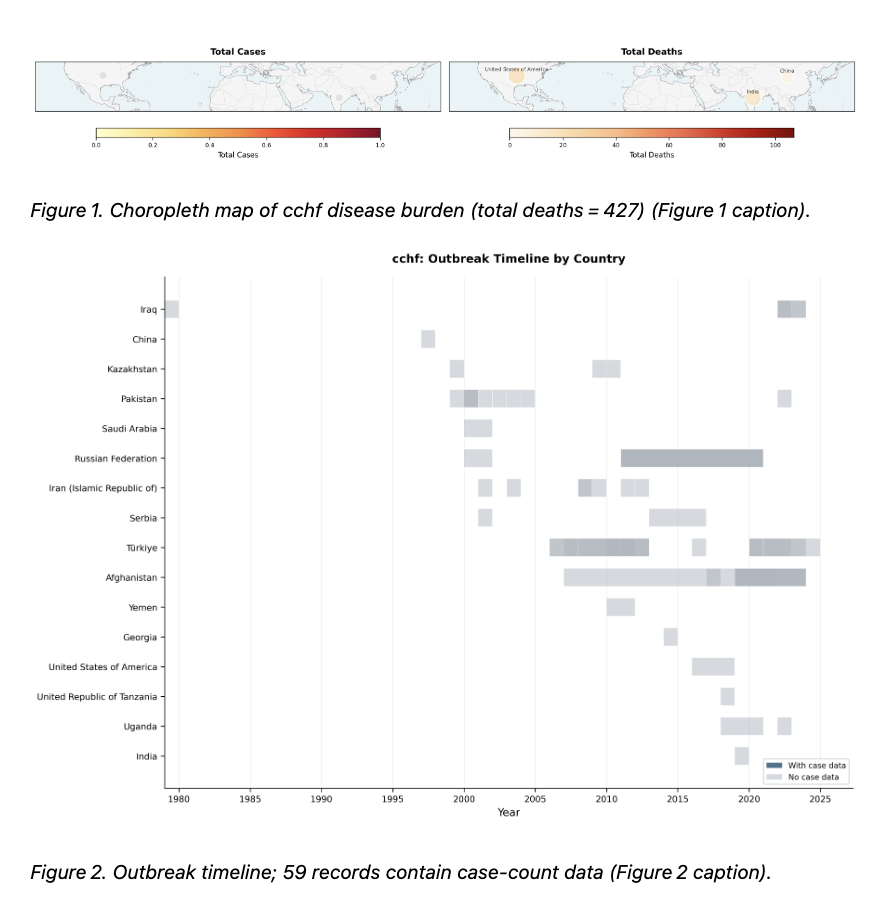}
        \caption{CCHF outbreak surveillance review presenting geographic burden (choropleth maps) and temporal distribution of $59$ outbreaks with case-count data.}
        \label{fig:cchf_outbreaks}
    \end{subfigure}
    \caption{\textbf{Preliminary living reviews for pathogens without completed PERG validation.} RVF and CCHF represent pathogens for which PERG has not yet commenced systematic screening.}
    \label{fig:unvalidated_reviews}
\end{figure}

For emerging or understudied pathogens, rapid synthesis of available evidence can inform outbreak preparedness even when comprehensive expert review remains infeasible. Figure~\ref{fig:unvalidated_reviews} presents excerpts from RVF and CCHF reviews, two pathogens for which PERG has not yet initiated systematic screening. The RVF transmission-modelling review (Figure~\ref{fig:rvf_models}) characterises $115$ models extracted from the retrieved literature, revealing a predominance of compartmental architectures and vector-to-human transmission pathways consistent with RVF's arboviral ecology. The CCHF outbreak surveillance review (Figure~\ref{fig:cchf_outbreaks}) maps $59$ outbreak records with quantitative case data, identifying geographic clusters and temporal patterns across affected regions. While these syntheses lack the validation rigour applied to Ebola, Lassa, SARS, and Zika, they demonstrate \name{} workflow's capacity to generate preliminary evidence summaries for resource allocation and hypothesis generation in under $48$ hours of wall-clock time. Once the evaluation scores are above a threshold, this same workflow can be utilised for assist in generate agile living reviews.

Table~\ref{tab:review_contents} summarises the standardised artefact structure maintained across all pathogen reviews. Both transmission-modelling and outbreak surveillance reports follow consistent schemas: model reviews characterise architecture distributions, stochasticity classifications, transmission pathways, and reproducibility indicators, whilst outbreak reviews present temporal timelines, geographic burden maps, detection methodology breakdowns, and case-count summaries. This structural consistency enables direct cross-pathogen comparison and ensures that future updates (as literature accumulates or as PERG completes validation for additional pathogens) maintain compatibility with existing syntheses.

\begin{table}[h]
\centering
\caption{\textbf{Key artefacts in \name{} workflow generated living reviews.} Each pathogen generates two review types with consistent visualisation and evidence table structures. The text-based LLM  uses manifests, with summary statistics of the figures to write its interpretation.}
\label{tab:review_contents}
\small
\begin{tabular}{lp{0.38\linewidth}p{0.38\linewidth}}
\toprule
\textbf{Type} & \textbf{Transmission-Modelling Review} & \textbf{Outbreak Surveillance Review} \\
\midrule
\textbf{Figures} & 
Model architecture distribution (compartmental, branching process, agent-based); Stochasticity classification; Transmission route breakdown; Code availability 
& 
Geographic burden (choropleth maps for cases and deaths); Outbreak timeline by country; Detection mode distribution \\
\midrule
\textbf{Tables} & 
Model type counts and proportions; Deterministic vs stochastic breakdown; Transmission routes with sample sizes; Modelling assumptions; Intervention categories; Spatial scale indicators; Code availability and language 
& 
Outbreak source categories; Detection methodology breakdown; Ongoing outbreaks at extraction; Case burden stratified by confirmation status; Sex disaggregation where reported \\
\bottomrule
\end{tabular}
\end{table}

\clearpage

\clearpage
\section{The PERG Review Pipeline (Human Reference Workflow)}
\label{app:perg_pipeline}

The \emph{Pathogen Epidemiology Review Group (PERG)} is an expert-led effort (started in 2019) whose goal is to maintain a definitive, curated source of epidemiological parameters for pathogens prioritised for epidemic preparedness. In practice, PERG delivers this through systematic literature reviews and meta-analyses targeting the WHO priority pathogens, with the explicit aim of supporting outbreak response and modelling when time is short and parameter choices matter. 

The scope is defined by the WHO priority pathogens framing: diseases that ``pose the greatest public health risk due to their epidemic potential and/or whether there is no or insufficient countermeasures." Examples highlighted in PERG onboarding include CCHF virus, Ebola virus, Marburg virus, Lassa virus, Middle East respiratory syndrome coronavirus (MERS-CoV), Severe Acute Respiratory Syndrome coronavirus 1 (SARS-CoV-1), Nipah virus, Rift Valley fever, and Zika virus. 

PERG’s workflow is end-to-end: it starts from a protocolised literature search, then moves through screening (title \& abstract, then full text), structured extraction into REDCap (including quality-assessment fields guided by the PERG wiki), meta-analysis, and finally the write-up of a review that can be used by modellers and public health teams. 

\subsection*{Step 1: Paper search (protocol-driven, pathogen-specific)}
PERG begins from a registered systematic review protocol (PROSPERO ID:  CRD42023393345), and uses a standardised query template that is then tailored to each pathogen. The core idea is to search broadly across the epidemiological concepts that tend to matter during outbreak response: transmission and epidemiology terms, transmission modelling (with explicit exclusion of imaging-related “model” matches), severity outcomes (e.g. CFR), key delays (e.g. incubation period, serial interval, generation time), transmission heterogeneity and superspreading/overdispersion, transmissibility measures (e.g. growth rate and reproduction numbers), serology/serosurveys, evolutionary signals (mutation/substitution/evolution), outbreak/cluster terminology and risk factors. The query is written with wildcards to capture term variants, and then adjusted where needed to avoid cross-contamination with neighbouring literatures (for example, excluding SARS-CoV-2 when the target is SARS-CoV-1).


\subsection*{Step 2: Title and abstract screening (broad triage against explicit criteria)}
The first screening pass is based on titles and abstracts. The emphasis here is not on perfect specificity, but on ensuring the pool remains wide enough to avoid missing relevant evidence that is only clearly described later in the paper. PERG’s \textbf{inclusion criteria} are simple but concrete: studies must be English-language, peer-reviewed original research (systematic reviews and meta-analyses are flagged rather than treated as primary extraction targets), and must involve human data. A paper is kept if it contains \emph{any one} of several types of useful information, including: quantitative descriptions of a human outbreak (size, year, location, duration, spatial scale), a mathematical or statistical model of transmission, estimates of key transmission or timing quantities (e.g., $R$, $R_0$, $R_t$, growth rate, generation time, serial interval, incubation or latent period, other delays), severity metrics (CFR, attack rate), evolutionary rates, overdispersion/superspreading, risk factors (together with the measure), seroprevalence, relative contributions of human-to-human vs zoonotic transmission, and, where relevant, vector-related quantities such as mosquito delays or mosquito reproduction numbers.

PERG’s \textbf{exclusion criteria} are equally explicit: non-English items; posters, conference proceedings, correspondence, and abstract-only records; in-vitro-only studies; solely animal studies (unless the paper provides clearly relevant transmission quantities); and small case studies with fewer than 10 cases.

\subsection*{Step 3: Full-text review (confirm ``extractability")}
Articles passing abstract screening move to full-text review. PERG applies the same conceptual criteria, but with a different mindset: reviewers scan the entire paper to confirm that there is \emph{something extractable}, i.e. not just that the topic is on-target. Importantly, PERG explicitly runs both title and abstract screening and this stage with \textbf{two reviewers}, reflecting the goal of consistency and defensible inclusion decisions when judgement calls are required.

\subsection*{Step 4: Parameter extraction (read, highlight, enter structured fields)}
Once a paper is included, PERG's extraction process is deliberately hands-on. Reviewers (i) check which papers they have been assigned, (ii) download and read the PDF, highlighting everything they may want to extract as they go, and then (iii) enter the extracted information into a REDCap web database (PERG maintains pathogen-specific REDCap projects).

PERG structures extraction into four broad blocks:
\vspace{-10pt}
\begin{itemize}[leftmargin=*, itemsep=-3pt]
    \item  \textit{Article metadata.} Basic bibliographic information such as title, DOI, journal, and related identifiers are recorded. 
    \item  \textit{What the paper contains:} outbreaks, models, parameters. PERG extracts (i) outbreak descriptions where present, (ii) mathematical models of transmission (these are not limited to SIR-type models; they can be theoretical and not necessarily fitted to data), and (iii) epidemiological parameter estimates. Parameter families include genomic/evolutionary quantities (mutation/evolution rates), reproduction numbers ($R_0$, $R_t$, and human-only or vector-related variants where relevant), human delays (serial interval, incubation period, time-to-death, etc.), severity (CFR/IFR), seroprevalence (e.g. IgG/IgM markers), risk factors (with attention to whether effects are statistically significant and adjusted), relative contributions (human-to-human vs animal-to-human), attack rates (including secondary attack rates), and overdispersion (e.g. the negative binomial $k$ parameter). 
    \item \textit{Associated context for interpretation.} PERG captures the contextual details that make parameter estimates comparable (or not): sex (male/female/both/unspecified), sample size, setting (general population vs hospital), subgroup (children, pregnant, etc.), age ranges, country and more specific location, study start/end dates, and whether the study was conducted before/mid/after an outbreak.
    \item \textit{Structured outbreak fields (when applicable)}. In addition to “is there an outbreak?”, PERG-style extraction treats outbreaks as structured entities. In our draft’s PERG-aligned outbreak guidance, outbreak characteristics include temporal bounds (start/end day/month/year; whether ongoing), geographic scope (country plus sub-location), outbreak source, mode of detection, case definition method, case counts by confirmation status (confirmed/probable/suspected/unspecified), asymptomatic and severe cases when reported, deaths, and (when available) demographic breakdown such as sex-disaggregated counts. A key principle is that these values are extracted \emph{as stated in the paper}, without calculating missing quantities or inferring unreported fields.
\end{itemize}

Across all extraction types, PERG points reviewers to the PERG wiki for “how to extract this specific thing” guidance, so that extraction decisions remain consistent across pathogens and across reviewers.


\subsection*{Step 5: Meta-analysis}
After extraction and quality assessment, PERG moves into synthesis and reporting. PERG maintains shared tooling for priority pathogens, including codebases that step through cleaning the extracted database, transforming quantities into a common format where needed, performing meta-analysis, and producing plots and summary tables. These outputs feed directly into the final PERG systematic review and meta-analysis write-up.

\subsection*{Step 6: Write-up}
The final stage is to turn the extracted REDCap database and the meta-analysis outputs into a PERG review that can be used in practice. In PERG, meta-analysis is implemented through shared, pathogen-focused tooling (the \texttt{priority-pathogens} and \texttt{epireview} codebases), which steps through cleaning and transforming the extracted data, running the statistical synthesis, and producing the figures and summary tables. These tables and plots then provide the backbone of the manuscript: the review documents what evidence was found for each parameter family (and in what contexts), presents the quantitative summaries produced by the meta-analysis, and translates them into a curated resource for outbreak modelling and public health decision-making. In PERG’s framing, this write-up is not just a paper draft: it is the mechanism by which extracted parameters become a stable, citable reference for the WHO priority pathogens, with the longer-term aim of supporting an evolving “live” resource as evidence accumulates.


\clearpage
\section{\name{} Annotation Tool (Beta)}
\label{appendix:annotation_tool}
This section documents the \name{} annotation and validation interface, a beta-stage prototype designed to facilitate systematic literature reviews (SLRs) through the integration of LLM-assisted information extraction and expert-led verification. This would  also allow evaluation of web and computer-use agents for SLRs beyond epidemiology.

\begin{figure}[h]
    \centering
    \includegraphics[width=0.95\textwidth]{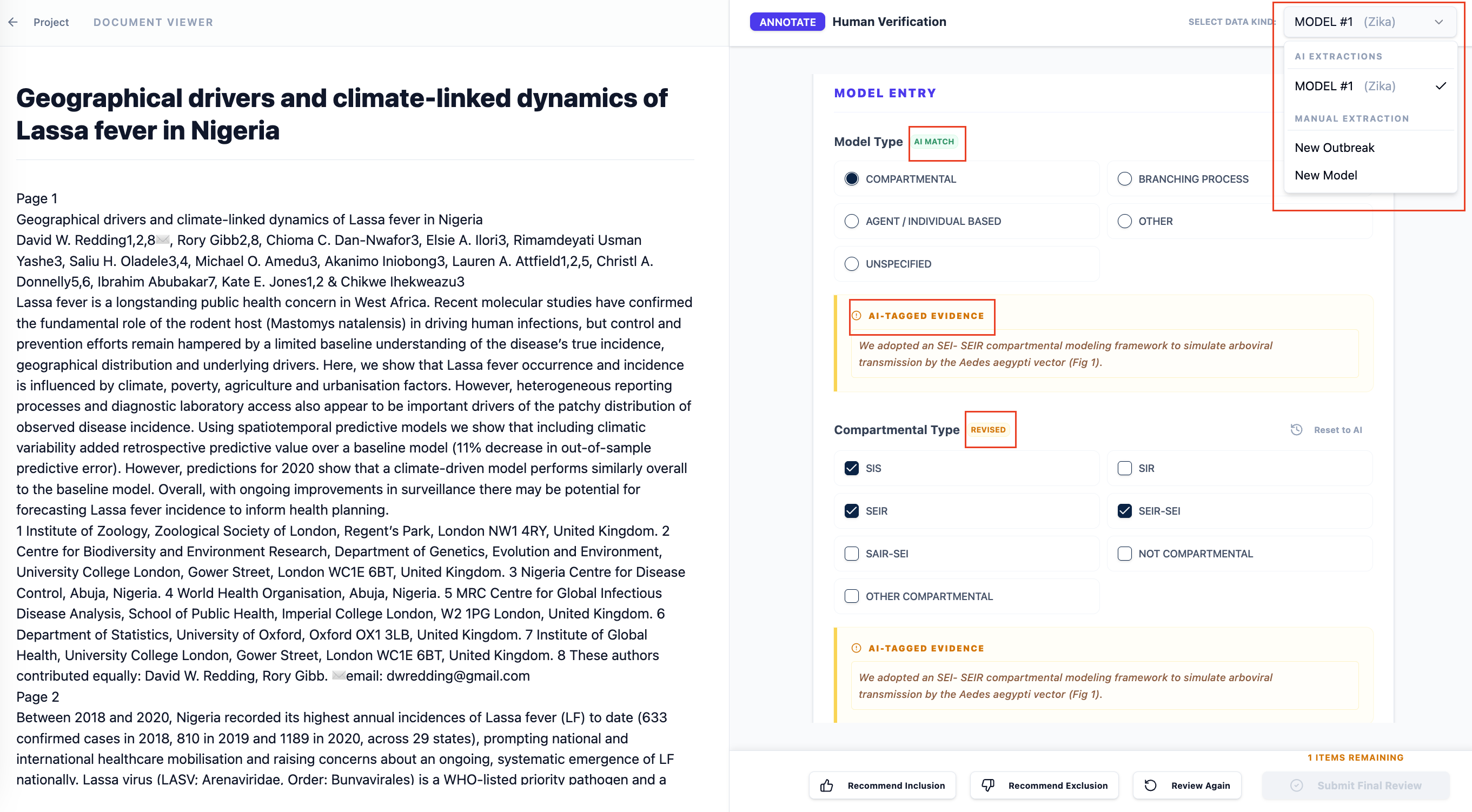}
    \caption{\textbf{The \name{} extraction and validation tool interface.} The dual-pane view presents the source document (\textbf{left}) alongside structured extraction fields (\textbf{right}). AI-predicted entries are pre-filled and accompanied by highlighted, AI-tagged evidence excerpts from the manuscript. Reviewers can accept, revise, or reject individual fields, with validation status indicators (e.g. ``AI Match'' or ``Revised'') reflecting whether human intervention was required.}
    \label{fig:ann_main}
\end{figure}

\paragraph{System Architecture and Core Functionality}
The \name{} annotation tool provides an interactive environment for technical validation of automated information extractions. The system utilises data gathered by LLMs equipped with structured tool-calling to identify and parse epidemiological parameters, transmission models, and outbreak characteristics. To ensure transparency and auditability, the tool implements a provenance layer that maps every extracted value to specific textual excerpts (AI-tagged evidence) within the source article.

\paragraph{User Interface Design}
The interface is optimised for high-throughput expert review via a dual-panel architecture:
\vspace{-10pt}
\begin{itemize}[leftmargin=*, itemsep=-3pt]
    \item \textit{Document Viewer (Left Panel):} Provides the original article text or rendered PDF, ensuring reviewers can verify the context of any extracted data point without context switching.
    \item \textit{Verification Interface (Right Panel):} Displays a form-based view of pre-filled fields generated by the \name{} pipeline. The extraction schema is dynamic, adapting based on the identified content type, such as compartmental model variables or spatio-temporal outbreak data.
\end{itemize}

\subsection{Human-in-the-Loop Validation}
The framework enforces a human-in-the-loop (HITL) protocol where automated extractions are audited by subject matter experts before being finalised for evidence synthesis. Within the interface, reviewers perform the following actions:
\vspace{-10pt}
\begin{itemize}[leftmargin=*, itemsep=-3pt]
    \item \textit{Verify:} Confirm the accuracy of the AI-extracted value and its linked evidence.
    \item \textit{Modify:} Correct extraction errors or refine data granularity; modified entries are flagged as ``Revised'' to facilitate system error analysis.
    \item \textit{Reject:} Entirely dismiss false positive extractions that do not meet inclusion criteria.
\end{itemize}

\subsection{Current Status and Field Testing}
The tool is currently in a beta development phase, with pilot testing conducted by epidemiologists focusing on WHO priority pathogens. While the current pilot utilises epidemiology-specific schemas, the architecture is designed to be domain-agnostic and can be adapted through schema reconfiguration and expert consultation.

Planned field testing is aligned with the Pathogen Epidemiology Review Group (PERG) workflow for remaining priority pathogens. This testing will utilise standardised extraction schemas for parameter, model, and outbreak data, and specifically target  systematic reviews for CCHF virus and Rift Valley fever virus.

\begin{figure}[t]
    \centering
    \begin{minipage}[b]{0.48\textwidth}
        \centering
        \includegraphics[height=5.7cm,keepaspectratio]{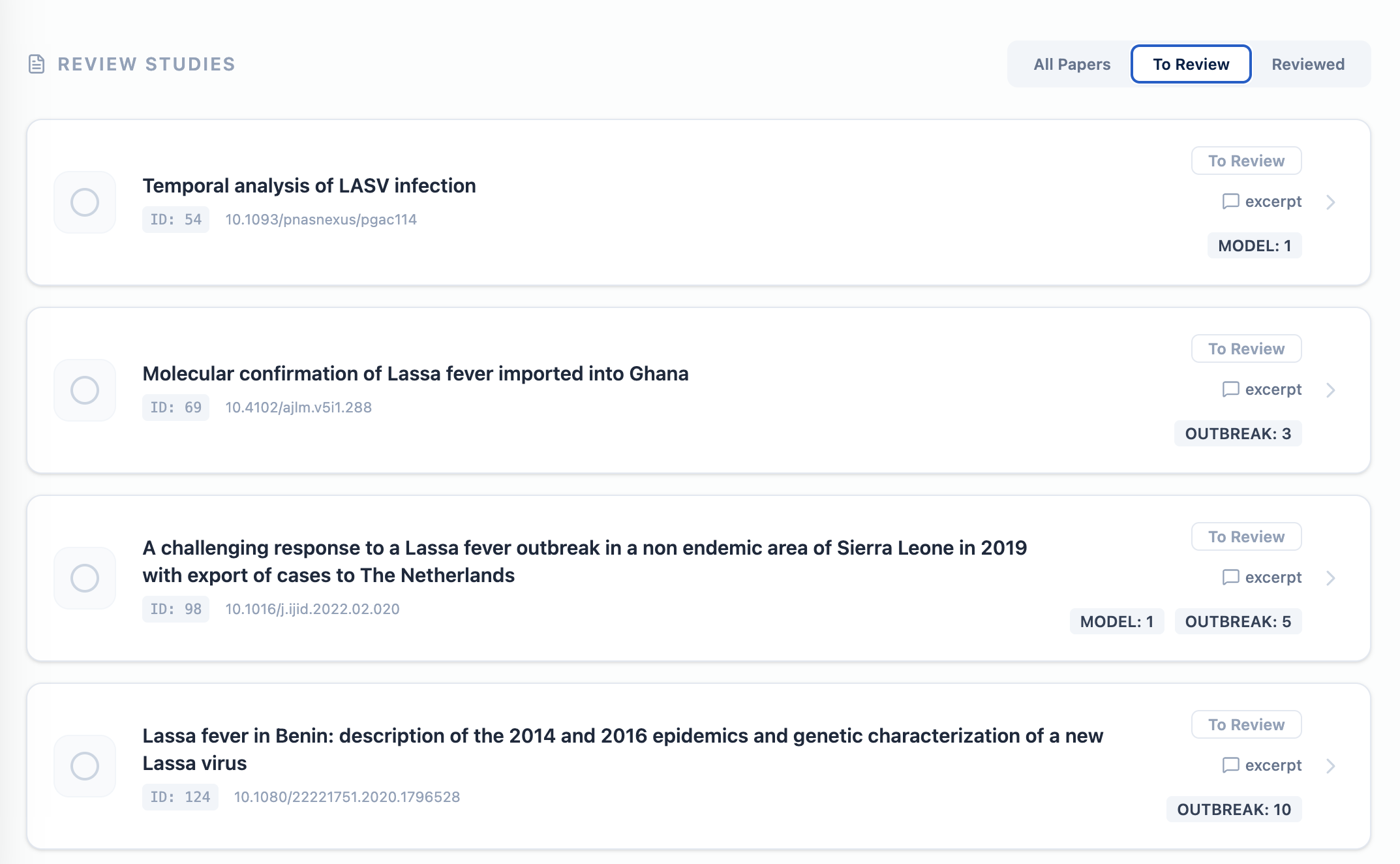}
        \caption*{(a) Tool Management Dashboard}
    \end{minipage}\hfill
    \begin{minipage}[b]{0.48\textwidth}
        \centering
        \includegraphics[height=5.2cm,keepaspectratio]{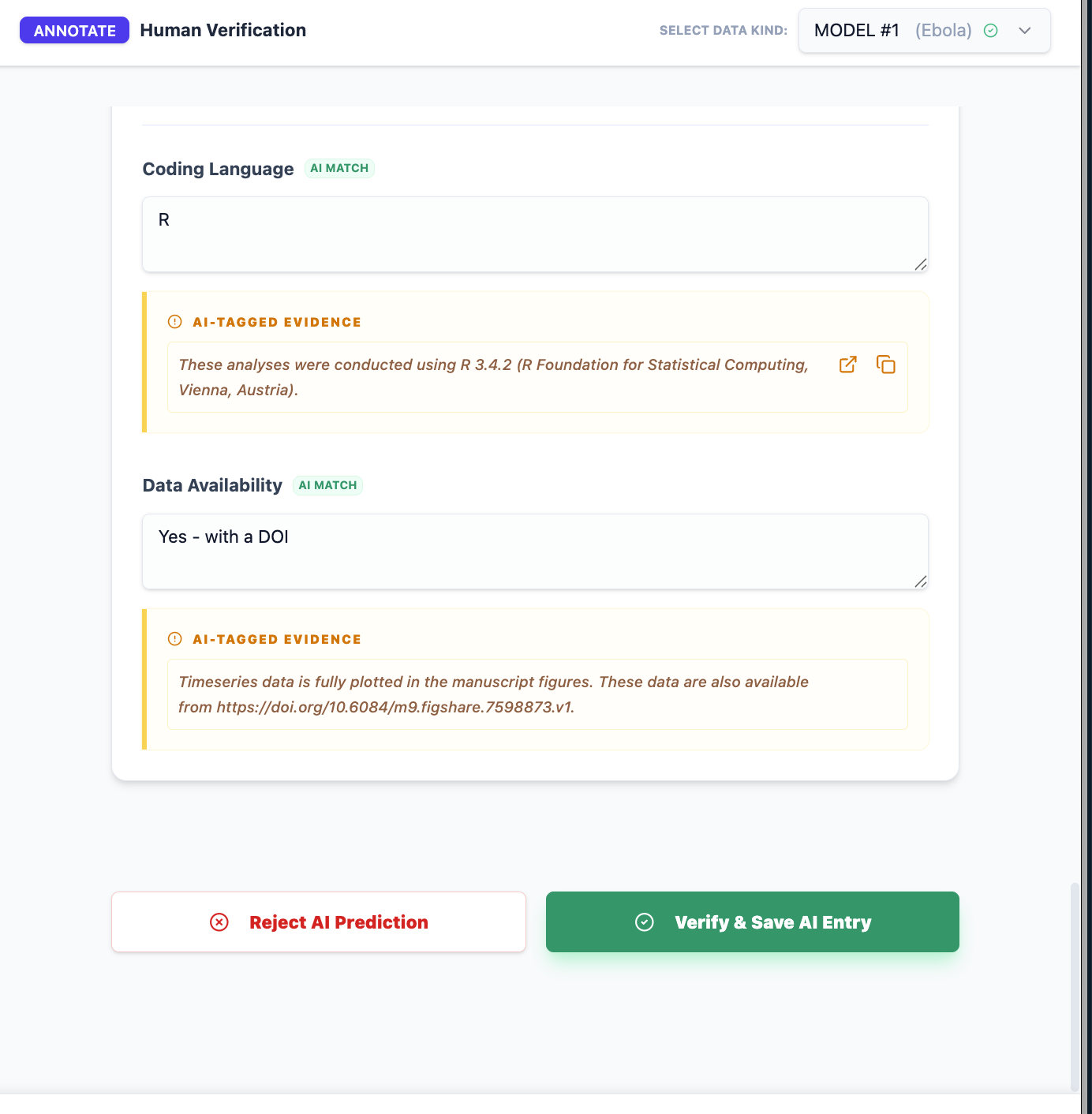}
        \caption*{(b) Verification and Submission}
    \end{minipage}
    \caption{\textbf{Review management and submission tracking interfaces.} (a) The study review list displays papers awaiting expert validation, along with associated model and outbreak counts and direct links to extracted excerpts. (b) The verification view presents finalised AI-assisted extractions, enabling reviewers to explicitly reject predictions or verify and save corrected entries, which are then recorded for downstream quality control and system evaluation.}
    \label{fig:ann_logistics}
\end{figure}

\subsection{Transparency and Reproducibility}
By maintaining a persistent link between the structured database and the source text, \name{} tool ensures that synthesised reports can be fully disaggregated. This audit trail is critical for scientific reproducibility, allowing researchers to trace every reported parameter, model and outbreak back to its exact location in the primary literature.

\clearpage

\end{document}